\documentclass{article}
\usepackage[a4paper, left=1in, right=1in, top=1in, bottom=1in]{geometry}
\usepackage{graphicx} 
\usepackage{hyperref}
\usepackage{amsmath}
\usepackage{amssymb}
\usepackage{listings}
\usepackage{tcolorbox}
\usepackage{xcolor}
\usepackage{booktabs}
\usepackage{braket}
\usepackage{graphicx}
\usepackage{subcaption}
\usepackage{float}            
\usepackage{booktabs}
\usepackage{array}
\usepackage{enumitem}
\usepackage{multirow}
\usepackage{colortbl}
\usepackage{tikz}
\usepackage{pgfplots}
\usepackage{tabularx}
\usepackage{cite}
\usepackage{authblk}
\usepackage{url}
\usepackage{doi}

\usetikzlibrary{patterns}

\definecolor{codegreen}{rgb}{0,0.6,0}
\definecolor{codegray}{rgb}{0.5,0.5,0.5}
\definecolor{codepurple}{rgb}{0.58,0,0.82}
\definecolor{backcolour}{rgb}{0.95,0.95,0.92}

\lstdefinestyle{mystyle}{
  backgroundcolor=\color{backcolour}, commentstyle=\color{codegreen},
  keywordstyle=\color{magenta},
  numberstyle=\tiny\color{codegray},
  stringstyle=\color{codepurple},
  basicstyle=\ttfamily\footnotesize,
  breakatwhitespace=false,         
  breaklines=true,                 
  captionpos=b,                    
  keepspaces=true,                 
  numbers=left,                    
  numbersep=5pt,                  
  showspaces=false,                
  showstringspaces=false,
  showtabs=false,                  
  tabsize=2
}
\title{Establishing Baselines for Photonic Quantum Machine Learning: Insights from an Open, Collaborative Initiative}

\begin{document}

\author[1]{Cassandre Notton\thanks{cassandre.notton@quandela.com}}
\author[2]{Vassilis Apostolou}
\author[3]{Agathe Senellart}
\author[2]{Anthony Walsh}
\author[2]{Daphne Wang}


\author[12]{Yichen Xie}

\author[7]{Songqinghao Yang}

\author[5,6]{Ilyass Mejdoub}
\author[4]{Oussama Zouhry}

\author[9,10]{Kuan-Cheng Chen}
\author[11]{Chen-Yu Liu}

\author[13]{Ankit Sharma}
\author[14]{Edara Yaswanth Balaji}
\author[15]{Soham Prithviraj Pawar}

\author[8]{Ludovic Le Frioux}
\author[8]{Valentin Macheret}
\author[8]{Antoine Radet}

\author[24]{Valentin Deumier}

\author[20]{Ashesh Kumar Gupta}
\author[22]{Gabriele Intoccia}
\author[19]{Dimitri Jordan Kenne}
\author[18]{Chiara Marullo}
\author[18]{Giovanni Massafra}
\author[21]{Nicolas Reinaldet}
\author[23]{Vincenzo Schiano Di Cola}

\author[16, 17]{Danylo Kolesnyk}
\author[16]{Yelyzaveta Vodovozova}

\author[2]{Rawad Mezher}
\author[2]{Pierre-Emmanuel Emeriau}
\author[2]{Alexia Salavrakos}
\author[2]{Jean Senellart}

\affil[1]{Quandela Quantique Inc., Montréal
, Canada}
\affil[2]{Quandela SAS, Massy, France}
\affil[3]{Universite Paris Cité, INRIA, Inserm, HeKA, Paris, France}
\affil[4]{Ecole Polytechnique, Palaiseau, France}
\affil[5]{Télécom Paris, Palaiseau, France}
\affil[6]{ENS Paris Saclay, Gif-sur-Yvette, France}
\affil[7]{Cavendish Lab., Department of Physics, University of Cambridge, Cambridge
, UK}
\affil[8]{Scaleway, Paris, France}
\affil[9]{QuEST, Imperial College, London, United Kingdom}
\affil[10]{I-X Centre for AI in Science, Imperial College London, London, United Kingdom}
\affil[11]{National Taiwan University, Taipei, Taiwan}
\affil[12]{La Salle College, Hong Kong}
\affil[13]{University of Delhi, New Delhi, India}
\affil[14]{Indian Institute of Technology Hyderabad, Telangana, India }
\affil[15]{International Institute of Information Technology, Bangalore, India}
\affil[16]{Technical University of Munich, Garching, Germany}
\affil[17]{Ludwig Maximilian University, Munich, Germany}
\affil[18]{ICAR–CNR
, Naples, Italy}
\affil[19]{Department of Mathematics and Physics, University of Campania
, Caserta, Italy}
\affil[20]{Dipartimento Interuniversitario di Fisica
, Università degli Studi di Bari Aldo Moro, Bari, Italy}
\affil[21]{Invent Vision, Brazil}
\affil[22]{Department of Mathematics and Applications
, University of Naples Federico II, Naples, Italy}
\affil[23]{Quantum2Pi Srl, Naples, Italy}
\affil[24]{Ecole des Mines de Paris - PSL, Paris, France}

\maketitle
\noindent

\begin{abstract}
The Perceval Challenge is an open, reproducible benchmark designed to assess the potential of photonic quantum computing for machine learning. Focusing on a reduced and hardware-feasible version of the MNIST digit classification task or near-term photonic processors, it offers a concrete framework to evaluate how photonic quantum circuits learn and generalize from limited data.
Conducted over more than three months, the challenge attracted 64 teams worldwide in its first phase. After an initial selection, 11 finalist teams were granted access to GPU resources for large-scale simulation and photonic hardware execution through cloud service. The results establish the first unified baseline of photonic machine-learning performance, revealing complementary strengths between variational, hardware-native, and hybrid approaches. This challenge also underscores the importance of open, reproducible experimentation and interdisciplinary collaboration, highlighting how shared benchmarks can accelerate progress in quantum-enhanced learning.
All implementations are publicly available in a single shared repository\footnote{\url{https://github.com/Quandela/HybridAIQuantum-Challenge}}, supporting transparent benchmarking and cumulative research. Beyond this specific task, the Perceval Challenge illustrates how systematic, collaborative experimentation can map the current landscape of photonic quantum machine learning and pave the way toward hybrid, quantum-augmented AI workflows.
\end{abstract}

\section{Introduction} 
Early research on Quantum Machine Learning (QML) centered on algorithms built from well-known quantum subroutines such as quantum phase estimation, aiming to demonstrate provable speedups over classical methods \cite{Harrow_2009, Lloyd_2014}. These approaches, however, assumed access to fault-tolerant quantum computers, which remain under development. With the advent of the NISQ era \cite{Preskill_2018}, and inspired by the widespread success of neural networks in classical machine learning, the QML community began shifting its interest towards more practical, hardware-compatible strategies, such as variational quantum algorithms \cite{Cerezo_2021} and quantum kernel methods \cite{ld2021supervisedquantummachinelearning, Havl_ek_2019}. 

As the field matured, a countercurrent emerged \cite{Wiebe_2020, Schuld_2022}. Rigorous benchmarking studies revealed that many widely cited QML models, even after extensive hyperparameter tuning, did not outperform classical baselines on standard tasks \cite{ bowles2024betterclassicalsubtleart, angrisani2024classically}. Subsequent efforts have extended this line of work to other domains, such as time-series prediction \cite{fellner2025quantumvsclassicalcomprehensive}. These findings recalibrated expectations and underscored the need for systematic evaluation methodologies, especially as models grow too large to be simulated classically. In precisely these regimes, where claims of quantum advantage are most compelling, rigorous benchmarking becomes the most challenging.

Alongside this methodological shift, advances in experimental platforms---including demonstrations of algorithms on superconducting devices \cite{Havl_ek_2019, Coyle_2021}, neutral atoms \cite{Albrecht_2023, kornjača2024largescalequantumreservoirlearning}, trapped ions \cite{Zhu2019, Suzuki2024}, and photonic platforms \cite{Saggio_2021, Peruzzo_2014}---have sparked interest in tailoring algorithms to hardware-specific constraints. On the one hand, noise, limited scale, and compilation challenges make hardware-aware design essential, with error mitigation strategies \cite{Endo_2018, MezherMills2024} and hybrid quantum–classical workflows emerging as standard tools. On the other hand, working directly with experimental devices also invites algorithm design around native quantum primitives—operations and encodings that arise naturally in a given platform—rather than imposing abstractions better suited to idealized fault-tolerant models. This dual motivation has given rise to a growing body of hardware-adapted QML approaches.


Photonics provides a particularly compelling testbed. Quandela recently introduced a linear-optical quantum processor \cite{Maring_2024} and its companion simulation framework, \emph{Perceval} \cite{Heurtel_2023}, enabling both cloud-based access and algorithm development. Beyond qubit encodings, linear optics can serve as a standalone computational paradigm, motivating the design of “photon-native” algorithms \cite{gan2022fock, Salavrakos_2025, mezher2023solving, Sedrakyan2024, Salavrakos_2025_2, Yin2024, Hoch_2025, lysaght2024quantumcircuitcompressionusing}. This creates an opportunity to explore QML models that are not only hardware-compatible, but hardware-driven.

In classical machine learning, progress has often been accelerated by open competitions and shared benchmarks, such as the Netflix Prize \cite{bennett2007netflix}, as well as challenges hosted on the platforms like Kaggle \cite{kaggle2025}, which provide datasets and leaderboards for broad community participation. Motivated by this tradition, we organized the {\bf Perceval Quest} \cite{perc_quest_repo}: a six-month-long hackathon dedicated to exploring QML within the framework of linear optics. We believe such events encourage innovation, broaden participation beyond the quantum community, and promote interdisciplinary collaboration with the wider AI ecosystem. Similar initiatives, like the 2024 Airbus–BMW Quantum Computing Challenge  \cite{airbus2024quantum} demonstrate how benchmarking competitions can accelerate innovation across quantum technologies.

To ground our challenge in a familiar benchmark, we chose the iconic MNIST dataset \cite{deng2012mnist}. MNIST has long served as a proving ground for computer vision models and provides a well-understood reference point for comparing quantum and classical approaches. Participants in the Perceval Quest were asked: \textit{what kinds of models can be built using linear optics and the \emph{Perceval} framework, and how do they compare to classical solutions---not only in terms of accuracy, but also in number of parameters and convergence speed?}

\vspace{0.3em}

\noindent The main contributions of this work are as follows:
\begin{itemize}
    \item Provide a {\bf systematic review} of the diverse approaches explored throughout the \emph{Perceval Quest}, organizing them into coherent methodological categories and identifying common design patterns: (i) photonic kernels, neural networks, and convolutional models, where the interferometer functions as an end-to-end feature extractor; (ii) enhanced CNNs and hybrid feature extractors, where it operates as a quantum annotator; and (iii) transfer learning and self-supervised learning (SSL) paradigms, where it supports model fine-tuning. We also propose a novel method that exploits the computational properties of photonic quantum processors through permanent-based computation.

    \item Highlight the {\bf migration of methods} from non-photonic QML paradigms to photonic implementations, noting that several of these ideas were developed further and submitted as independent articles \cite{chen2025distributed, xie2025quantum};
    \item Place a strong emphasis on reproducibility, consolidating all implementations into a unified open-source repository \cite{perc_quest_repo} to ensure transparency and facilitate further research;
    \item Establish benchmarking practices by comparing photonic models against one another and, whenever possible, against classical baselines—not only in terms of accuracy, but also model size, parameter efficiency, and convergence speed.
\end{itemize}

\noindent Although no clear heuristic quantum advantage was observed at this stage, the challenge provides a systematic foundation through reproducible code, benchmarking protocols, and diverse algorithmic strategies, paving the way for more decisive tests as photonic hardware matures.


\section{Problem definition and classical solutions} 
\label{sec:mnist_and_classical}

As outlined above, the participants were asked to design algorithms that integrate machine learning with linear-optical quantum computing to classify the MNIST dataset, making use of the Perceval framework to simulate the quantum components and interface with available hardware. The MNIST dataset \cite{deng2012mnist} consists of black-and-white images of handwritten digits, normalized to fit within a $28 \times 28$ pixel box. The dataset contains 70{,}000 images in total: 60{,}000 for training and 10{,}000 for testing. 

MNIST has played a crucial role in the development and validation of computer vision models, from traditional machine learning techniques to modern deep neural networks. State-of-the-art convolutional neural networks have achieved near-perfect performance, with test errors as low as $0.6\%$, while simpler architectures, such as multilayer perceptrons with ReLU activations, report errors around $1.1\%$ \cite{hinton2022forward}.

This historical role makes MNIST a natural candidate for benchmarking QML approaches. However, several important caveats must be considered. One of the main challenges is the relatively high dimensionality of the dataset—each image has 784 features—which exceeds the effective dimensionality accessible to most current quantum hardware and simulators (whether measured in qubits, modes, or feasible circuit depth). As noted in \cite{bowles2024betterclassicalsubtleart}, most QML models using MNIST as a benchmark therefore apply classical preprocessing techniques, such as PCA, to reduce the dimensionality of the dataset. Another common simplification is to restrict the dataset to only two out of the ten digits \cite{hur2022quantum}, thereby reducing the task to binary classification, which is substantially less complex than the full multi-class problem. Such simplifications make it difficult to draw meaningful conclusions about the specific role of the quantum component or the intrinsic value of QML models. For this reason, in the present challenge we opted for a reduced version of the dataset that retains all ten classes and the full image resolution, while limiting the number of samples to make the task more challenging for classical models. Participants could still apply classical preprocessing techniques but were required to provide detailed comparisons with classical machine learning models trained and evaluated on the same dataset.

Moreover, participants were encouraged to conduct ablation studies to systematically assess the impact of the quantum components within their approaches. To ensure fair evaluation, a classical benchmark model was provided as a reference point. This benchmark, based on a convolutional neural network (CNN), achieved a test error of approximately $3\%$—a reasonable baseline given the reduced size of the challenge dataset. Together, these requirements emphasized reproducibility and rigorous benchmarking, enabling meaningful comparisons between photonic and classical models.


To push the boundaries of current technological capabilities, participants were provided with access to high-performance computing resources. These included powerful graphics processing units (GPUs) capable of handling large-scale simulations via Scaleway’s Quantum-as-a-Service platform \cite{scaleway}, as well as access to Quandela’s Quantum Processing Units (QPUs) through its cloud platform \cite{quandela_cloud}. This infrastructure ensured that participants could explore models beyond the limits of standard hardware, thereby enabling stress test of both algorithms and simulation frameworks.


\section{Related work} 
\noindent {\bf MNIST as a benchmark.} 
The MNIST dataset has long been one of the most widely used testbeds in machine learning research due to its simplicity, accessibility, and well-understood properties~\cite{lecun1998mnist, deng2012mnist}. 
It provides a balanced classification task that is neither trivial nor excessively complex, making it an attractive reference point for methodological comparisons across decades of work. 
A further advantage is that MNIST lends itself to \emph{bidirectional complexity shaping}. Reducing the dataset can make classification easier (e.g., using well-separated binary pairs), but it can also create harder problems when the chosen digits are visually confusable (such as 3/5 or 4/9). Similarly, dimensionality reduction with PCA may improve efficiency at moderate levels, yet aggressive compression can remove discriminative structure and degrade accuracy, while expanded encodings push the problem into higher-dimensional regimes. Community variants extend this idea by explicitly increasing task difficulty through rotations, affine distortions, or clutter (Rotated/Cluttered MNIST, affNIST, MNIST-C), and through drop-in replacements that are empirically more challenging (Fashion-MNIST, EMNIST, Kuzushiji-MNIST). These knobs make MNIST a flexible benchmark that can be tuned both below and beyond its original complexity, allowing researchers to systematically probe learning models under controlled variations~\cite{hinton2006reducing,sutskever2011generating,jaderberg2015spatial,tieleman2013affnist,mu2019mnistc,
xiao2017fashionmnist,cohen2017emnist,clanuwat2018deep}.
This adaptability has allowed MNIST to play a central role in systematically probing learning architectures in a controlled, progressive manner, culminating in landmark results such as the multi-column deep neural networks of Cire{\c{s}}an et al.~\cite{ciregan2012multi}, which were among the first to achieve near-human performance.

\begin{figure}[ht]
    \centering
    \begin{tikzpicture}
    \begin{axis}[
      width=15cm, height=8cm,
      ybar stacked,
      bar width=14pt,
      ymin=0,
      xlabel={Year}, ylabel={Number of Papers},
      xtick=data,
      x tick label style={/pgf/number format/1000 sep=},
      enlarge x limits=0.06,
      grid=major,
      legend style={
        at={(0.02,0.98)}, 
        anchor=north west,
        draw=none, fill=white, 
        fill opacity=0.7,
        font=\small
      },
      legend cell align={left}
    ]

    \pgfplotstableread[row sep=\\]{
    year gate anneal photonic qinsp analog nonqml \\
    2015  0    1      0        0     0      0 \\
    2016  0    0      0        0     0      0 \\
    2017  0    1      0        1     0      0 \\
    2018  2    3      0        1     0      0 \\
    2019  6    6      1        2     0      0 \\
    2020 11    2      2        4     0      2 \\
    2021 12    1      1        2     0      0 \\
    2022 28    0      1        3     0      0 \\
    2023 24    4      1        1     1      2 \\
    2024 43    1      2        1     0      1 \\
    2025 54    3      6        3     3      1 \\
    }\datatable

    \addplot+[draw=none, fill=blue!60]    table[x=year,y=gate]     {\datatable}; \addlegendentry{Gate-based}
    \addplot+[draw=none, fill=orange!70]  table[x=year,y=photonic] {\datatable}; \addlegendentry{Photonic}
    \addplot+[draw=none, fill=green!60]   table[x=year,y=anneal]   {\datatable}; \addlegendentry{Annealing}
    \addplot+[draw=none, fill=cyan!60]    table[x=year,y=qinsp]    {\datatable}; \addlegendentry{Quantum-inspired}
    \addplot+[draw=none, fill=purple!50]  table[x=year,y=analog]   {\datatable}; \addlegendentry{Analogic}
    \addplot+[draw=red,   fill=red!12, pattern=north east lines, pattern color=red!50]
            table[x=year,y=nonqml]{\datatable}; \addlegendentry{Non-QML (excluded)}

    \end{axis}
    \end{tikzpicture}

    \caption{Number of QML+MNIST papers per year, broken down by modality. There is a clear increasing trend in the number of publications that include the terms MNIST and quantum machine learning on their title or abstract over the years. This rise reflects the adoption of MNIST as a benchmark for testing novel quantum approaches.}
    \label{fig:qml_mnist_stack}
\end{figure}
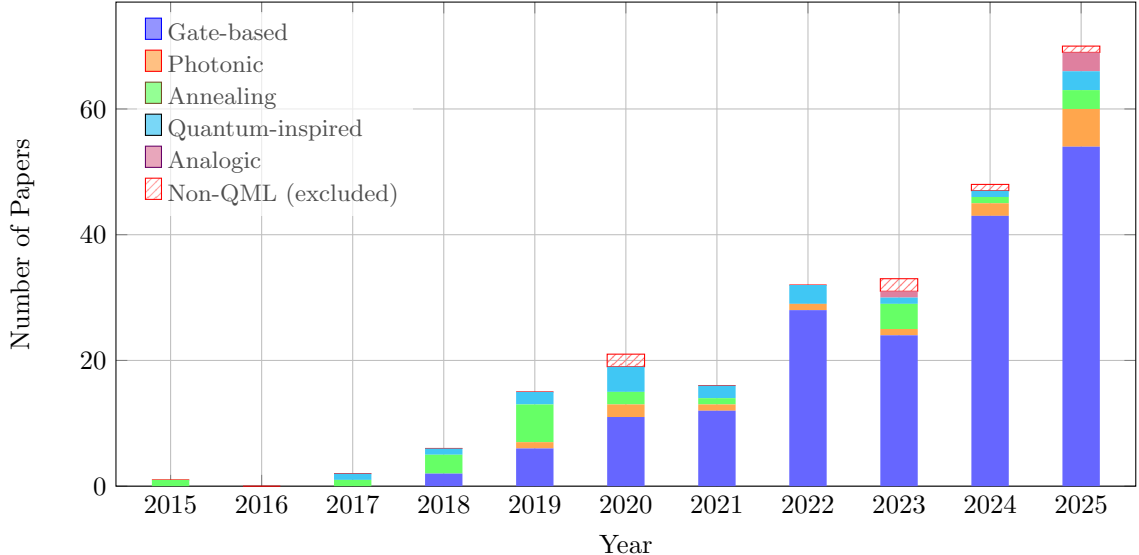

\vspace{0.15\baselineskip}
\noindent  {\bf Landscape of QML+MNIST}. To contextualize our study, we assembled a keyword-filtered snapshot of arXiv papers that mention both quantum machine learning and MNIST in the title or abstract ($n=244$, 2015–2025). Figure \ref{fig:qml_mnist_stack} shows the temporal trend, broken down by modality: gate-based approaches dominate, with smaller but sustained activity in annealing, photonic, and quantum-inspired models. Table \ref{tab:qml_mnist_summary} quantifies this distribution and adds further indicators such as binary vs.\ multiclass usage, code availability, and hardware execution.

\vspace{0.15\baselineskip}
\noindent {\bf System dimensionality}. Table~\ref{tab:qml_mnist_summary} also reports the range and variability of input dimensionalities used in QML+MNIST studies. While raw MNIST has 784 features, most works operate on substantially reduced embeddings, with mean dimensionalities of only 30–100 across modalities. At the same time, some studies expand inputs into the thousand-dimensional regime (up to 3530 for gate-based and 1550 for photonic). The large variances (on the order of $10^4$–$10^5$ for gate-based and photonic) highlight the heterogeneity of preprocessing choices. This confirms that dimensionality reduction and encoding strategies are major uncontrolled variables in the literature, complicating fair comparison. Our challenge therefore standardizes the full 784-dimensional task while tracking parameter efficiency.

\vspace{0.15\baselineskip}
\noindent {\bf Photonic contributions}. Photonic approaches remain underrepresented in our snapshot (14 papers, $\sim$6\%), though they place stronger emphasis on multiclass classification (12/14) and exhibit above-average code availability (43\%). Most of these works are simulator-based, with only three reporting hardware-only results. Notably, a recent parallel effort by Sakurai \emph{et al.} (2025) introduces a boson-sampling–powered quantum optical reservoir computing model and applies it to MNIST, signaling growing interest in more sophisticated photonic-native methodologies \cite{sakurai2025quantum}. This trajectory underscores both the opportunity and rising need for standardized photonic benchmarks.

\vspace{0.15\baselineskip}
\noindent {\bf Positioning of the present work}. Existing studies mostly establish feasibility under reduced datasets and without consistent baselines or ablations. By contrast, our challenge is explicitly photonic-native, uses the full ten-class MNIST task (rather than downsampled or binary subsets), and evaluates not only accuracy but also parameter efficiency, FLOPs, and convergence speed. With all implementations released in a unified repository, we aim to provide a reproducible benchmark suite that directly addresses the gaps evident in Table~\ref{tab:qml_mnist_summary} and Figure~\ref{fig:qml_mnist_stack}.

\vspace{0.3cm}
\noindent The surveyed literature demonstrates both the breadth of modalities and the heterogeneity of experimental setups (dimensionality, task type, hardware vs.\ simulator). However, it also reveals a striking lack of consistency: most works rely on strong dataset simplifications, custom encodings, or narrow binary tasks, making cross-comparison difficult. This aligns with the broader concerns articulated by Schuld and Killoran~\cite{Schuld_2022}: the field often frames itself in terms of ``quantum advantage'' over classical ML, yet current tools, datasets, and hardware only support highly restricted experiments. They advocate shifting the emphasis away from outperforming classical ML toward model building, theoretical frameworks, and software infrastructure that prepare QML for realistic scales. Our present work follows this spirit. Rather than seeking immediate advantage, we aim to establish reproducible, photonic-native benchmarks on a full multiclass MNIST task—a step toward standardized evaluation practices that can ground future debates on expressivity, efficiency, and scalability.

\begin{table}[ht]
\centering
\begin{tabularx}{\textwidth}{lrrrrrr}
\toprule
 & Gate & {\bf Photonic} & Anneal & Q-inspired & Analog & Non-QML \\
\midrule
Number of papers      & 180 & {\bf 14} & 22 & 18 & 4 & 6 \\
\% of total           & 73.8\% & {\bf 5.7\%} & 9.0\% & 7.4\% & 1.6\% & 2.5\% \\
\midrule
Binary tasks (\#)     & 85  & {\bf 2}  & 5  & 1 & 2 & 0 \\
Multiclass (\#)       & 95  & {\bf 12} & 17 & 17 & 2 & 6 \\
\midrule
Code available (\#)   & 48  & {\bf 6}  & 3  & 3 & 0 & 2 \\
\% with code          & 27\% & {\bf 43\%} & 14\% & 17\% & 0\% & 33\% \\
\midrule
Hardware only (\#)    & 8   & {\bf 3}  & 9  & 4 & 1 & - \\
Simulator+HW (\#)     & 78  & {\bf 5}  & 8  & 1 & 1 & - \\
Simulator only (\#)   & 94  & {\bf 6}  & 5  & 13 & 2 & - \\
\midrule
Min.–Max. dimension   & 1–3530 & {\bf 1–1550} & 1–784 & 1–784 & 1–255 & 1–196 \\
Mean [Var OOM]        & 72 [10$^{4}$] & {\bf 104 [10$^{4}$]} & 34 [10$^{4}$] & 31 [10$^{3}$] & - & - \\
\bottomrule
\end{tabularx}
\caption{Summary of 244 arXiv papers (2015–2025) mentioning “QML+MNIST” in title/abstract. 
The reported dimensionality refers to the effective size of the quantum system used for encoding: 
for gate-based models this typically corresponds to number of qubits (and, where specified, expanded feature maps), 
for photonic models to the number of modes and/or photons, and for annealing or quantum-inspired models to effective variable counts. 
Values are those explicitly reported in the papers. Means and variances are computed per modality.}

\label{tab:qml_mnist_summary}
\end{table}

\section{Preliminaries: linear optical quantum computing}
\label{sec:preliminaries}


In quantum linear optics, information is encoded in the Fock states of photons distributed among spatial or temporal modes. For a system of $n$ photons in $m$ modes, the input state can be written as
$\lvert \vec{n}_{\text{in}} \rangle = \lvert n^{\text{in}}_1, n^{\text{in}}_2, \ldots, n^{\text{in}}_m \rangle$
where $n^{\text{in}}_i$ denotes the number of photons in mode $i$ and $\sum_i n^{\text{in}}_i = n$. The input state is propagated through an interferometer and then measured by photon number detectors (or threshold detectors). This yields a vector $\ \vec{n}_{\text{out}} = ( n^{\text{out}}_1, n^{\text{out}}_2, \ldots, n^{\text{out}}_m)$, which describes the arrangement of $n$ photons in $m$ modes, and where $\sum_i n^{\text{out}}_i = n$ in the absence of loss.

Transformations between input and output Fock states are governed by the evolution of the creation operators under the unitary describing the linear optical network. The fundamental gates in such a network are:
\begin{itemize}
    \item Phase shifters which are U(1) transformations acting on a single mode: $\hat{P}_{\phi} = \left( e^{i\phi} \right)$, where $\phi \in [0,2\pi]$,
    \item Beam splitters which are U(2) transformations acting on pairs of modes, described by\footnote{Note that other beam splitter conventions exist that can involve additional parameters.}: \[
\hat{U}_{\mathrm{BS}}(\theta) =
\begin{pmatrix}
    \cos\!\left(\tfrac{\theta}{2}\right) & i\,\sin\!\left(\tfrac{\theta}{2}\right) \\[4pt]
    i\,\sin\!\left(\tfrac{\theta}{2}\right) & \cos\!\left(\tfrac{\theta}{2}\right)
\end{pmatrix},\] where $\theta$ is related to reflectivity of the coupler.
\end{itemize}
Any arbitrary unitary transformation $U\in U(m)$ over the optical modes can be decomposed into a sequence of such beam splitters and phase shifters, following the triangular \cite{reck1994experimental} or rectangular \cite{clements2016optimal} decompositions, also known as Reck and Clements decompositions, respectively. The resulting interferometer is a universal linear-optical processor, in the sense that it can perform any linear-optical operation.

The probability of observing an output state $\vec{n}_{\text{out}}$ given input state $\lvert \vec{n}_{\text{in}} \rangle$ and unitary $U$ is given by:
\[
p(\vec{n}_{\text{out}}) = \frac{|\text{Perm}(U_{\vec{n}_{\text{in}} \rightarrow \vec{n}_{\text{out}}})|^2}{n^{\text{in}}_1 ! \cdots n^{\text{in}}_m ! \text{ } n^{\text{out}}_1 !  \cdots n^{\text{out}}_m !},
\]
where $U_{\vec{n}_{\text{in}} \rightarrow \vec{n}_{\text{out}}}$ is a submatrix of $U$ obtained by taking $n^{\text{in}}_i$ times the $i$th row of U, then $n^{\text{out}}_j$ times the $j$th row of that matrix.  $\text{Perm}(\cdot)$ denotes the matrix permanent, which is defined for a matrix $M$ as:
\[
\operatorname{Perm}(M) = \sum_{\sigma \in S_n} \prod_{i=1}^{n} M_{i, \sigma(i)}.
\]
The problem of sampling from such an output distribution corresponds to the definition of boson sampling which was proposed by Aaronson and Arkhipov \cite{aaronson2011}.

Quandela's current hardware is based on quantum linear optics. The Ascella QPU described in \cite{Maring_2024} consists of a deterministic single-photon source based on a quantum dot, followed by a demultiplexer, thus preparing input Fock states. They are sent to a chip corresponding to a 12-mode interferometer containing thermo-optic phase shifters and directional couplers, which enable reconfigurable unitary transformations. Output states are measured with SNSPDs. On top of that, an active stabilization and machine-learned transpilation is implemented to compensate for fabrication imperfections and phase shifts \cite{fyrillas2024scalable}. This architecture provides a fully programmable and controllable linear optical processor on which quantum circuits defined in Perceval can be executed natively.

At the software level, \texttt{Perceval} serves as the primary interface between algorithm design, numerical simulation, and hardware execution. It allows users to define photonic circuits through a high-level Python API, which autmatically translates them into optical networks composed of phase shifters and beam splitters. It supports circuit composition and visualization, as well as multiple simulation back-ends, i.e. algorithms optimized for different tasks given a specific input state: \texttt{CliffordClifford2017} efficiently samples individual single output states; \texttt{Naive} based on Ryser algorithm \cite{ryser1963combinatorial}, computes the probability or probability amplitude of obtaining a given output state; and \texttt{SLOS} and \text{Stepper} describe the exact complete output state either by evaluating the entire circuit or by evolving the quantum state incrementally through circuit gates. \texttt{Perceval} also support different types of photon detectors, and include noise modeling for photon loss, imperfect components, photon distinguishability, and single-photon purity, as well as hardware connectivity. For the purpose of this challenge, it was extended with ad-hoc gradient backpropagation capabilities, enabling integration into machine-learning workflows.

When designing an algorithm based on a programmable interferometer, the parameters will correspond to the phase shifts applied by the phase shifters and the mixing angles of the beam splitters. In a fully simulated model, both sets of parameters can be freely optimized in order to, for instance, minimize a task-dependent loss function. However, in practical photonic hardware, the beam splitters are usually implemented as fixed 50:50 couplers, while only the phase shifters are tunable. Consequently, the optimization effectively acts on the set of controllable phases which are sufficient to modulate the overall interferometer unitary and hence the output photon-count statistics.

The Quandela framework unifies a programmable photonic interferometer and the \texttt{Perceval} software library into a consistent environment for quantum machine-learning experiments.
Its photonic-native representation allows researchers to benchmark realistic quantum algorithms on both simulated and physical hardware. In the context of this challenge, this stack enabled the implementation and training of variational photonic circuits to classify MNIST digits.


\section{Model proposals and results}
\newcommand{\Proposal}{%
  \noindent\textbf{Proposal. }%
}
\newcommand{\Results}{%
  \vspace{0.15\baselineskip}
  \noindent\textbf{Results. }%
}
\newcommand{\RealQPUValidation}{%
  \vspace{0.15\baselineskip}
  \noindent\textbf{Real QPU Validation. }%
}

\begin{figure}[ht]
    \centering
    \includegraphics[width=0.8\linewidth]{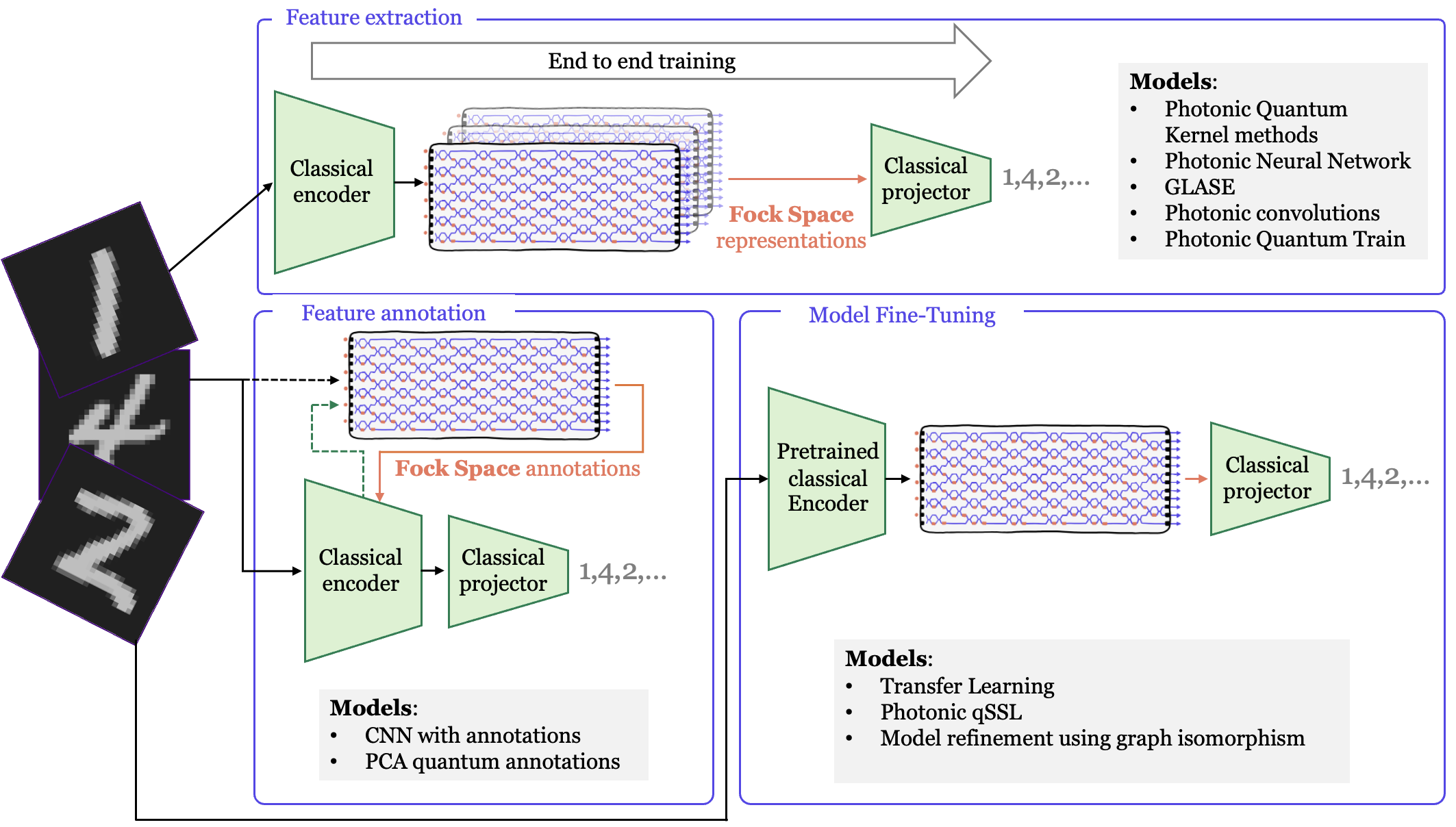}
    \caption{Three hybrid circuits trends observed in the challenge. When the photonic interferometer 
is used as a \textbf{feature extractor}, the model is trained end-to-end. When the photonic interferometer is used as a \textbf{feature annotator}, image or representations are passed through the encoder and their representations through the interferometer are fed as annotations to the encoder. In the case of \textbf{model fine-tuning}, a pretrained encoder is used and the photonic interferometer is used in the projection head, either for transfer learning, model refinement or self-supervised learning.}
    \label{fig:all_approaches}
\end{figure}

\label{sec:models}
In this section, we present thirteen methods developed by the participants, which can be grouped within three main approaches, as is shown in Figure \ref{fig:all_approaches}. In the first approach, models are trained end-to-end, utilizing the interferometer for feature extraction. In the second approach, models  employ the quantum interferometer for annotation purposes. In the third approach, the proposals use the interferometer for fine-tuning, either through transfer learning or for refining and correcting the model.

Most of the results stem from numerical simulations, both ideal (noise-free) and noisy. Additionally, several experiments were executed on hardware, more specifically on Quandela's QPUs accessible through a cloud platform~\cite{quandela_cloud}, and many simulations also leveraged GPUs through Scaleway’s Quantum-as-a-Service platform~\cite{scaleway}, as highlighted in Section \ref{sec:mnist_and_classical}. All quantum models are evaluated against classical baseline models, with comparisons made in terms of training and testing accuracy, number of trainable parameters, and computational cost measured in floating-point operations per second (FLOPS).

For each proposal, the details on interferometer architecture, data encoding, and hyperparameter values can be found in the Supplementary Materials. 

\subsection{Photonic interferometer for feature extraction}
For the first category of models, the photonic interferometer is used to extract meaningful features from the data: the photonic interferometer acts as a feature extractor. It is sometimes combined with a classical encoder to enhance performance. 
\subsubsection{A quantum kernel method}
\label{sec:qool_proposal}
\Proposal
This first model is a photonic quantum-kernel whose classical counterpart is a Support Vector Machine (SVM) \cite{burges1998tutorial}. 

\begin{figure}[ht!]
\centering
\includegraphics[width=0.5\linewidth]{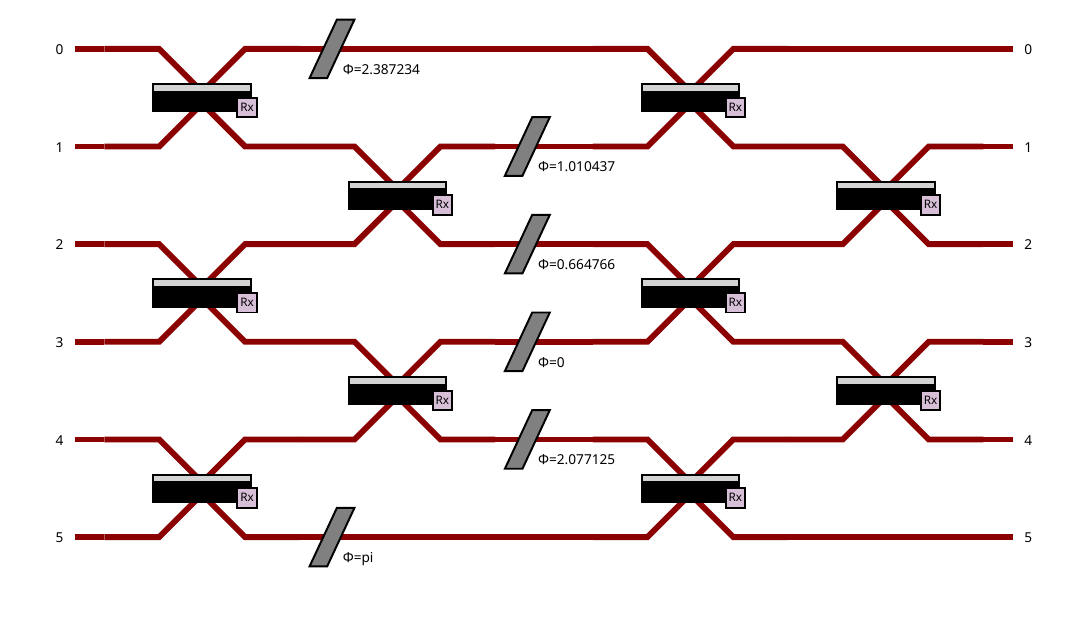}
\caption{Example photonic circuit diagram with $m=6$ modes. Photons are injected, pass through repeated beam-splitter (BS) layers, and accumulate phase shifts set by the PCA features. The measurement yields an $m$-mode photon-number distribution that encodes the feature vector.}
\label{fig:qool-circuit}
\end{figure}

In this first approach, the photonic circuit is a $m$-mode photonic interferometer whose design follows \cite{ding2025quantum}. Beginning from a Fock state \(\ket{n_1,n_2,\dots,n_m}\), alternating layers of beam splitters and phase shifters are applied (Fig.~\ref{fig:qool-circuit}).
The resulting multi-mode photonic state \(\ket{\psi_{\vec\varphi}}\) captures the structure of the data in a  space of dimension ${m+n-1 \choose n}$ .

We estimate each kernel
$ \kappa(\vec x_i,\vec x_j)\;=\;\bigl|\braket{\psi_{\vec\varphi_i},{\psi_{\vec\varphi_j}}}\bigr|^2
$
by repeated photon‐number measurements on each of the two circuits, yielding an \(N\times N\) kernel matrix.  Optionally, we apply a nonlinear post‐processing step—either a sigmoid transform \(\tanh(\alpha\,\kappa + \beta)\) or a polynomial map \((\gamma\,\kappa + c)^d\)—to adjust the kernel geometry before supplying it to a classical one‐versus‐all multiclass SVM solver.


\Results
We benchmark on $600$ training and $60$ validation MNIST samples,
balanced across digits.  Each image is center-cropped and then downscaled to $14\times14$ using Principal Component Analysis (PCA). This is equivalent to retaining $m=20$ principal components. We construct a circuit with $20$ modes and $5$ photons.

Our validation results, summarized in Table~\ref{tab:qool_results}, show that the linear classical kernel achieved the highest accuracy of 90.00\%, while both the sigmoid and polynomial classical kernels reached 88.33\%.  The photonic quantum‐kernel SVM, simulated without noise, using a sigmoid‐transformed fidelity, attained 85.00\% accuracy.

\begin{table}[ht!]
\centering
\begin{tabular}{lcc}
\toprule
\textbf{Model}         & \textbf{Kernel}   & \textbf{Val.~acc.\ (\%)}\\
\midrule
Classical SVM          & Linear            & 90.0\\
Classical SVM          & Sigmoid           & 88.3\\
Classical SVM          & Polynomial        & 88.3\\
\midrule
Photonic Q-SVM         & Linear            & 82.0\\
Photonic Q-SVM         & Polynomial        & 83.0\\
Photonic Q-SVM         & Sigmoid           & \textbf{85.0}\\
\bottomrule
\end{tabular}
\caption{Validation accuracy on reduced MNIST ($600$/train, $60$/val).}
\label{tab:qool_results}
\end{table}

\begin{figure}[ht!]
    \centering    \includegraphics[width=0.6\linewidth]{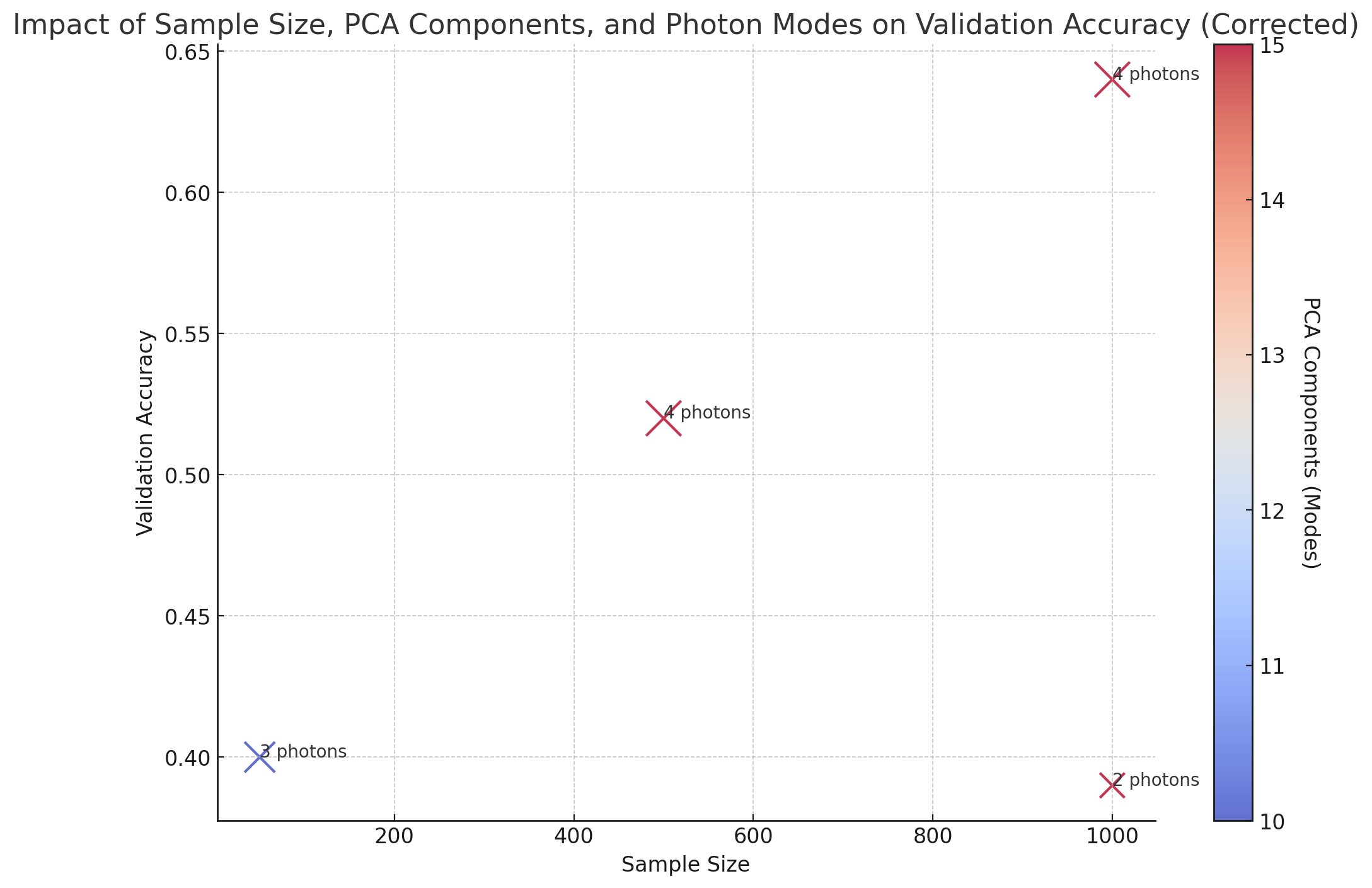}
    \caption{Accuracy vs.\ number of photonic modes $m$ for the
             sigmoid-transformed kernel.}
    \label{fig:qool_scaling}
\end{figure}

As illustrated in Figure~\ref{fig:qool_scaling}, increasing the number of injected photons \(n\) leads to a clear improvement in classification accuracy, indicating that larger photon counts can further narrow the gap with classical approaches.

\subsubsection{Leveraging the unitary dilation matrix for feature extraction}
\label{sec:lancelot_model}

\Proposal
This method leverages the \textit{unitary dilation theorem} \cite{mezher2023solving} that states that if $A \in M_n(\mathbb{C})$ of bounded norm, and $||A||\leq 1$ then \[U_a:=\begin{pmatrix}
    A & \sqrt{\mathbb{I}_{n\times n}-AA^\dagger} \\ \sqrt{\mathbb{I}_{n\times n}-A^\dagger A} & -A^\dagger
\end{pmatrix}\]

\begin{figure}[h]
    \centering   \includegraphics[width=0.6\linewidth]{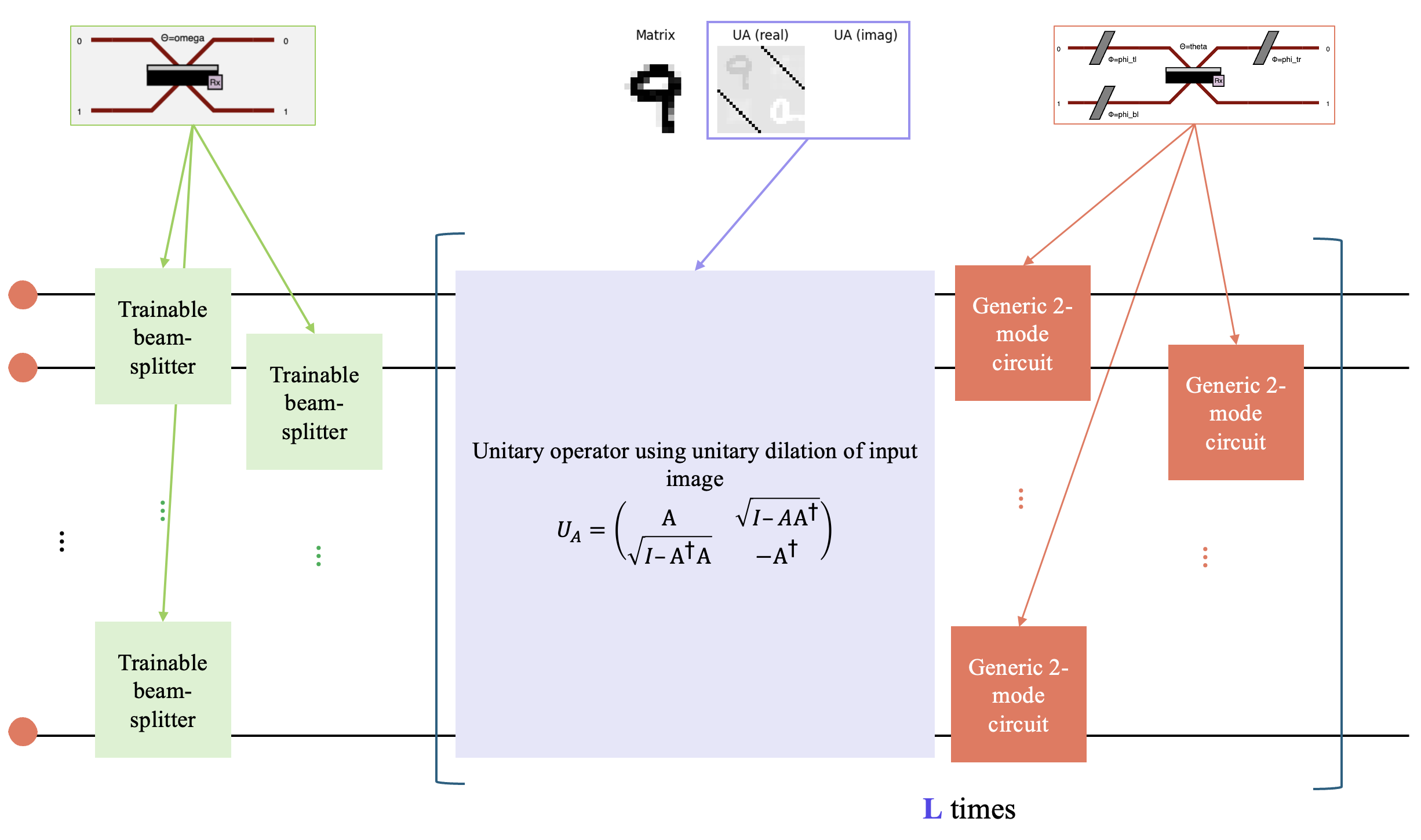}
    \caption{The trainable circuit is made of a first row of beam splitters, then $L$ blocks of Unitary matrices followed by circuits made with generic 2-modes circuits}
    \label{fig:lancelot_blocks}
\end{figure}

This $2n\times 2n$ unitary matrix, twice as big as the original data ($n\times n$), shows a way of encoding a (scaled-down version of) $A$ into a linear optical circuit. Following \cite{mezher2023solving}, the post-selection consists in observing $n$ photons in the first $n$ output modes. Therefore, to build the Unitary Dilation Encoding Neural Network (UDENN), a row of beam splitters is applied on the first $n$ output modes. Following this row, the trainable model consists of $L$ layers of a trainable unitary matrix followed by a brickwork construction of generic 2-modes circuit, to reproduce the effect of a classical convolutional kernel, extracting features from neighboring input values. One observation is that the associated circuit tends to shift the photons from one half of the circuit to another: the brickwork are therefore placed accordingly, as shown in Figure \ref{fig:lancelot_blocks}. A final post-selection condition is applied so that every photons end up in the same half of the circuit.

The output of this optical circuit is therefore made of the number of photons detected in each selected mode. A classical linear layer interprets this output and maps it to the 10 labels of the MNIST dataset. During training, the classical components were optimized using the Adam optimizer, while the optical components employed Simultaneous Perturbation Stochastic Approximation (SPSA). The two subsystems were trained in an alternating fashion. Here, a model with $L=6$ blocks is chosen and the classical part is made of one hidden layer, for a total of $565$ ($=325+240$) trainable parameters.
\label{sec:lancelot_exp}

\Results
For fair comparison, the Unitary Dilation Encoding Neural Network (UDENN) quantum model with unitary dilation encoding and the classical baseline were matched for trainable parameters and input dimension ($14\times 14$). Validation set performance was assessed using accuracy and confusion matrix metrics. More information on the training dynamics is given in Appendix \ref{sec:lancelot_supp}. During the challenge, the implementation was done with slow optimization using SPSA and therefore, the training of the hybrid model is slower than the training of the classical model (1 sec/epoch versus 1.4 hours/epoch). This is mainly due to the postselection involved in the hybrid model. However, we can envision improvements to this training time by using probability boosting techniques in \cite{mezher2023solving}. Another potential improvement lies in optimization of the back-end so the model can run faster.  Table \ref{tab:lancelot_results} presents the best validation accuracy across the training for the classical and hybrid models. Performance evaluation revealed that the Hybrid UDENN achieved 46.73\% accuracy on the validation set. While this represents a performance gap compared to the classical CNN, the result substantially exceeds random baseline performance, validating the efficacy of the unitary dilation encoding approach for feature extraction in this quantum-classical hybrid framework.
\begin{table}[ht]
    \centering
    \begin{tabular}{|c|c|}
        \hline
        Model & Validation accuracy (\%) \\ \hline\hline
        Classical CNN & 52.33 \\ 
        Hybrid UDNN & 46.73 \\ \hline
    \end{tabular}
    \caption{Best validation accuracy for a 5-epoch training for the UDENN}
    \label{tab:lancelot_results}
\end{table}
\subsubsection{A photonic quantum neural network}
\label{sec:qnn_proposal}
\begin{figure}[h]
    \centering
    \includegraphics[width=0.7\linewidth]{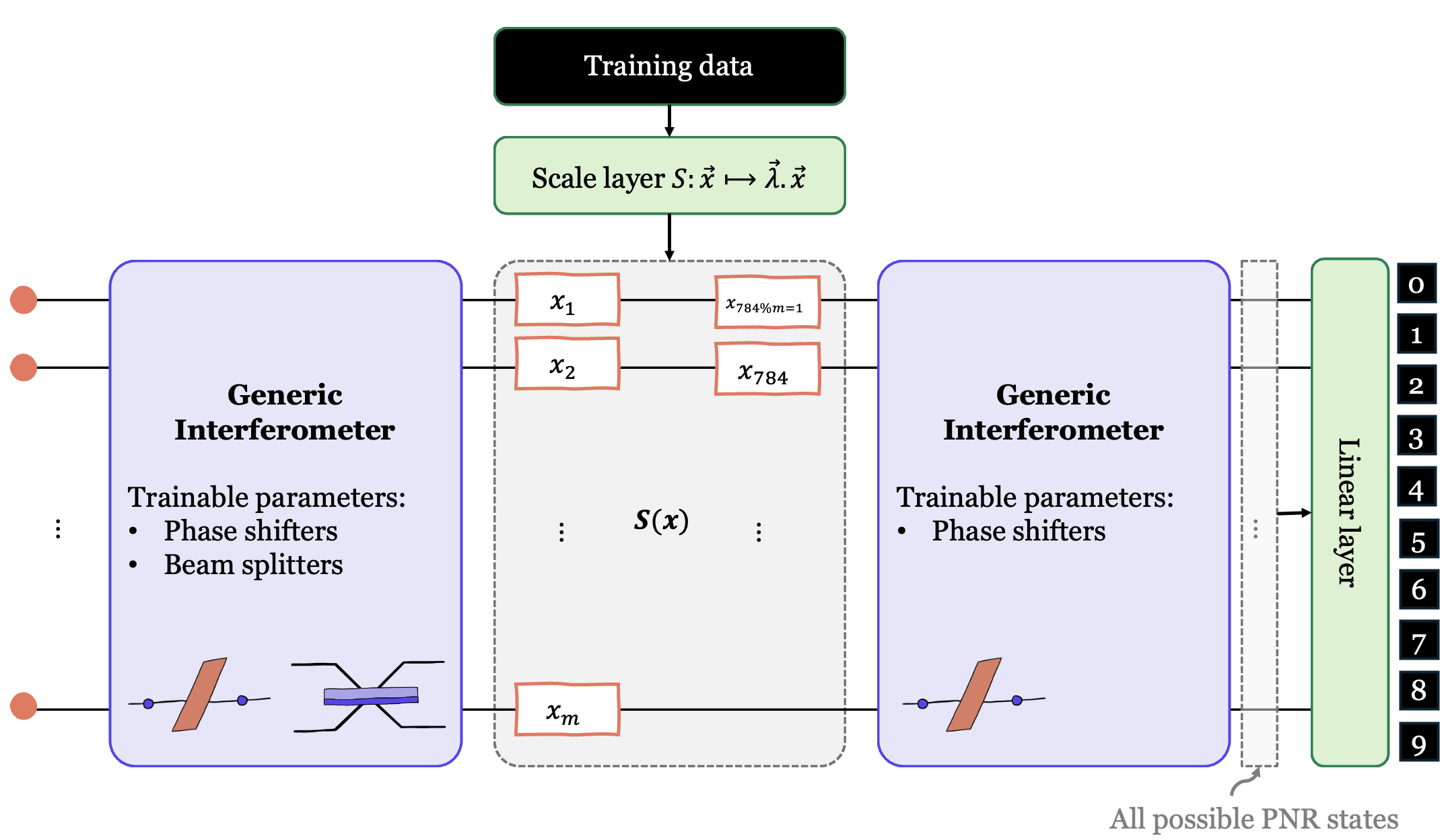}
    \caption{Photonic Quantum Neural Network, composed of two trainable generic interferometers \cite{clements2016optimal}, in purple, of an encoding layer (in gray) and two classical trainable layers (in green)}
    \label{fig:CodeQalibur_pqNN}
\end{figure}

\Proposal
This model consists of a \textbf{photonic neural network}. In this approach, the inputs of the interferometer are the $28\times 28$ images whose pixels values are scaled using learnable weights. These scaled pixels are encoded in phase shifters, following a circuit set-up as proposed in \cite{gan2022fock}, shown in Figure \ref{fig:CodeQalibur_pqNN}. It is composed of 2 trainable generic interferometers from \cite{clements2016optimal} and an encoding layer. We follow an encoding strategy such as $\forall x \in \mathbb{R}^{784}, S(x)=\lambda x $, where $\lambda \in \mathbb{R}^{784}$ is learned through gradient descent.
Further details on this model design and ablation studies are presented in Appendix \ref{sec:supp_pqNN}.

The photonic neural network outputs are classified into the 10 MNIST digit classes through a trainable linear classification layer. The complete model is trained end-to-end in simulation using the Adam optimizer \cite{kingma2014adam} with cross-entropy loss as the objective function.
\label{sec:pqNN_experiments}

\Results
Based on the ablation study explained in Appendix \ref{sec:supp_pqNN}, we find that an interferometer of 10 modes provide satisfying results. Increasing the number of modes would not provide valuable increase in performance compared to increase computational complexity. To provide a fair comparison with a classical model, in terms of number of trainable parameters, we chose a 2-layer MLP with \verb|ReLU| activation that has 21475 parameters. We reproduce our experiments 5 times and our results are shown in Table \ref{tab:pqNN_results}.

\begin{table}[h]
    \centering
    \begin{tabular}{|c|c|c|}
        \hline
         \textbf{Model} & \textbf{Test Accuracy} & \textbf{Number of parameters}\\ \hline \hline
         Hybrid Photonic qNN & $81.31 \pm 2.04$ & 21084 \\
         Classical MLP & $94.14 \pm 0.4$ & 21475 \\
         SVM & $95.38$ & 7850 \\ \hline
    \end{tabular}
    \caption{Test Accuracy and Number of parameters for the different models trained}
    \label{tab:pqNN_results}
\end{table}

While the hybrid photonic qNN does not demonstrate superior performance compared to classical approach for this classification task, it achieves a respectable 81.31\% accuracy on MNIST, indicating the viability of quantum-photonic architectures for machine learning applications. This result suggests that hybrid photonic systems, though not yet competitive with state-of-the-art classical methods, represent a promising foundation for future quantum machine learning implementations.
\subsubsection{GLASE: Gradient-free Light-based Adaptive
Surrogate Ensemble}
\label{sec_glase_proposal}
\Proposal
In this approach, we leverage a surrogate model to enable end-to-end backpropagation through a photonic quantum neural network (QNN). This model is a photonic adaptation of previous work that was designed for qubit-based circuits \cite{xie2025quantum}. In classical machine learning, the backpropagation algorithm is key to efficient training of deep neural networks. Typically for QNNs, training schemes rely on gradient estimation through the finite-difference method or the parameter-shift rule \cite{mitarai2018quantum, Pappalardo_2025} -- however, they both require a much higher cost in circuit evaluations and can be unstable in some settings. Our method circumvents this challenge by introducing a neural-network-based surrogate model that learns from data generated by the photonic backend. This allows integration into standard deep learning workflows while maintaining compatibility with quantum optical circuits simulated via boson sampling.

Our pipeline begins with a lightweight convolutional neural network (CNN) that extracts feature embeddings $\mathbf{z} \in \mathbb{R}^{256}$ from each $28 \times 28$ MNIST image $\mathbf{x}$. These features are mapped to quantum optical circuit parameters via a classical encoder $\boldsymbol{\phi} = \Pi(\mathbf{z})$, where $\boldsymbol{\phi} \in \mathbb{R}^{M}$ represents the programmable phase shifts in an $M$-mode photonic interferometer. The photonic circuit used is based on a boson sampling setup, where $N$ indistinguishable photons propagate through a fixed linear optical network governed by a unitary transformation $U(\boldsymbol{\phi})$. To enable differentiable learning, we introduce a surrogate neural network $g_{\alpha}(\boldsymbol{\phi})$ trained to approximate the expected photon count per mode $\langle \hat{\mathbf{n}} \rangle$ : $
g_{\alpha}(\boldsymbol{\phi}) \approx \langle \hat{\mathbf{n}} \rangle$. These values then go through a softmax layer to perform the classification task. More details on the mathematical background of this approximation as well as on the encoding strategy are given in Appendix \ref{sec:supp_glase}.

The surrogate model is periodically updated by minimizing the squared error loss
\[
\mathcal{L}_{\text{sur}} = \left\| g_{\alpha}(\boldsymbol{\phi}) - \langle \hat{\mathbf{n}} \rangle \right\|^2,
\]
using simulated outputs from the quantum photonic backend. During training, this surrogate replaces the quantum layer in the backpropagation pass, allowing gradients to flow from the output loss $\ell_{\text{CE}}(\hat{y}, y)$ to the CNN encoder parameters $\theta$. The full training loss becomes
\[
\mathcal{L}_{\text{total}} = \ell_{\text{CE}}(\hat{y}, y) + \lambda \cdot \mathcal{L}_{\text{sur}},
\]
where $\lambda$ controls the regularization strength for the surrogate fit (empirically set to $0.5$ in our experiments).

\begin{figure}[htb!]
    \centering
    \includegraphics[width=0.8\linewidth]{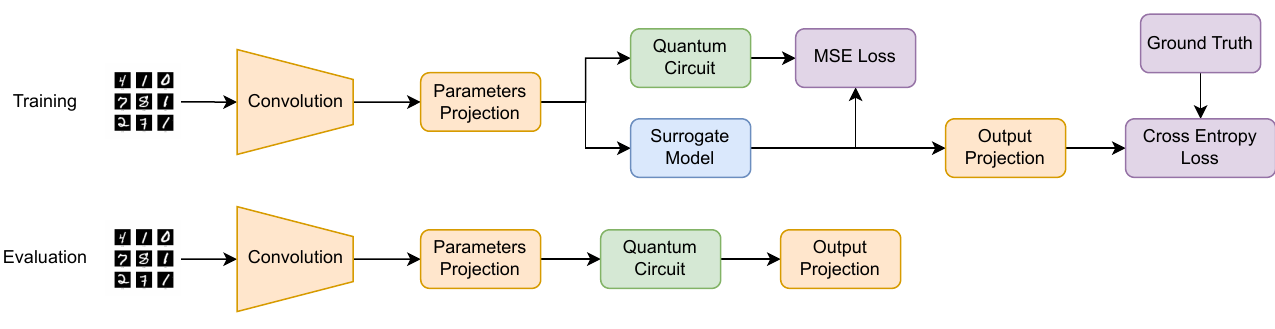}
    \caption{The GLASE architecture. A CNN encodes MNIST images into latent features which are mapped to photonic circuit parameters. A surrogate model approximates the photonic circuit’s output statistics to enable backpropagation during training. During the evaluation phase, the surrogate is not required and quantum hardware can be employed.}
    \label{fig:glase}
\end{figure}

On top of enabling more efficient training, we believe that the surrogate network may also avoid optimization issues like barren plateaus \cite{mcclean2018barren}, or those stemming from discrete hardware constraints. Indeed, in our numerical experiments, we found that the surrogate-assisted optimization remained stable throughout training. Overall, the surrogate model serves as a differentiable proxy that is periodically re-trained to match the behaviour of the true quantum photonic simulator, allowing for end-to-end optimization of the pipeline. Once trained, the surrogate can even be deployed as a fast approximate emulator for photonic inference when real hardware is unavailable. 

\label{sec_glase_results}

\Results
We conduct experiments on both simulated photonic backends and real photonic hardware to evaluate the performance of GLASE on MNIST classification. Training is performed for 50 epochs using a subset of 6,000 training and 1,000 validation images. The hybrid model integrates a photonic QNN backend with $P=3$ photons over $M=20$ optical modes. We report both training metrics and classification accuracy under different settings.

We first benchmark GLASE using the \texttt{CliffordClifford2017} boson sampling simulator provided by the Perceval platform. Two GLASE variants with different network capacities are compared to classical baselines, including a mini ResNet and a vanilla multilayer perceptron (MLP). We find that both GLASE models outperform their classical counterparts in validation accuracy while using fewer parameters. 

\begin{table}[htb!]
    \centering
    \begin{tabular}{@{}lrrrr@{}}
        \toprule
        \textbf{Model} & \textbf{Params} & \textbf{Train Acc} & \textbf{Val Acc} & \textbf{Val Loss} \\
        \midrule
        Vanilla MLP & 670k & 100.00\% & 97.00\% & 0.2192 \\
        Mini ResNet & 716k & 100.00\% & 98.17\% & 0.1405 \\
        QNN (380k) & 380k & 100.00\% & \textbf{99.00\%} & \textbf{0.1044} \\
        QNN (517k) & 517k & 100.00\% & \textbf{99.33\%} & \textbf{0.0961} \\
        \bottomrule
    \end{tabular}
    \caption{Performance of GLASE and classical baselines on simulated MNIST subset.}
    \label{tab:sim_results}
\end{table}

\RealQPUValidation
To validate GLASE under hardware constraints, we deploy a compressed version using 16 modes and 3-photons on Quandela's photonic QPU (Ascella), which supports real-time boson sampling experiments. Inference is performed with 1,000 shots per image on 150 test samples and results can be found in Table \ref{tab:qpu_results}.

\begin{table}[htb!]
    \centering
    \begin{tabular}{@{}lcc@{}}
        \toprule
        \textbf{Backend} & \textbf{Val Acc (\%)} & \textbf{Val Loss} \\
        \midrule
        CliffordClifford2017 (sim) & 91.00 & 0.3780 \\
        sim:sampling:h100 & 94.17 & 0.3125 \\
        qpu:ascella & 76.79 & 0.8070 \\
        \bottomrule
    \end{tabular}
    \caption{Validation accuracy and loss on real and simulated photonic backends.}
    \label{tab:qpu_results}
\end{table}

We attribute the accuracy drop on real hardware to noise such as photon loss and distinguishability, shot uncertainty, and hardware imperfections such as mode mismatch. Despite this, the result significantly outperforms random guessing (10\%) and demonstrates the surrogate model’s potential for generalization across simulator and physical implementations.

Overall, GLASE’s performance on simulators is competitive with state-of-the-art classical models while using fewer parameters. 
On real QPUs, performance is limited by current hardware constraints, but shows encouraging trends, suggesting that the framework will scale well as quantum photonic processors mature. The modular surrogate-assisted training can be adapted to new interferometer configurations, making GLASE highly flexible for future hardware generations. In Appendix \ref{sec:supp_glase}, we share further insights about our results.
\subsubsection{A photonic native quantum convolutional neural network}
\label{sec:qcnn_proposal}
\Proposal
The introduction of Convolutional Neural Networks (CNNs) in classical machine learning has revolutionised the field of computer vision. At the heart of the success of CNNs is an important inductive bias, namely \emph{translation invariance}. Now, translation invariance is highly dependent on the strategy used to encode the images as quantum state. Here, we will use amplitude encoding on qudits. More specifically, for an $N\times N$ greyscale (i.e. 2-dimensional) image $\mathbf{x} = \left(x_{i,j}\right)_{i, j=0}^{N-1}$, we will use $2N$ modes and $2$ photons and encode it as:
\begin{equation}
    \ket{\psi_{in}} = \sum_{i, j=0}^{N-1} \frac{x_{i,j}}{\left|\left|\mathbf{x}\right|\right|_2} \ket{e_i}\ket{e_j}
\end{equation}
where the state $\ket{e_{i}}$ (resp. $\ket{e_{i}}$) is a state over $N$ mode and a single photon such that the photon is located at the position $i$ (resp. $j$). This choice of encoding will allow us to define translation invariant operations with respect to the input state.

\begin{figure}[htb!]
    \centering
    \includegraphics[width=0.5\linewidth]{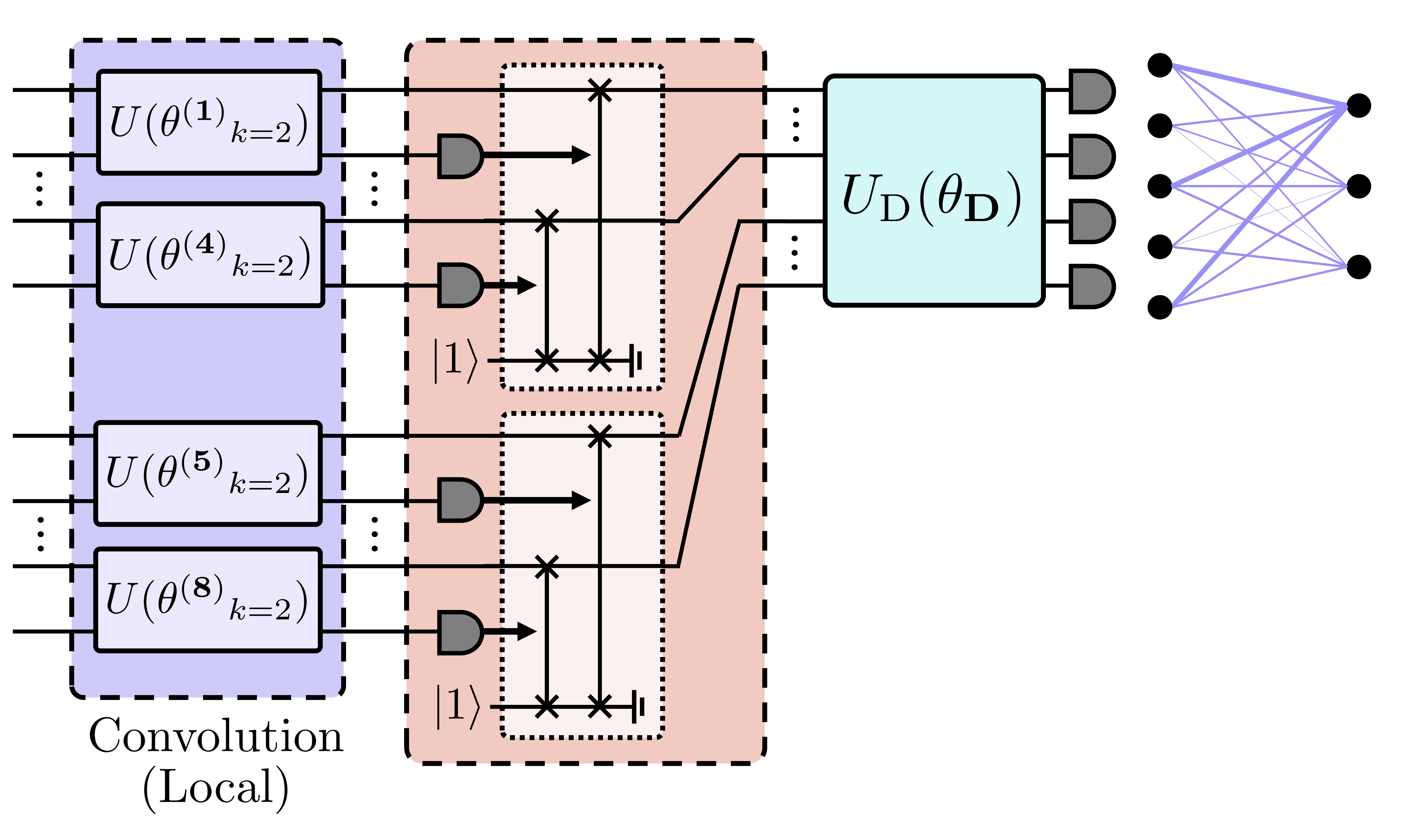}
    \caption{Architecture of the photonic QCNN}
    \label{fig:qloqroachArchi}
\end{figure}

In our approach, we produced a photonic analogue of the LeNet architecture~\cite{lecun1998mnist} which consists:
\begin{itemize}
    \item Photonic \emph{convolutional layers} that implement several (dependent) local convolutions using repeated local interferometers. This operation is translation invariant (with respect to specific translations, see Appendix~\ref{app:qloqroachConv})
    \item \emph{Pooling layers} which reduces the dimension of the image using adaptive photon injections~\cite{monbroussou2024_TowardsQuantumAdvantage_kashefi}. As for the photonic convolutions, this operation is also translation invariant
    \item A photonic \emph{dense layer} which is simply a global universal interferometers on all the modes. This operation is the analogue of a classical dense layer which dismisses the 2D structure of the image and processes all of the obtained features together.
\end{itemize}
The output of the photonic circuit is a probability distribution over the possible detection patterns; for example, using photon-number resolving detectors, this therefore gives us $\begin{pmatrix}m_r + 1\\2\end{pmatrix}$ different probabilities where $m_r$ is the number of remaining modes after pooling (recall that we only have 2 photons throughout the circuit). Since we required to discriminate between 10 different classes for the MNIST classification task, we treat the obtained probability distribution as a vector, and feed it to a classical (trained) linear transformation to obtain 10 different scores, followed by a softmax which will normalise those scores. The output of the full process will be the probability distribution over the ten different classes. The overall architecture is depicted in Figure~\ref{fig:qloqroachArchi}.

This approach was largely inspired from the qubit-based Hamming-weight preserving QCNN~\cite{subspaceQCNN} and was developed at the same time as the very similar framework presented in~\cite{monbroussou2025photonicquantumconvolutionalneural}.

\Results
To train the overall model, we converted Perceval circuits (with feed-forward for the pooling layers) into a PyTorch module, therefore allowing us to train the model using backpropagation and the Adam algorithm~\cite{kingma2014adam}. The models were trained for 40 epochs (with a batch size of 100) with the cross entropy loss. We then compared the performance of our model with a classical CNN containing the same number of layers, the same kernel sizes, and the same final classical linear layer. 

\begin{figure}[htb!]
    \centering
    \begin{subfigure}[c]{.4\linewidth}
    \includegraphics[width=\linewidth]{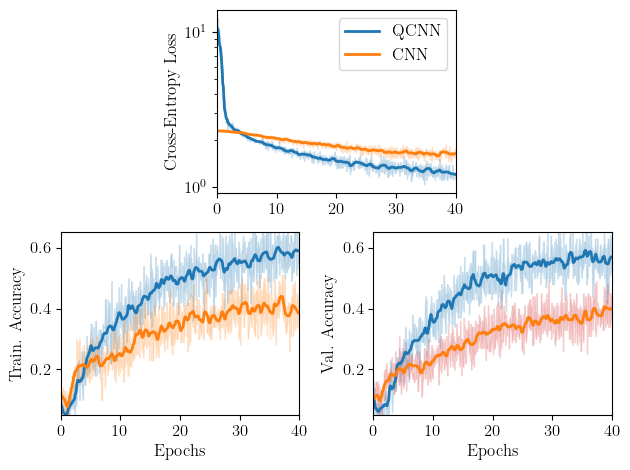}
    \caption{$4\times 4$ images}
    \end{subfigure}\quad%
    \begin{subfigure}[c]{.4\linewidth}
    \includegraphics[width=\linewidth]{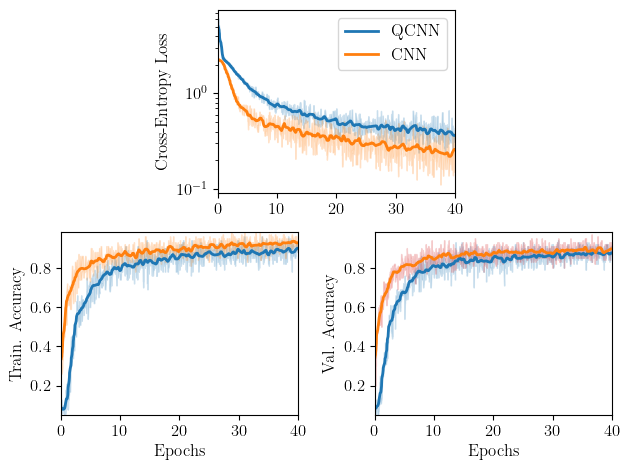}
    \caption{$12\times 12$ images}
    \end{subfigure}
    \caption{Comparison of the comparison of the QCNN and  classical CNN.}
    \label{fig:qloqroachResults}
\end{figure}

For the simulation to be tractable, we reduced the size of the images from $28\times 28$ to $4\times 4$ and $12\times 12$ images, and restricted ourselves to only have one convolutional layer (and a single pooling layer). The evolution of the loss function and accuracies are shown in Figure~\ref{fig:qloqroachResults}. We observed that, for $4\times 4$ images, the Quantum CNN (QCNN) clearly outperforms the classical equivalent, reaching a $58\%$ test accuracy as opposed to $40\%$ for the classical CNN, while having less trainable parameters (126 trainable parameters for the QCNN and 165 parameters for the classical CNN). For larger images, namely the $12\times 12$ images, the classical CNN reaches a higher final accuracy (93\% train accuracy, 90\% test accuracy) compared to the QCNN (89\% train accuracy, 88\% test accuracy). However, the performance of the QCNN still remains close to the classical equivalent, while having significantly less parameters to train (926 parameters to train for the QCNN as opposed to 3681 parameters for the classical CNN). 

\subsubsection{A convolutional layer using a photonic quantum kernel}
\label{sec:qubiteers_proposal}
\Proposal
We propose a hybrid quantum-classical image classification model that leverages a \emph{Photonic Quantum Kernel} (PQK) alongside a classical CNN. The PQK acts like a convolutional kernel by processing each $2\times2$ image patch through a photonic interferometer. Pixel intensities are normalized and encoded as phase shifts, modulating the path of single photons through beam splitter layers. The output detection pattern becomes the quantum feature vector for that patch. The full image is scanned with stride 2, producing a feature map composed of 5 or 20 channels per patch. Boson sampling is implemented using \texttt{Perceval}. This method leverages quantum interference in high-dimensional Hilbert spaces to encode complex local correlations that classical kernels may miss.

Here, 2 types of PQK are proposed following different kernel strategies: a reservoir one and a trainable one. For a $k\times k$ kernel, $m = \left\lceil kernel\_size^2/2 \right\rceil$ modes are used. The first $m$ pixels are encoded in a phase shifter on each mode. This first layer of phase shifters is followed by a row of beam splitters and followed by the remaining $m-1$ pixels to be encoded.

\begin{figure}[h!]
\centering
\includegraphics[width=0.7\columnwidth]{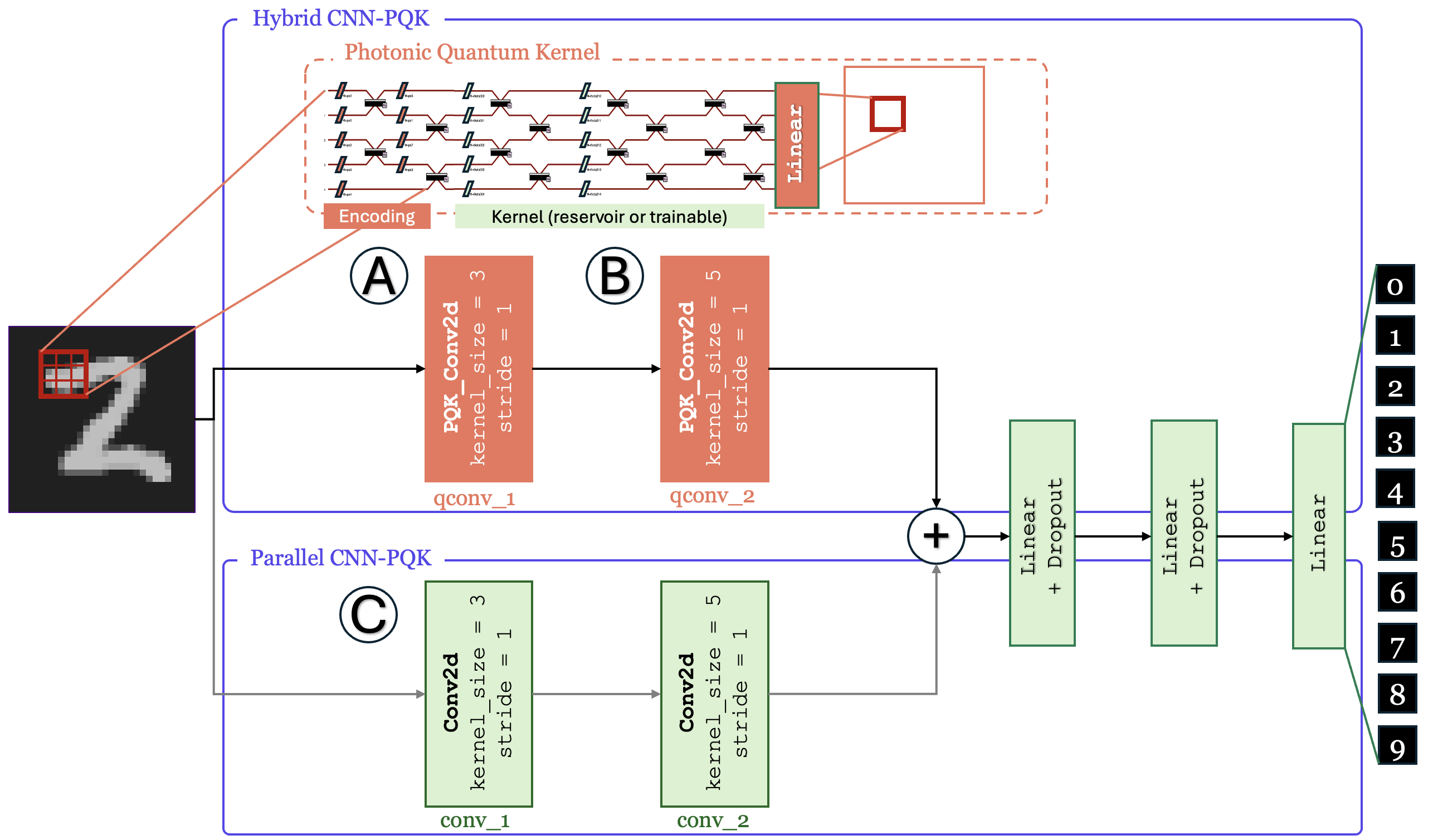}
\caption{CNN with PCK architectures. In the single layer hybrid architecture \textbf{(A)}, only one PQK convolution is used and then the output is forwarded to a MLP for final classification. For the two-layer hybrid architecture \textbf{(B)}, the image is forwarded through 2 PQK convolutions with increasing kernel size ($3\times 3$ then $5\times 5$), then through a MLP. For the Parallel CNN-PCK \textbf{(C)}, the image is forwarded through 2 quantum convolutions on the one hand, and through 2 classical convolutions with similar kernel size ($3\times 3$ then $5\times 5$) on the other end. Then, both outputs are concatenated and fowarded through a MLP for final classification}
\label{fig:PQK_model}
\end{figure}

Two types of models are then built. The \textbf{hybrid model} leverages one (Model \textbf{A} on Figure \ref{fig:PQK_model}) or two PQK convolutions (Model \textbf{B}) with ReLU activation, followed by a MLP. The first convolution as a $3\times 3$ kernel, a stride of 1 and an ouput size of 16, while the second one has a $5\times 5$ kernel, a stride of 1 and an output size of 32.
The\textbf{ parallel model} consists of two parallel branches. The first branch is a classical pathway where a conventional CNN with ReLU activation is applied to the grayscale input. The second branch is a quantum pathway (either trainable or non-trainable) that applies the two PQK-based convolutions with ReLU activation. The outputs from both branches are then concatenated and fed to a dense layer that maps them to the 10 classes. Further details about these branches are provided in Appendix \ref{sec:supp_qubiteers}.

\label{sec:qubiteers_results}

\Results
We present our findings comparing the baseline CNN model and our hybrid quantum-classical architectures enhanced with Photonic Quantum Kernels (PQK). Our study highlights how the dimensionality and encoding strategy of PQKs influence model accuracy, learning speed, and generalization. Table~\ref{tab:PQK_results} presents training and validation accuracy, while the figures in Appendix \ref{sec:supp_qubiteers} support our qualitative analysis of the experimental outcomes.

For the hybrid models, it seems that 2 layers 
\begin{table}[ht]
\centering
\begin{tabularx}{\textwidth}{lcccc}
\toprule
Model & Epochs & Quantum Encoding & \shortstack{Val\\Accuracy (\%)} & \shortstack{Train\\Accuracy (\%)} \\
\midrule
Full Baseline CNN (no quantum) & 5 & None & 97.67 & 98.05 \\
Hybrid PQK 1 layer & 5 & Reservoir & 11.08 & 12.33 \\
Hybrid PQK 1 layer & 5 & Trained & 92.58 & 92 \\
Hybrid PQK 2 layers & 5 & Reservoir & 10.58 & 12.33 \\
Hybrid PQK 2 layers & 5 & Trained & 11.8 & 12.33 \\
Parallel PQK & 5 & Reservoir & 97.12 & 95.33 \\
Parallel PQK & 5 & Trained & 96.77 & 95.67 \\
\bottomrule
\end{tabularx}
\caption{Validation and Test accuracies on MNIST (in \%) after 5 epochs of training.}
\label{tab:PQK_results}
\end{table}

The experimental results presented in Table \ref{tab:PQK_results} highlight the effectiveness of integrating quantum components into classical architectures for image classification. While the full classical CNN achieved the highest validation accuracy (97.67\%), hybrid models with trained quantum encodings showed competitive performance, particularly the 1-layer hybrid PQK (92.58\%) and the parallel PQK configurations (up to 97.12\%). In contrast, models using untrained reservoir encodings, especially in sequential (hybrid) architectures, performed poorly (~11\%), suggesting that quantum circuits require optimization to extract meaningful features. Notably, parallel quantum-classical integration appears more robust, preserving performance even with untrained quantum layers.

\textbf{Conclusion:}
This analysis confirms that photonic quantum kernels can benefit classical models when used in a hybrid framework. Effectiveness depends heavily on the encoding scheme and feature dimensionality. Well-designed quantum circuits — particularly Type~1 encoding with sufficient entanglement — yield useful features that complement classical learning. These insights open the path to extending PQK-based models to more complex datasets and deeper architectures.

\textbf{Remark:} the encoding strategy described above presents a significant flaw: the phase shifters in the first row (initial phase) have no influence on the output distribution of the interferometer. Moreover, no light propagates through some of those elements due to the periodic input state. Consequently, for a $3 \times 3$ kernel, only 4 of the 9 pixels actually contribute to the transformation. While this is clearly a limitation of the current design, it also uncovers an interesting phenomenon: despite this masked behaviour, the system remains capable of classifying MNIST effectively. This suggests that MNIST can be processed with a partially masked kernel. Moreover, with a stride of 1, the masked kernel eventually covers the entire image, allowing all pixels to be considered over the course of the convolutional operation.
\subsubsection{A convolutional layer using a photonic feature map}
\label{sec:qaradoq_model}

\Proposal
Here, we introduce \verb|qconv2d|, a parametric quantum convolution. Two-dimensional convolution is a common layer in neural networks for visual applications. Inspired by \cite{shi2023hybrid}, the dot product of traditional convolution is replaced by a quantum circuit. The sliding window principle is preserved. Here, the quantum convolution is made of a photonic circuit with two parts: a fixed and untrained feature map to encode the data, followed by a trainable ansatz. Different architectures are considered for these 2 components and are depicted in Table \ref{tab:qaradoq_circuit_config}. These models differ in terms of number of input photons, fixed and trainable components. 

\begin{table}[ht]
\centering
\small
\begin{tabular}{p{1cm}| p{3cm} p{5cm} p{5cm}}
\toprule
& \textbf{Input State} & \textbf{Feature Map - Circuit} & \textbf{Ansatz - Circuit} \\
\midrule
1) Tristan & $|1,0,1,0,0,1,0,0,1\rangle$ & 
\begin{itemize}[itemsep=0pt, topsep=0pt, parsep=0pt]
\item 27 components (BS.H, BS.v and PS), in triangle-like shape
\item Set $9 \times 3$ angles using pixel value, with 3 different transformations: $-2\pi x$, $2\pi x$, $\sin(2\pi x)$
\end{itemize} & 
\begin{itemize}[itemsep=0pt, topsep=0pt, parsep=0pt]
\item MZI mesh with a depth of 6
\item 96 components, 48 learnable parameters
\end{itemize}
 \\
\midrule
2) Dagonet & $|0,0,1,0,0,0,1,0,0\rangle$  &
\begin{itemize}[itemsep=0pt, topsep=0pt, parsep=0pt]
\item 15 BS in a cross arrangement
\item 9 phase shifters (one per mode) with angles set using pixel values: $(0.5 - x) \frac{\pi}{2}$
\end{itemize} & 
\begin{itemize}[itemsep=0pt, topsep=0pt, parsep=0pt]
\item Set of 1 BS.H + 2PS, and 1 BS.v + 2PS
\item 96 components, 96 learnable parameters
\end{itemize} \\
\bottomrule
\end{tabular}
\caption{Quantum circuit configurations for feature mapping and ansatz. The first configuration is made of a mix feature map [\textit{Achilles}] and a MZI ansatz [\textit{Penarddun}] while the second is made of a dispatch feature map [\textit{Odysseus}] with a custom ansatz [\textit{Gofanon}]}
\label{tab:qaradoq_circuit_config}
\end{table}

The output of such circuit is made of the probability of each possible output Fock state. The quantum convolution layer outputs $m$ matrices, where $m$ is the number of modes.

\begin{figure}[h]
    \centering
    \includegraphics[width=0.7\linewidth]{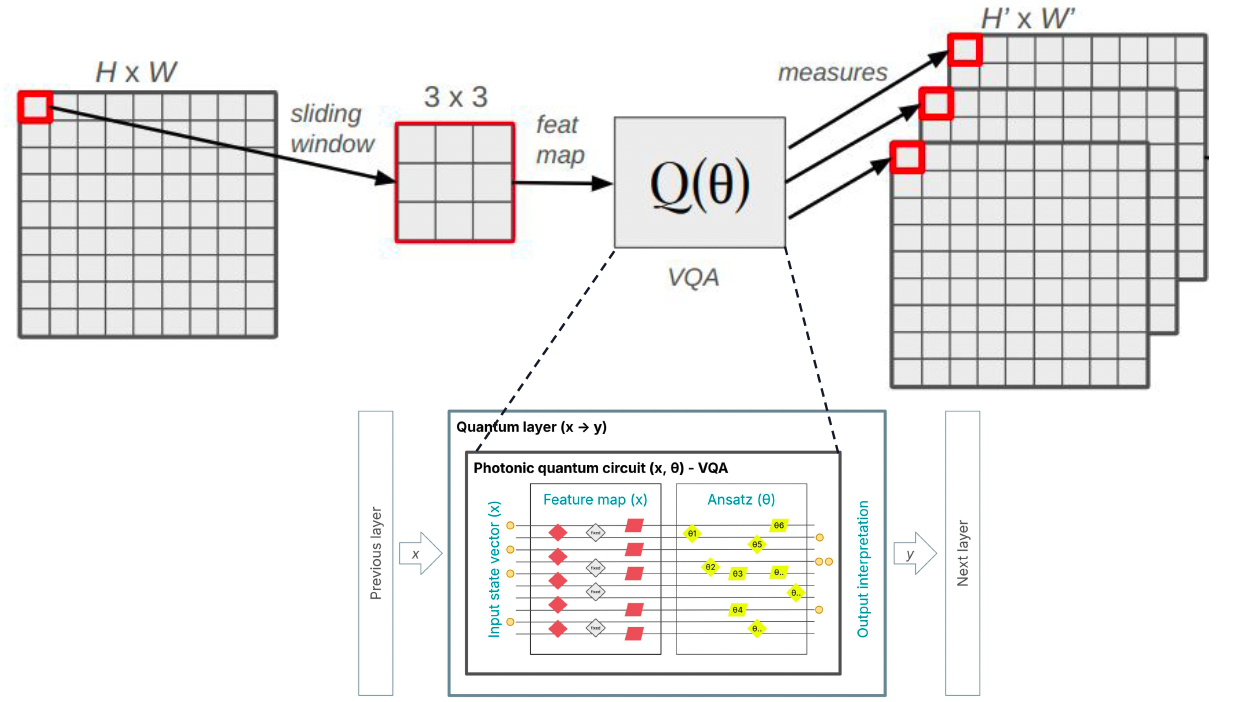}
    \caption{The two-dimension quantum convolution with $3\times 3$ kernels and a stride of $2$, with no dilatation and no padding. The output consists of 9 matrices of $13\times 13$ coordinates}
    \label{fig:qaradoq_model}
\end{figure}

Figure \ref{fig:qaradoq_model} presents the overall framework while more details on the feature maps and ansatz are given in the Appendix \ref{sec:qaradoq_supp}.

\label{sec:qaradoq_exp}

\begin{figure}[h]
\centering
\begin{subfigure}[c]{0.5\textwidth}
\centering
\begin{tabular}{|p{2cm} | p{2cm} |p{2cm}|}
\hline
\textbf{Model} & \textbf{Validation accuracy} & \textbf{Time per epoch} \\
\hline\hline
linear & 88 & 00'01'' \\
conv2d ($s=3$) + linear & 90 & 00'01'' \\
\hline
Tristan & 84 & 07'20'' \\
Dagonet & 78 & 07'05'' \\

\hline
\end{tabular}
\caption{Validation accuracy after 5 epochs for the classical and hybrid models}
\label{tab:qaradoc_results}
\end{subfigure}%
\hfill
\begin{subfigure}[c]{0.4\textwidth}
\centering
\includegraphics[width=\textwidth]{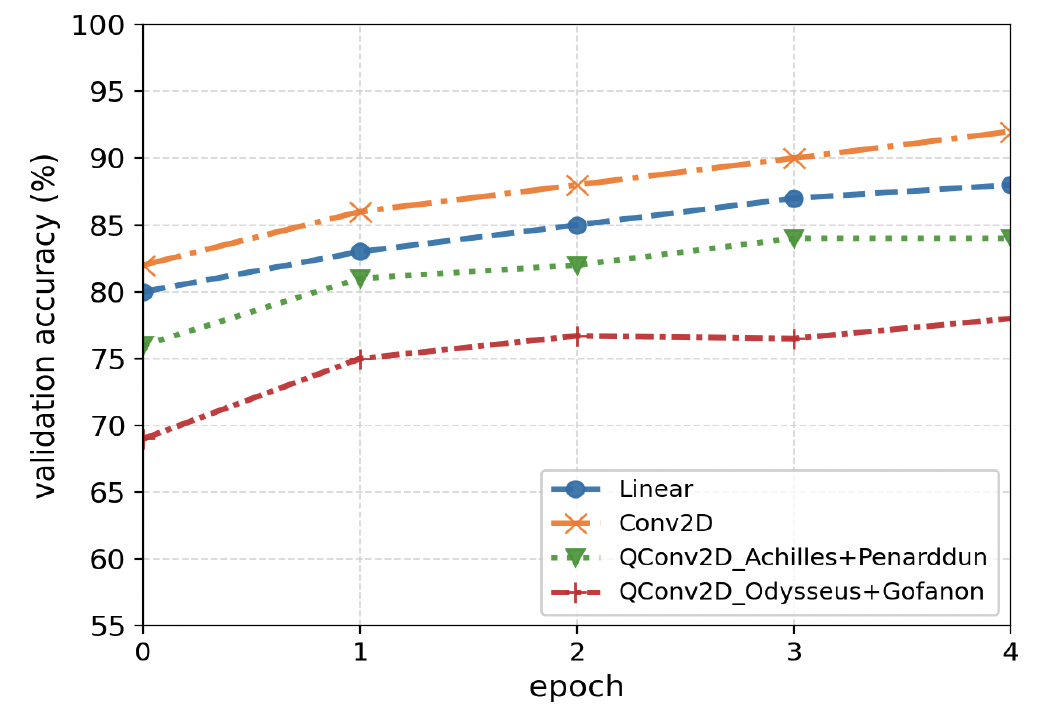}
\caption{Validation accuracy over 5 training epochs}
\label{fig:qaradoq_results}
\end{subfigure}
\caption{Comparison of classical and hybrid quantum models}
\end{figure}

\Results
The photonic circuit was simulated using a differentiable quantum layer. From results shown in Table \ref{tab:qaradoc_results}, quantum implementations lead to lower validation accuracy, but converges faster than classical models as shown on Figure \ref{fig:qaradoq_results}. Moreover, it is also highlighting that there is plenty room for improvement since classical machine learning does 10\% better.
Furthermore, the 10\% performance gap between the hybrid model and classical CNN indicates room for improvement in the quantum-classical architecture. This differential suggests that with refined quantum circuit design, the hybrid approach may achieve competitive performance with conventional methods.

\vspace{0.5cm}
\subsubsection{Photonic Quantum-Train}
\label{sec:qtx_prop}
\Proposal
At the core of the photonic Quantum-Train (QT) framework\cite{chen2025distributed}, a parametrized quantum circuit (PQC) is used as a \emph{parameter generator} for a classical neural network (NN). The key observation is that a modest number of PQC controls can induce a joint distribution over exponentially (or combinatorially) many computational-basis outcomes, with the number of degrees of freedom dictated by the Hilbert-space sector being addressed. We exploit this by mapping measurement probabilities to real-valued NN weights through a learned, low-rank tensor-network map.  

We instantiate two photonic quantum neural networks (QNNs),
\(
\mathrm{QNN}_{1}(\vec{\theta}^{(1)}) \) and \( \mathrm{QNN}_{2}(\vec{\theta}^{(2)})
\),
with $M_1$ and $M_2$ optical modes, respectively. Each device is operated in a fixed-excitation (Hamming-weight) subspace with $N_1$ and $N_2$ excitations. In Appendix \ref{sec:qtx_supp_math}, we show that, by steering the QNN controls one can populate at least $m$ effective degrees of freedom for the target NN. A learnable \emph{mapping model} $G_{\boldsymbol{v}}$ based on a matrix-product state (MPS) \cite{liu2024quantumTN} allows to map the outputs of the QNN to the weights of the NN.
This hybrid models can be trained using gradient descent as demonstrated in Appendix \ref{sec:qtx_supp_gradient}. Figure \ref{fig:qtx_scheme} depicts the photonic quantum train scheme.

We implement the photonic QNN with a programmable multi–mode interferometer realized as a rectangular mesh of nearest–neighbour two–mode Mach–Zehnder Interferometers (MZIs), each composed of two balanced beam splitters and internal/external phase shifters, following the decomposition of Clements \emph{et al.}~\cite{clements2016optimal}. More details of this decomposition are given in Appendix \ref{sec:qtx_supp_photons}.

\begin{figure}[h]
    \centering
    \includegraphics[width=0.9\linewidth]{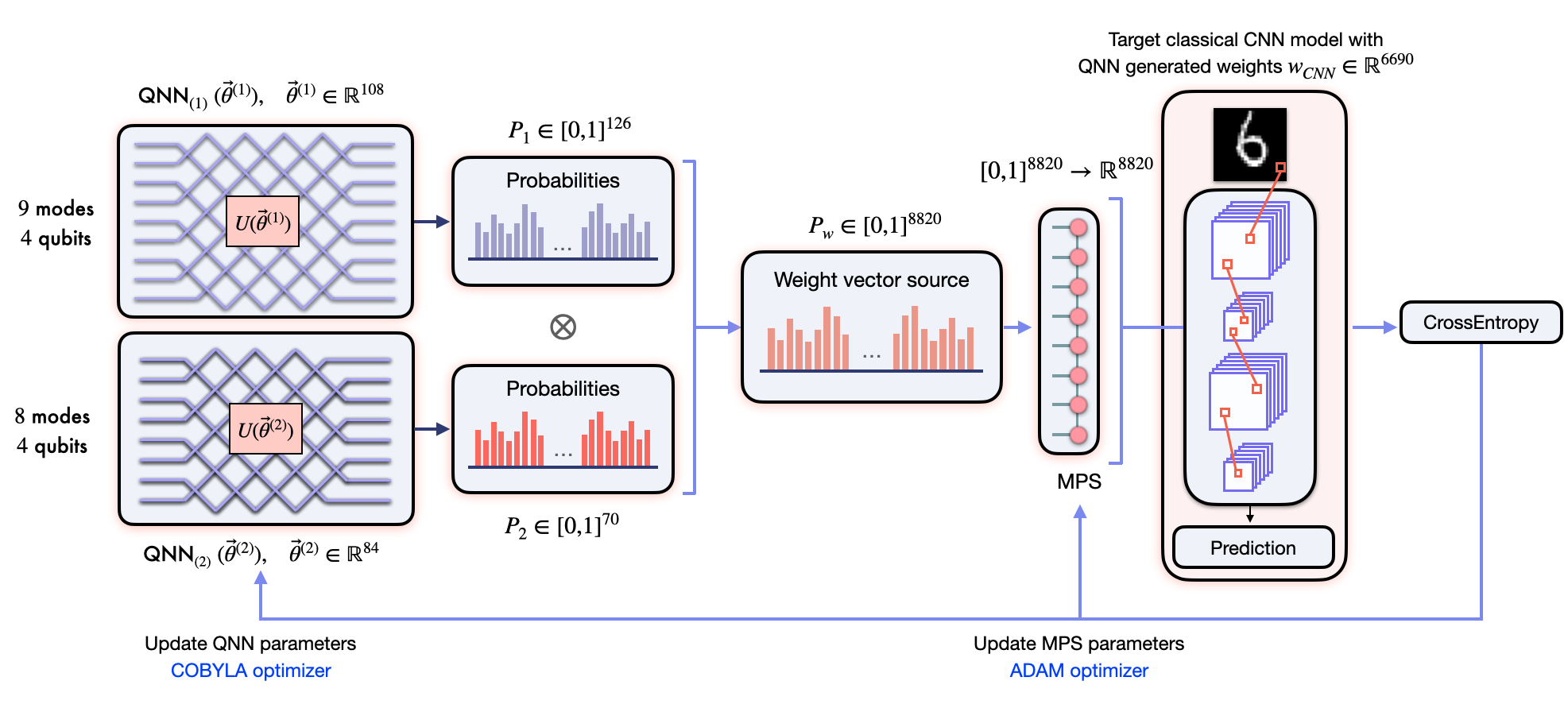}
    \caption{Overview of the photonic quantum-train scheme: 2 QNNs are trained with COBYLA optimizers so their outputs are mapped to the weights of a classical NN using a classical matrix-product state, trained using Adam optimizer.}
    \label{fig:qtx_scheme}
\end{figure}

\label{sec:qtx_exp}

\Results
Our implementation (adapted from the \texttt{Perceval} library \cite{Heurtel_2023}) programmatically assigns and updates the interferometer parameters. Users may supply explicit lists $\{\theta_\ell\}$ and $\{\phi_\ell\}$ or initialize them randomly. On hardware, thermo–optic or electro–optic modulators provide continuous, real–time tuning of internal and external phases, supporting closed–loop optimization to minimize a task–specific cost. This decomposition yields a hardware–friendly layout with minimal optical depth, low mode–dependent loss accumulation, and robust reconfigurability—features that are advantageous for boson sampling, quantum–enhanced machine learning, and other protocols relying on low–noise, large–scale multi–mode interference.

A thorough study of the parameter efficiency in photonic QT (presented in Appendix \ref{sec:qtx_supp_parameter}) allows us to conclude that models with higher bond dimensions achieve consistently lower training loss and higher training accuracy, indicating enhanced expressiveness and optimization. 

\begin{figure}[t]
\centering
\includegraphics[width=\linewidth]{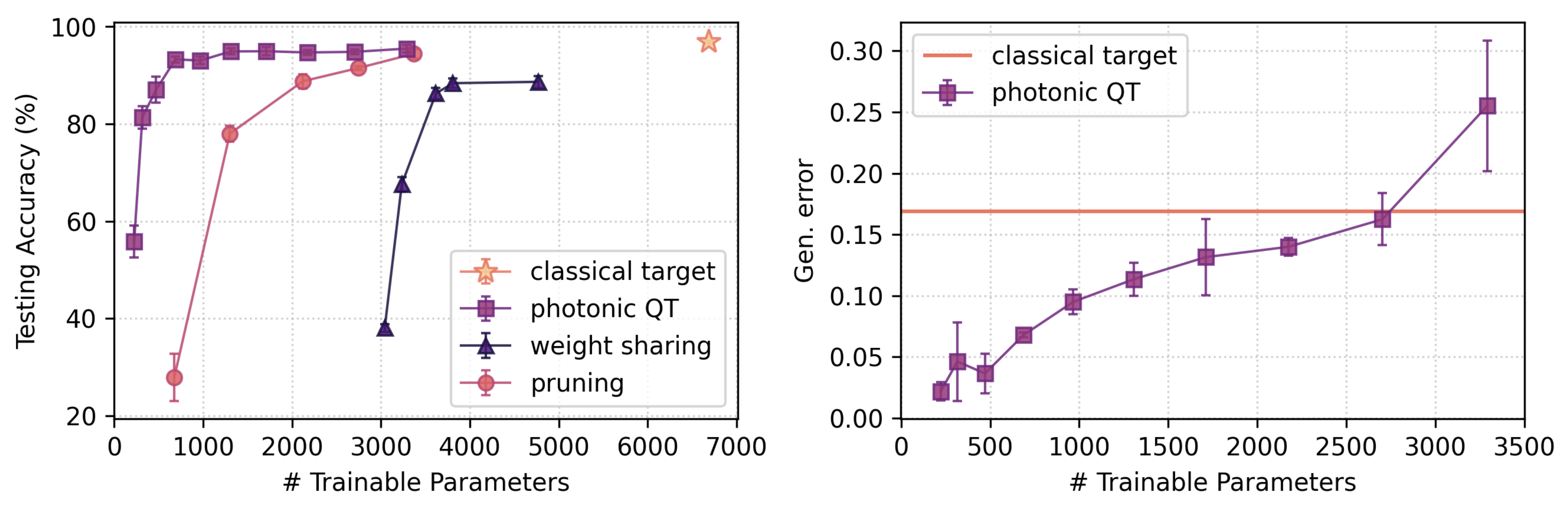}
\caption{
(Left) Testing accuracy vs.\ number of trainable parameters. 
(Right) Generalization error. Photonic QT outperforms weight sharing and pruning in accuracy, but shows an increasing generalization error as parameter count rises\cite{chen2025distributed}}.
\label{fig:acc_gen_error}
\end{figure}

Figure~\ref{fig:acc_gen_error} compares the photonic QT framework with classical model compression techniques, including weight sharing and pruning. The left panel shows testing accuracy versus model size, while the right panel plots the generalization error. QT offers superior accuracy for a given parameter count and outperforms classical baselines in the low-parameter regime.

Table~\ref{tab:mm} summarizes the number of trainable parameters needed to achieve comparable testing accuracy across different methods. The photonic QT model with bond dimension $D=10$ achieves $95.5\%$ accuracy using just 3292 parameters, less than half the size of the full classical CNN. At $D=4$, QT requires only 688 parameters to achieve over $93\%$ accuracy—surpassing both pruning and weight sharing at similar sizes.

\begin{table}[htbp]
\centering
\caption{Number of trainable parameters required to achieve comparable test accuracy\cite{chen2025distributed}.}
\vspace{6pt}
\begin{tabular}{lcc}
\toprule
\textbf{Method} & \textbf{\# Parameters} & \textbf{Test Acc. (\%)} \\
\midrule
Original CNN & 6690 & $96.89 \pm 0.31$ \\
Weight Sharing & 4770 & $88.67 \pm 1.21$ \\
Pruning & 3370 & $94.44 \pm 0.92$ \\
Photonic QT ($D=10$) & 3292 & $95.50 \pm 0.84$ \\
Photonic QT ($D=4$) & 688 & $93.29 \pm 0.62$ \\
\bottomrule
\end{tabular}
\label{tab:mm}
\end{table}

In summary, the photonic QT framework exhibits strong parameter efficiency by achieving competitive performance with substantially fewer parameters. While higher bond dimensions improve accuracy, they also increase generalization error, indicating a trade-off that must be balanced. Classical compression techniques offer alternative strategies, but their performance saturates below that of the quantum-enhanced model. These results underscore the potential of photonic quantum systems for efficient neural network training and invite further exploration into hybrid training and regularization techniques to mitigate overfitting in high-capacity QNNs.

\subsection{Photonic interferometer for quantum annotations}
The following models present innovative frameworks that utilize photonic systems to perform feature annotation and enhancement. Here, the use of trainable classical models is necessary -- the classical models are combined with photonic interferometers in the hope of improving the overall performance.
\subsubsection{Enrich classical CNN representations} %
\label{sec:naples_proposal}
\Proposal
In this method, we combine a photonic quantum system which is employed as a feature extractor, with a classical machine learning model. The core of our approach is a boson-sampler-based quantum embedding \cite{gan2022_FockStateEnhanced_angelakis,monbroussou2024_TowardsQuantumAdvantage_kashefi}, which transforms each image into a unique Fock-state probability distribution. The quantum embedding is used either as input to a classical neural network or concatenated with the original pixel data to enhance the feature set.


We encode the input data directly into the properties of the photonic circuit. Each MNIST image ($28\times 28$) is flattened and reduced to $d = 126$ dimensions using PCA, retaining approximately 93.7\% of the variance. This dimension matches the number of programmable phase parameters in a 12-mode interferometer.

\begin{figure}[htb!]
    \centering
    \includegraphics[width=1\linewidth]{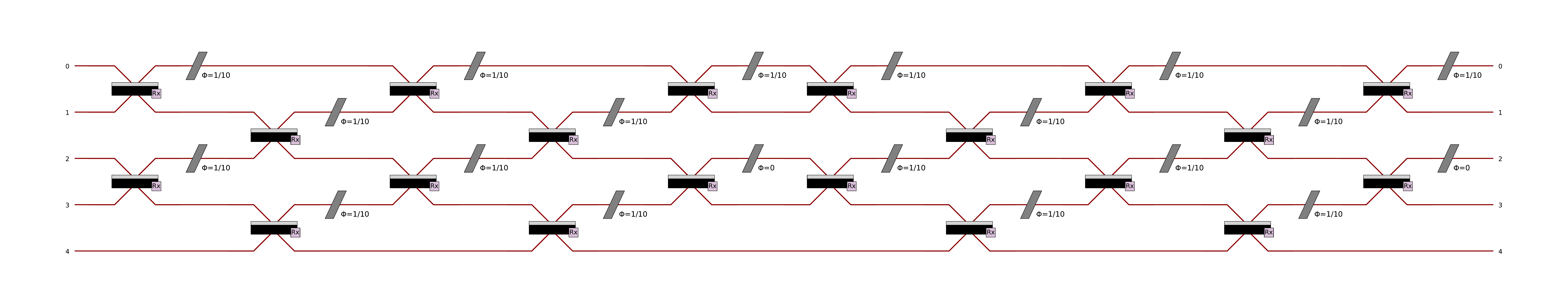}
    \caption{Interferometer consisting of two sequential rectangular meshes of beam splitters (BS) and phase shifters (PS), as implemented in the Perceval framework. Rectangular meshes~\cite{clements2016optimal} offer the same universality as triangular meshes but with reduced optical depth, which can lower noise accumulation and ease calibration.}
    \label{fig:phaseknight_mesh}
\end{figure}

The reduced feature vector is directly mapped to the phase shifters of the 12-mode photonic interferometer. We explored multiple mesh configurations (triangular, rectangular, and convolution-inspired), but our final model uses two sequential rectangular meshes. 
Single photons are injected into predefined modes, and the interferometer transforms the state via beam splitter and phase shifter layers. The output photon count distribution (dimension $\binom{12}{2} = 66$ for $n=2$ photons) forms the quantum embedding.
Figure~\ref{fig:phaseknight_mesh} is the layout used, two sequential rectangular meshes with programmable PS layers between BS stages. 

We believe that a boson sampler can act as a nonlinear feature map: small differences in the PCA-reduced input can yield strongly decorrelated permanents and thus nearly orthogonal output distributions. This can improve class separability in the downstream classifier. We elaborate on this question in Appendix \ref{sec:supp_q2pi}.

To optimize the model's performance, we utilize Tree-structured Parzen Estimators \cite{nguyen2022_QuantumEmbeddingSearch_chen} for hyperparameter optimization. This method systematically searches for the best values for key parameters, such as the learning rate, number of photons, and number of modes, ensuring an efficient training process. We also developed a ``fast approach" to address the long simulation times encountered with the full dataset. This strategy uses a smaller subset of the MNIST data and fewer shots, allowing us to quickly prototype and test the model's feasibility and learning capabilities. This rapid experimentation enables us to gather preliminary results and insights in a fraction of the time, paving the way for more extensive, full-scale tests on real QPUs.

\Results
We performed our simulations using Perceval’s \texttt{SLOS} backend (noiseless) and, for benchmarking, we used the GPU-enabled \texttt{sim:sampling:2l4} backend. Remote execution using \texttt{sim:sampling:2l4} was slower than local SLOS but provided a more realistic evaluation setting. We note that backend choice affected runtime but not accuracy. For the classical baseline, we employed an MLP trained on the 126 PCA features, and both models are matched for comparable parameter counts.

Our approach reaches a validation accuracy of 96.50\% with a final cross-entropy loss of 0.1239, surpassing the classical PCA baseline (93.8\%). Macro-averaged F1-score is 0.9636, with per-class F1 ranging from 0.942 to 0.984. The gains of our hybrid model were largest for digits \texttt{3}, \texttt{5}, and \texttt{9}.

\subsubsection{Hybrid feature extractor}
\label{sec:proposal_solal}

\begin{figure}[h]
    \centering
    \includegraphics[width=0.7\linewidth, trim=0cm 2.5cm 0cm 2.5cm, clip]{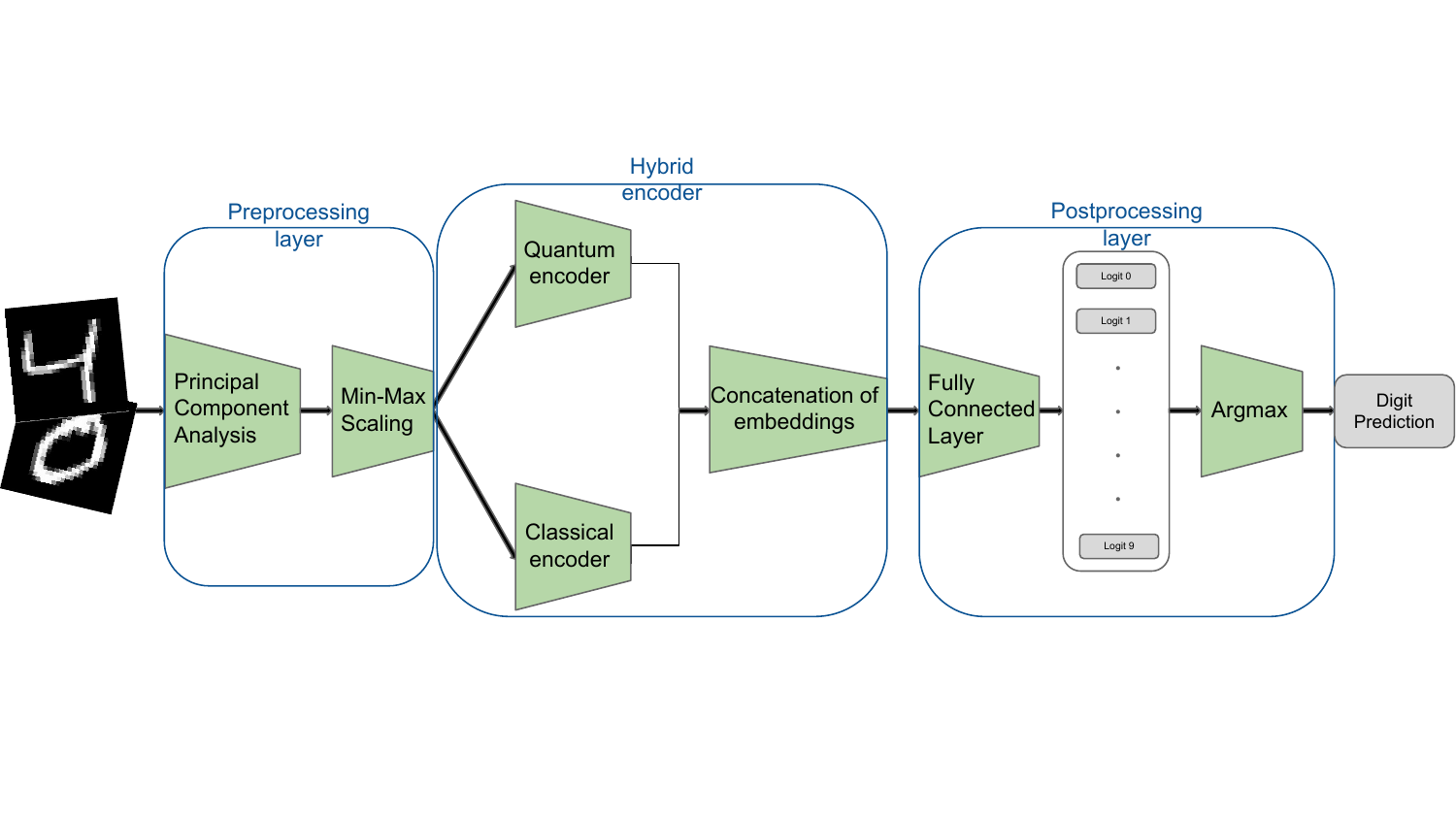}
    \caption{Architecture of the hybrid feature extractor. We can distinguish three main parts: a preprocessing layer where PCA is applied to the data, a hybrid encoder made of photonic interferometer and a classical layer, and a post-processing layer to map to the 10 classes of MNIST.}
    \label{fig:solal_model_figure}
\end{figure}

\Proposal
The following model is a hybrid feature extractor consisting of 2 main components depicted in Figure \ref{fig:solal_model_figure}:
\begin{enumerate}
    \item A trainable \textbf{Quantum Block} encodes the classical data into the Fock space. We consider input states that are Fock states of the form $\ket{1,0,1,0,\dots}$. We define a quantum circuit alternating encoding and processing layers. The goal of the encoding layers is to inject the input data as phases in the circuit, while the preprocessing layers define the trainable quantum parameters. Specifically, the encoding layers are formed by at most one phase shifter per mode, with a total number of phase shifters being equal to the input data dimension $d$. This dimension verifies by construction $d \leq m$. The processing layers consist of fixed beam splitters and trainable phase shifters in a triangular configuration. A grouping output strategy is used to partition the $m$ modes into $10$ disjoint groups. 
    
    \item \textbf{Classical Block.} In addition to the quantum block, we define a feedforward classical layer with a ReLU activation. This layer outputs a classical embedding of the $d$-dimensional input vector.

    \item \textbf{Postprocessing layer.} This last layer consists in a learnable fully connected layer which maps the concatenated embeddings from the Quantum and Classical blocks to the 10 classes
    
\end{enumerate}

\label{sec:exp_solal}

\Results
To investigate the impact of each component in our architecture, we compare three models:

\begin{itemize}
    \item Classical-only model: the quantum blocks are removed.

    \item Almost-fully-quantum (AFQ) model:  In the encoder, we consider only the quantum encoder, without the classical encoding.
    
    \item The proposed hybrid model: the full approach described above with both classical and quantum parts.
\end{itemize}

The classical and hybrid models have a fairly similar number of parameters (2290 and 2122) and FLOPs (852.48 KFLOPS and 750.72 KFLOPS), while the AFQ model only had 112 parameters due to computational constraints. 

\begin{figure}[h]
    \centering
    \includegraphics[width=0.4\linewidth]{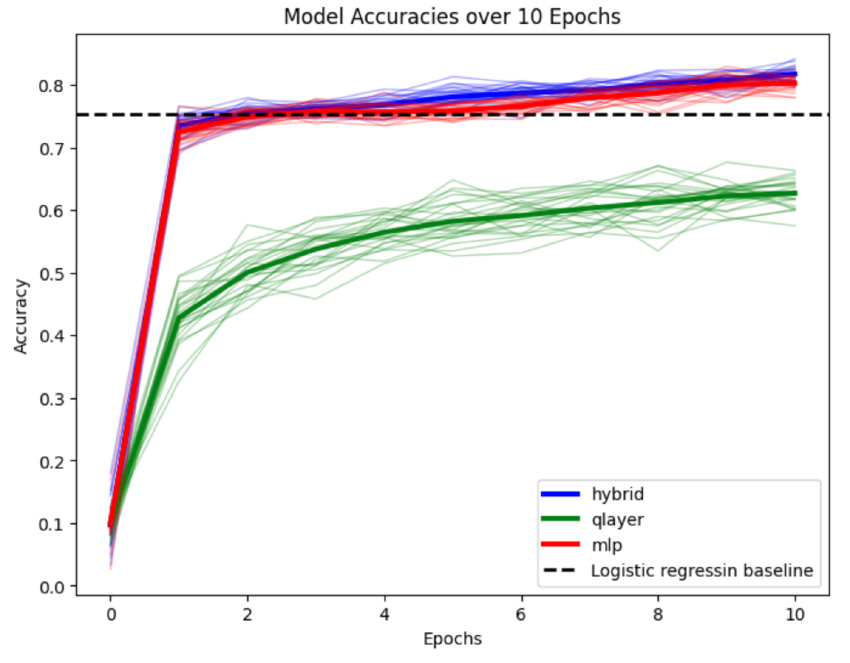}
    \caption{Average test accuracies over 25 runs.}
    \label{fig:solal_test_acc}
\end{figure}

In Figure \ref{fig:solal_test_acc}, we present the mean test accuracies computed over 25 independent runs. Overall, the AFQ model demonstrates inferior performance compared to both the classical and hybrid counterparts. Additionally, the hybrid model exhibits a marginal improvement over the classical model, suggesting that the inclusion of the QuantumLayer may contribute to extracting features beneficial for the classification task.

\subsection{Photonic interferometer for model fine tuning}
In this category of models, photonic quantum circuits are using for fine-tuning, and combined with trainable classical encoders.
\subsubsection{Transfer learning}
\label{sec:tl_prop}
\Proposal
We propose a hybrid quantum–classical transfer-learning (TL) architecture based on a photonic interferometer. This technique is inspired by \cite{mari2020transfer} where the authors highlight the possibility of transferring some pre-acquired knowledge at the classical-quantum interface. 

\begin{figure}[htb!]
    \centering
    \includegraphics[width=0.8\linewidth]{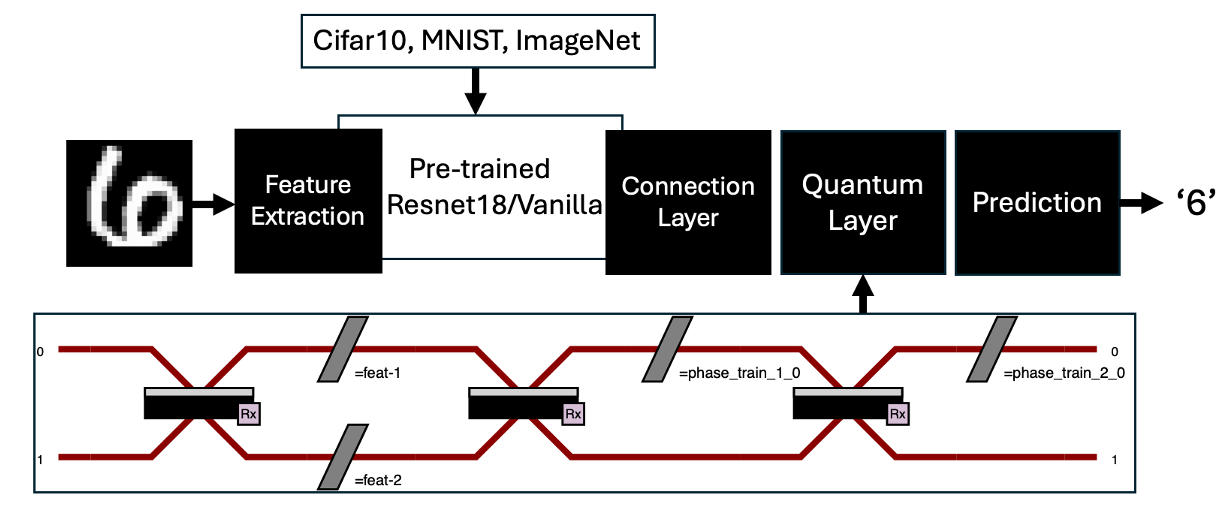}
    \caption{Schematic of the hybrid photonic transfer-learning architecture. The pretrained CNN extracts a 256‐dimensional feature vector from each MNIST image. These features are encoded into optical modes, which then propagate through a programmable photonic interferometer consisting of layered MZI blocks. Photon-counting detectors at each of the 10 output waveguides produce class scores.}
    \label{fig:model_architecture}
\end{figure}

The transfer learning framework consists of two stages: a classical feature‐extractor pretrained on a large-scale image dataset, followed by a photonic neural network that refines and classifies MNIST features in the optical domain. The classical feature extractor is the backbone of a pretrained convolutional neural network (e.g., ResNet-18 \cite{he2016deep}) trained on CIFAR-10. The learned representation $z \in \mathbb{R}^{256}$ is then encoded using two consecutive methods: first, a classical linear encoding maps the 256-dimensional representation vector to the target input dimension; second, quantum feature embedding maps the classical features to phase shifters applied to specific modes. These methods are detailed in Appendix \ref{sec:supp_nomad}. The photonic circuit itself is composed of a series of Mach–Zehnder Interferometer (MZI) blocks arranged in a cascaded Fourier‐mesh topology (Figure \ref{fig:model_architecture}), parameterized by programmable phase‐shifters \(\{\theta_i\}\). Through proper sequential arrangement, these interferometers can realize the complete set of \verb|SU(2)| operations necessary for universal optical quantum processing~\cite{clements2016optimal}. The output layer measures the number of photons at each of the $m$ modes channels after processing through the optical network. These numbers are then mapped classically to the 10 classes.

\Results
To investigate the viability of quantum-enhanced classifiers in a transfer learning (TL) setting, we performed a series of experiments using both classical and quantum post-processing. Specifically, we tested several TL pipelines involving ResNet18 (pretrained on ImageNet or CIFAR-10) and a simple CNN trained on CIFAR-10. The goal was to evaluate the statistical effectiveness of quantum encodings with respect to classical models under a variety of transfer conditions.

Each TL strategy used either full MNIST or selected MNIST classes as the target dataset. For the quantum models, the final classification step was replaced by a photonic encoding and simulated boson sampling circuit. In contrast, classical baselines retained fully classical linear classification heads.

We conducted a series of transfer learning experiments to evaluate the effectiveness of classical versus quantum classifiers, using ResNet18 and a vanilla CNN across several source-target configurations. All results are presented in Table \ref{tab:tl-results}. The first experiment reproduced the setup from Schuld et al \cite{mari2020transfer}, where a ResNet18 model pretrained on ImageNet was used as a feature extractor for MNIST classification. As in the original paper, the classical model achieved over 90\% accuracy without retraining the backbone. The quantum classifier, built using a boson sampling layer, managed to reach accuracy of $\sim$67\%. This result is better than random guessing in a 10-class setting but still it's not close to the accuracy achieved by the classical model. Extending this to general 10-class MNIST classification with the same ImageNet-trained ResNet18, we observed similarly high classical performance (over 92\%). The performance of the quantum model is also similar to the previous experiment ($\sim$67\%). We then evaluated binary TL tasks by selecting visually distinct MNIST digits (e.g., ``1'' vs.\ ``8''), where classical models achieved near-perfect classification accuracy ($>$99\%). The quantum models managed to performed equally well in this simplified task. The achieved accuracy was $\sim$98\%. A similar outcome emerged when classifying visually similar digits like ``3'' and ``5''; although the task was more difficult, the classical model still surpassed 97\% accuracy, and the quantum classifier managed to reach $\sim$96\%. 

To isolate the benefit of feature alignment, we trained ResNet18 directly on full MNIST and transferred it to a 2-class MNIST subset (``3'' vs ``5''), to better map the method in \cite{mari2020transfer}; unsurprisingly, the classical model reached $>$99\% accuracy, whereas the quantum model reached 98\%. We also tested domain transfer by using ResNet18 trained on CIFAR-10 and applying it to MNIST. Despite the domain mismatch, classical performance remained relatively high ($\sim$86\%), reflecting the general utility of early convolutional layers. The quantum classifier yielded 46\%. Both models were evaluated on the 10-class case. Finally, we implemented the transfer from a shallow CNN (either one or two convolutional layers followed by a fully connected layer) trained on CIFAR-10. Even with this simpler architecture, classical accuracy exceeded 95\% (in the 10-class case), while quantum performance was around 55\% in the 10-class case and $>$99\% in the binary case. Across all settings, the classical models benefited substantially from transfer learning, while the quantum models managed to match the performance of classical methods in some tasks. However, in more complex tasks where the dataset contains 10 classes, even though their performance was better than random guessing, it was much worse compared to classical models.

\textbf{Mode selection:} In all experiments involving the boson sampling layer, we explored a range of photon and mode numbers to assess their impact on the model performance. Specifically, we varied the number of modes from 10 (the expected minimum needed to potentially encode digit identity in a 10-class task) up to 24 (our largest tested system size for a suitable runtime). For the number of photons, we experimented with values between 2 and 4 to make it compatible with the first linear layer. These choices were guided by the need to balance computational tractability with the expressive capacity of the quantum system. However, despite this range of configurations, none of the tested combinations led to a noticeable improvement in classification accuracy.


\begin{table}[h]
\centering
\begin{tabular}{@{} lcc @{}}
\toprule
\textbf{Experiment} & \textbf{Classical Accuracy} & \textbf{Quantum Accuracy} \\
\midrule
ResNet18 (ImageNet) $\rightarrow$ MNIST (Reproduction) & $\sim$93\% & $\sim$67\% \\
ResNet18 (ImageNet) $\rightarrow$ MNIST (10-class)     & $>$92\%     & $\sim$67\% \\
ResNet18 (ImageNet) $\rightarrow$ MNIST (1 vs. 8)       & $>$99\%     & $\sim$98\% \\
ResNet18 (ImageNet) $\rightarrow$ MNIST (3 vs. 5)       & $>$97\%     & $\sim$96\% \\
ResNet18 (Full MNIST) $\rightarrow$ MNIST (2-class)     & $>$99\%     & $\sim$98\% \\
ResNet18 (CIFAR-10) $\rightarrow$ MNIST                 & $\sim$86\%  & $\sim$46\% \\
Vanilla CNN (CIFAR-10) $\rightarrow$ MNIST              & $>$95\%     & $\sim$55\% \\
\bottomrule
\end{tabular}
\caption{Comparison of classical and quantum transfer learning accuracy across different source-target setups. Classical models consistently outperform quantum counterparts, often by a significant margin.}
\label{tab:tl-results}
\end{table}

\subsubsection{Self-supervised learning}
\label{sec:qssl_proposal}
\Proposal
This model employs Self-Supervised Learning (SSL) to extract meaningful feature representations through pretext tasks, thereby eliminating the need for labeled data during backbone training. SSL represents a significant paradigm in machine learning that enables the exploitation of vast unlabeled datasets for representation learning. Prominent frameworks in this domain include SimCLR \cite{chen2020simple} and Barlow Twins \cite{zbontar2021barlow}. More precisely, the objective is to leverage photonic quantum computing to learn from unlabelled data. A previous work \cite{jaderberg2022quantum} leverages a gate-based framework as a representation network in a SSL framework. Here, we propose to use a photonic interferometer as a projector network from the representation space to the loss space. The self-supervised framework is as follows: an encoder is taking the $28\times28$ MNIST images as inputs and map them to a representation space of dimension $\mathbb{R}^{r}$, then, these representations are encoded in phase shifters following \cite{gan2022fock} implementation. The interferometer outputs are subsequently compared using a similarity metric $sim$, which serves as the self-supervised loss function for the system. The underlying principle is to enforce invariance in the learned representations: augmented views derived from the same input image should yield similar representations when processed through the interferometer, thereby encouraging the model to learn transformation-invariant features

\begin{figure}[h]
    \centering
    \includegraphics[width=0.7\linewidth]{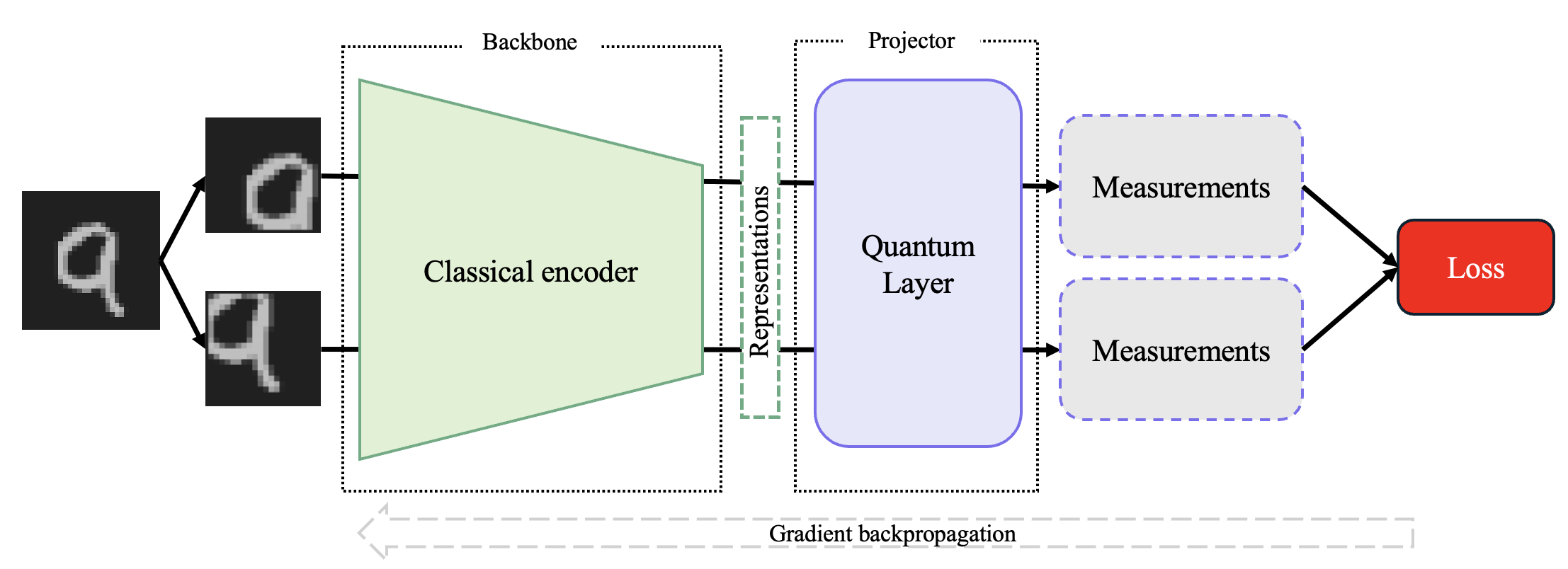}
    \caption{Self-Supervised Learning framework made of a classical backbone, a quantum projector and a classical loss.}
    \label{fig:CodeQalibur_qSSL_model}
\end{figure}

A fundamental component of the SSL paradigm is data augmentation \cite{chen2020simple}: given an input image $\vec{x_i}$, two distinct augmentation functions are applied to generate transformed versions. Common augmentation techniques include cropping, rotation, blurring, and color distortion, as performed in \cite{jaderberg2022quantum}. However, these transformations must preserve the visual distinguishability of the underlying object, which constrains the applicable augmentation strategies for certain datasets. Specifically, for MNIST, the grayscale nature and orientation-dependent semantic content preclude the use of rotation and color distortion. MNIST is not a great candidate for SSL learning but our goal here is to provide a proof of concept for photonic quantum SSL. Therefore, we perform crops at the top left and bottom right of the MNIST image. Gaussian blurring with a low probability was investigated but it was not benefiting the training. Figure \ref{fig:CodeQalibur_qSSL_model} depicts the overall SSL framework.

\Results
For fair comparison with a classical baseline, the quantum layer can be replaced by a Linear layer that maps the representations to a loss space of similar dimension as a Quantum Layer would (i.e. $\binom{m+n-1}{n}$ with photon number resolving detectors or  $\binom{m}{n}$ with single photon receptors, where $m$ and $n$ stands for the number of modes and photons). To evaluate the learned representations, we perform a linear evaluation: the trained backbone is frozen and a fully connected layer is trained on top of it to map the representations to the correct number of classes. To assess the utility of the SSL training, we also perform a linear evaluation on a frozen random backbone. Table \ref{tab:qssl_results} shows the results after training different backbones for 100 epochs and performing linear evaluation during 50 epochs. All experiments are reproduced 5 times. It is important to highlight that these results are not comparable with a fully trained model: for instance, a MLP without activation (\textsc{Linear}: $400 \rightarrow 8 \rightarrow 10$) can provide a validation accuracy of 91.13\%. To evaluate the efficacy of our approach, we conduct comparative analysis across three backbone configurations: those trained using the proposed qSSL pipeline, those trained via classical SSL methods, and untrained networks with random weight initialization serving as a baseline.

\begin{table}[h]
\centering
\begin{tabular}{|p{2.5cm}|c|c|c|c|c|c|}
\hline
\centering \textbf{Model} & \textbf{Loss} & \textbf{Hidden dim} & \textbf{Quantum} & \textbf{Trained} & \shortstack{\bf No\\\bf bunching}& \textbf{Val. ACC. }\\
\hline\hline
\multirow{9}{*}{\centering \parbox{2.5cm}{\centering MLP\\(\textsc{Linear}: $400 \rightarrow 8$)}} & \multirow{8}{*}{\textsc{InfoNCE}} & \multirow{8}{*}{8} & \multirow{4}{*}{Yes} & \multirow{2}{*}{Yes} & False & $35.03\pm3.47$ \\
\cline{6-7}
& & & & & True & $42.8\pm5.83$ \\
\cline{5-7}
& & & & \multirow{2}{*}{No} & False & \cellcolor{green!20}{$47.43\pm3.15$} \\
\cline{6-7}
& & & & & True & $45.70\pm4.03$ \\
\cline{4-7}
& & & \multirow{4}{*}{No} & \multirow{2}{*}{Yes} & False & $37.9\pm3.48$ \\
\cline{6-7}
& & & & & True & $39.97\pm1.66$ \\
\cline{5-7}
& & & & \multirow{2}{*}{No} & False & $40.77\pm4.58$ \\
\cline{6-7}
& & & & & True &  $39.9\pm2.81$  \\
\cline{2-7}
& \multicolumn{5}{c|}{Random} & $40.83\pm 2.14$ \\
\hline 
\multirow{9}{*}{\centering \parbox{2.5cm}{\centering MLP\\(\textsc{Linear}: $400 \rightarrow 32 \rightarrow 8$)}} & \multirow{8}{*}{\textsc{InfoNCE}} & \multirow{8}{*}{32-8} & \multirow{4}{*}{Yes} & \multirow{2}{*}{Yes} & False & $28.2\pm1.24$ \\
\cline{6-7}
& & & & & True & $32.17\pm2.69$ \\
\cline{5-7}
& & & & \multirow{2}{*}{No} & False & $29.27\pm 1.03$ \\
\cline{6-7}
& & & & & True & $30.03\pm3.78$ \\
\cline{4-7}
& & & \multirow{4}{*}{No} & \multirow{2}{*}{Yes} & False & $34.57\pm5.56$ \\
\cline{6-7}
& & & & & True & $24.4\pm 4.43$ \\
\cline{5-7}
& & & & \multirow{2}{*}{No} & False & $32.57\pm7.24$ \\
\cline{6-7}
& & & & & True &  $27.8\pm2.28$  \\
\cline{2-7}
& \multicolumn{5}{c|}{Random} & $23.67\pm 3.46$ \\
\hline
\end{tabular}
\caption{Validation Accuracy for different models after linear evaluation}
\label{tab:qssl_results}
\end{table}
\subsubsection{Future Direction: Leveraging graph isomorphism to classify digits}
\label{sec:mnist_graph}
\Proposal
In this section we outline a novel approach for handwritten digit classification that leverages the computational properties of photonic quantum processors. Although a detailed evaluation  of its performance is left for future work, we present here the main steps of the proposed method.

Most traditional approaches rely on convolutional neural networks~\cite{rawat2017deep, sun2019evolving} or support vector machines~\cite{chapelle1999support, tang2013deep} operating on pixel intensities. Our method transforms MNIST images into graphs through superpixel segmentation~\cite{achanta2010slic, wang2017superpixel, yang2020superpixel, li2015superpixel}, then uses the matrix permanent of the adjacency  matrix of each graph (as well as various of its subgraphs) as a quantum feature. These quantities are particularly relevant in the context of photonic quantum computing, as the matrix permanent governs the probability amplitudes of multiphoton interference events and therefore constitutes the core algorithmic primitive of linear-optical quantum computing~\cite{aaronson2011computational, mezher2023solving}. 

In this approach, the underlying intuition is to represent MNIST images as graphs and to hypothesize that graphs corresponding to the same digit share identical permanent values. Interestingly, this representation is inherently robust to standard image transformations such as rotations or horizontal flips, since these operations preserve the graph structure and therefore its permanent. As a result, the method will fail to distinguish between digits such as 6 and 9, which are related by such transformations. This invariance, however, also indicates that the representation captures a fundamentally different type of information than traditional image-analysis methods based on local pixel intensities. As an illustrative example of its discriminative power, consider two images represented by graphs $G_1$ and $G_2$ of equal size. If these graphs satisfy Theorem~2 in~\cite{mezher2023solving}—that is, if all subgraphs of $G_1$ and $G_2$ have the same permanent under some fixed bijection of vertices—then $G_1$ and $G_2$ are isomorphic, and consequently represent the same underlying graph structure.

Each digit is thus represented by a set of numbers being the permanents of the adjacency matrix of its graph and some randomly chosen subgraphs. These features capture both geometric and topological features of the handwritten character. This approach demonstrates how quantum photonic systems can extract meaningful complementary information that classical models may overlook. Through their intrinsic ability to estimate matrix permanents via boson-sampling protocols, photonic devices can enrich classical neural networks with quantum-derived features relevant to computer vision. Rather than competing with standard encodings, these quantum-estimated features provide orthogonal information that can be leveraged by classical networks.


\begin{figure}[htb!]
    \centering
    \includegraphics[width=0.8\linewidth]{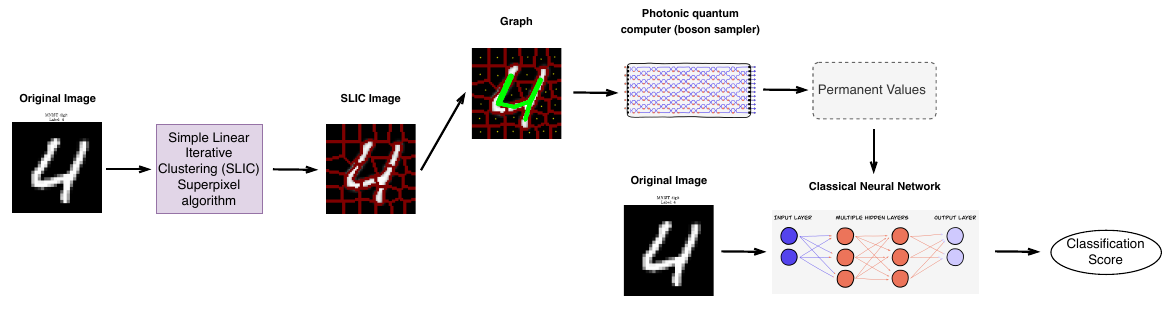}
    \caption{Workflow of the proposed hybrid quantum-classical model for classifying MNIST digit images by transforming them into graphs and using photonic quantum computers to compute their permanent. First, each image is divided into superpixels using SLIC, and the centroid of each superpixel can be treated as a node in the corresponding graph. To ensure a fixed-size graph representation, we select $K$ nodes and edges are constructed between nodes associated with high-intensity regions (illustrated in green). The resulting graph structures serve as inputs to a photonic quantum processor, which is used to evaluate permanent values. Finally, the quantum-generated features are combined with a classical neural network for the final classification step.}
    \label{fig:graph_workflow}
\end{figure}

We will now describe our protocol for constructing a graph from a MINST image. 
The first step  is the \textit{superpixel segmentation}~\cite{wang2017superpixel}. Each 576 pixel
MNIST image is transformed into an \emph{coarse-grained} image of $M<576$ pixels (called \emph{superpixels}) using the Simple Linear Iterative Clustering (SLIC) algorithm~\cite{achanta2010slic, achanta2012slic}. Then, we choose the $K$ superpixels of highest intensities, discarding all other superpixels. The centroids of these $K$ superpixels correspond to the $K$ nodes of our graph, we label these nodes $1, \dots, K$ according to some arbitrary ordering. Our rule for constructing the $K$ edges of our graphs is 

\begin{enumerate}
    \item For all nodes $i=1$, sweep over all nodes $j \neq i$ and choose $j_m$ such that the geometric distance $|i-j_m|$ is minimized. 
    \item Connect the pair $(i,j_m)$ by an edge.
    \item Repeat steps 1 and 2 for all values $i \in \{2, \dots, K\}$ with an additional constraint : discard $j_m$ and restart the search if $(i,j_m)$ are already connected by an edge in a previous step.
\end{enumerate}

In this way, we obtain for each image an associated graph $G$ with $K$ vertices and edges. More complex rules for constructing $G$ are possible—for instance, one could ensure that an edge connecting two high-density regions does not cross a large low-density area, thereby preserving the spatial coherence of digit strokes. However, we believe the above simple rule already already enables non trivial performances. 

For each image $G$, we feed into the classical neural network a set $\mathcal{S}$ of values corresponding to the permanents of the various subgraphs of $G$. In the worst case, as described in Theorem 2 in \cite{mezher2023solving}, $|\mathcal{S}|$ is the number of all possible subgraphs of $G$ of any size. It would be interesting to investigate whether we obtain meaningful performance enhancements when using a smaller number of subgraphs, selected at random. Indeed, the flexibility of our approach lies in the fact that $M$, $K$, and $|\mathcal{S}|$ can all be adjusted to optimize performance. 

In order to compute the various permanents of the subgraphs using a linear optical circuit, we use the approach of \cite{mezher2023solving} (see also section \ref{sec:lancelot_model}) where the adjacency matrix of $G$ is encoded onto the linear optical circuit, by unitary dilation, and relevant output events are post-selected on to estimate $\mathsf{Per}^2(A_{G_s})$ (and consequently $\mathsf{Per}(A_{G_s})$ ), where $A_{G_s}$ is a $(0,1)$ matrix representing the adjacency matrix of a subgraph $G_s$ of $G$.

\begin{figure}[htb!]
    \centering
    \includegraphics[width=0.6\linewidth]{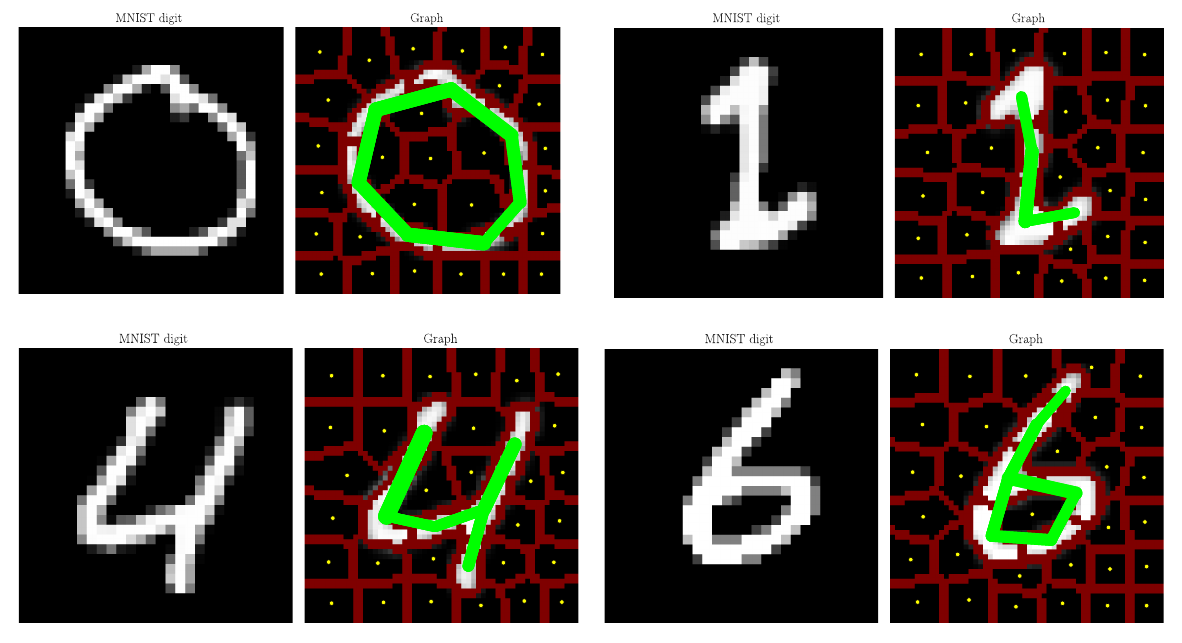}
    \caption{Examples of MNIST digit images transformed into fixed-size graph representations. Black segments denote superpixels corresponding to background regions, while white segments highlight digit strokes. Green nodes and edges represent high-intensity areas capturing the main digit structure. The number of nodes and edges is fixed to ensure uniform graph size, enabling their use as inputs to the photonic quantum processor.}
    \label{fig:graph_mnist}
\end{figure}

In Figure~\ref{fig:graph_mnist}, we provide examples of original MNIST images being transformed into graphs using the technique described previously. Also, Figure~\ref{fig:graph_workflow} contains a diagram that describes the workflow of the model.

\section{Discussion} 
The results obtained from the Perceval Challenge did not reveal any clear evidence of a heuristic quantum advantage on the studied task. This outcome is both expected and informative: the classification problem considered here is already fully solved by classical machine-learning techniques and does not exhibit any structural features for which quantum computation would be expected to provide an edge. Rather than seeking to outperform classical methods, this work provides the first unified set of baseline performances for a wide range of photonic machine-learning (ML) strategies. These results offer a foundation upon which future studies can build, in line with the argument formulated in \cite{Schuld_2022}. The challenge demonstrates that systematic benchmarking and reproducible experimentation are more valuable at this stage than isolated claims of superiority. In this sense, the Perceval Challenge is less a competition than a collective exploration of what photonic learning systems can currently achieve.

\subsection{Lessons from the collective experiment}
Gathering thirteen independently developed methods within a common evaluation framework provides a rare snapshot of the current design space of photonic machine learning. The approaches covered a broad spectrum—from fully end-to-end quantum models (kernels, neural networks and CNN), to architectures where the interferometer served as a feature extractor (enhanced CNNs, enhanced MLP), to methods exploring fine-tuning or transfer-learning strategies (Transfer Learning and quantum Self Supervided Learning (SSL)). While these experiments covered diverse uses of photonic components, none of the approaches explicitly followed a hybrid strategy or aimed at augmenting a state-of-the-art classical model with a quantum module. Exploring such combinations—where quantum photonic circuits could enrich or specialize parts of a classical learning pipeline—could therefore represent an important avenue for future research. This perspective aligns with the idea that near-term quantum ML may benefit less from full replacement of classical architectures than from targeted integration within them.

The Challenge also highlights that hardware-native approaches, such as those relying on the application of the intrinsic permanent computation at the core of boson sampling, should be viewed as complementary to more generic variational strategies. While variational models offer flexibility and trainability, hardware-native circuits embody the physical expressivity of the underlying photonic platform. Comparing both families under a unified benchmark provides insight into how future architectures might blend these paradigms—leveraging hardware efficiency while retaining algorithmic adaptability. The coexistence of these approaches within the same challenge illustrates the field’s diversity and the need for frameworks that allow fair comparison between fundamentally different learning paradigms.

Crucially, the Perceval Challenge represents the first initiative of its kind and scale in photonic quantum computing, bringing together a wide range of teams and methodologies within a common experimental setting. Participants from diverse backgrounds—spanning quantum information, quantum optics, computer science, and artificial intelligence—worked over several months to develop, test, and refine their approaches. This diversity of expertise fostered cross-pollination of ideas and demonstrated that significant progress in quantum machine learning can emerge from open, interdisciplinary collaboration. 

Finally, this collective effort underscored the need for scalable frameworks and robust tooling capable of supporting even larger and more complex benchmarking activities. Conducting the Challenge required the development of dedicated infrastructure to manage submissions, training, and evaluation in a reproducible way. Building on this experience, expanding such infrastructure to handle larger datasets, deeper circuits, and hardware-in-the-loop testing will be essential to sustain community-scale progress. Establishing a standardized, open framework for photonic ML experimentation could make collaborative challenges routine rather than exceptional—accelerating discovery and solidifying best practices across the field.

\subsection{Reproducibility and methodological convergence}
An essential contribution of this work lies in its commitment to reproducibility. The full codebase of all the thirteen implementations is publicly available in a single repository (\url{https://github.com/Quandela/HybridAIQuantum-Challenge}), allowing others to rerun, modify, or extend the experiments. This level of transparency remains uncommon in quantum ML research, where bespoke setups and restricted access to hardware often hinder replication. In addition to the open availability of the code, reproducibility in this Challenge was also reinforced by the fact that most of the proposed approaches were hardware-compliant, designed with the constraints of photonic quantum processors in mind. Several groups even confirmed their results on actual quantum photonic hardware (QPU), demonstrating that the reported performances are not limited to simulation environments. This combination of software transparency and experimental validation represents a strong step toward reproducible and verifiable research in photonic machine learning.

The Perceval Challenge thus establishes a practical standard for openness—similar in spirit to the role of ImageNet or GLUE in the AI community, where progress depends on shared baselines and community-wide evaluation protocols.

Equally significant is the interdisciplinary diffusion of ideas observed throughout the Challenge. Several participants came from classical AI backgrounds, bringing with them rigorous practices in hyperparameter tuning, ablation studies, and validation methodology. Their contributions demonstrate how methodological rigour from the AI world can directly benefit quantum research, ensuring that claims are statistically grounded and experimentally reproducible. This exchange exemplifies a broader cultural shift: quantum ML is evolving from proof-of-principle demonstrations toward data-driven, engineering-oriented experimentation.

\subsection{Limitations and path forward}
A consistent limitation reported by nearly all participants was the computational cost of photonic simulation. The time required to simulate quantum interferometers constrained the exploration of larger architectures and datasets. Consequently, no “large-system” experiments were attempted. Yet, even within these constraints, several photonic approaches achieved performances close to classical baselines—an encouraging sign that small-scale quantum circuits can already encode nontrivial structure. This suggests that with dedicated hardware acceleration and improved simulation tools, progress could follow the same trajectory as classical AI, which took more than a decade to master MNIST but advanced rapidly once reproducible pipelines became standard.

Moreover, while no quantum advantage has been demonstrated, some of the results point to directions worth further exploration. Certain methods exhibited promising behaviors, such as indications that performance might improve with more photons  that comparable accuracy could be achieved with significantly fewer parameters, or that training could converge faster in specific setups. These preliminary signs of potential merit systematic investigation through larger-scale experiments and hardware-based studies, which we leave for future work.

Looking ahead, the Challenge results emphasize the importance of systematic discovery over serendipitous breakthroughs. Faster simulators, standardized datasets, and accessible hardware backends could collectively accelerate iteration cycles, enabling a more data-driven exploration of photonic architectures. Extending future challenges to larger or more diverse datasets, or including dedicated hardware tracks, would help bridge the simulation–experiment gap and provide more realistic performance estimates. Notably, two of the approaches presented here have already led to independent scientific publications\footnote{one accepted to a major conference and another released as an open preprint on arXiv} \cite{xie2025quantum, chen2025distributed} and two additional works are currently under submission. This outcome further attests to the scientific value and lasting impact of this collective effort.

\subsection{Conclusion remarks}
In summary, the Perceval Challenge did not uncover a heuristic quantum advantage—but it has achieved something arguably more fundamental: it has mapped the baseline landscape of photonic machine learning and established the infrastructure for cumulative progress. The field now benefits from open, reproducible implementations spanning a range of hybrid and hardware-native paradigms. Together, these results suggest that quantum photonics can meaningfully contribute to learning tasks, provided it is integrated into hybrid pipelines and studied with methodological rigor.
Echoing Schuld’s perspective, the key question shifts from “{\it Can quantum models outperform classical ones?}” to “{\it How might quantum systems enrich the process of learning itself?}”. The challenge outcomes point toward a phase of convergence—between physics-based and data-driven paradigms—where photonic computing, guided by reproducible methodology and community collaboration, could accelerate the next generation of hybrid intelligence.

\section{Acknowledgments}
This work has been co-funded by the UFOQO Project financed by the French State as part of France 2030.

\section{Author contributions}
The Quandela team wishes to thank all participants of the challenge for their active involvement and contributions to the project.
Team contributions are detailed below:
\begin{itemize}
    \setlength\itemsep{0.2em}
    \setlength\parskip{0pt}   
    \setlength\parsep{0pt}    
    \item Y. Xie formed the Quantum Tree team and developed the surrogate approach described in Section \ref{sec_glase_proposal}. He was the winner of the challenge;
    \item P. Yang formed the Quantum Nomad team and developed the photonic transfer learning approach described in \ref{sec:tl_prop}. He ranked second in our challenge;
    \item O. Zouhry and I. Mejdoub formed the Solal team and developed the feature engineering approach described in Sections \ref{sec:proposal_solal} and \ref{sec:exp_solal}. They obtained the third place in our challenge;
    \item A. Sharma, E.Y. Balaji and S.P. Pawar formed the Qubiteers team and developed the convolutional kernel described in Section \ref{sec:qubiteers_proposal}. They received a special prize for their findings;
    \item K.C. Chen and Chen-Yu Liu formed the QTX team and developed the distributed approach described in Section \ref{sec:qtx_prop};
    \item V. Deumier formed the Lancelot team and developed the unitary dilation approach described in Section \ref{sec:lancelot_model};
    \item C. Marullo, G. Massafra, D.J. Kenne, A.K. Gupta, N. Reinaldet, G. Intoccia and V. Schiano Di Cola formed the Quantum Naples team and developed the feature annotation approach described in Section \ref{sec:naples_proposal};
    \item D. Kolesnyk and Y. Vodovozova formed the Qool team and developed the photonic kernel in Section \ref{sec:qool_proposal};
\end{itemize}

Additionally,
\begin{itemize}
    \setlength\itemsep{0.2em}
    \setlength\parskip{0pt}   
    \setlength\parsep{0pt}    
    \item C. Notton drafted the manuscript and assembled all code in the repository;
    \item V. Apostolou was a member of the Quandela CodeQalibur team and helped develop the approaches in Sections \ref{sec:qnn_proposal} and \ref{sec:qssl_proposal}. He helped revised this manuscript;
    \item A. Senellart was a member of the Quandela CodeQalibur team and helped developed the approaches in Sections \ref{sec:qnn_proposal} and \ref{sec:qssl_proposal};
    \item D. Wang and A. Walsh were members of the Quandela QLOQroaches team and developed the photonic QCNN presented in Section \ref{sec:qcnn_proposal};
    \item R. Mezher was a member of the Quandela CodeQalibur team and  revised this manuscript;
    \item P.E. Emeriau was a mentor in the challenge and revised this manuscript;
    \item A. Salavrakos contributed to the writing and organization of this manuscript;
    \item J. Senellart conceived and supervised the challenge, and guided the overall logic and narrative of the manuscript, notably in the introduction and discussion sections.
\end{itemize}

\bibliographystyle{unsrt}
\bibliography{bibliography}

\newpage
\appendix
\label{sec:supp}
\section{A Quantum Kernel method}
\label{sec:supp_qool}

\subsection*{Implementation details}
We now describe our approach to classifying the MNIST dataset using classical and quantum kernels. All images of the dataset are first reduced from their original 784‐dimensional pixel representation to the top \(m=20\) principal components via principal component analysis (PCA); each component is then normalized to lie in \([0,1]\) and rescaled by a factor of \(\pi\) to form the feature vector \(\vec\varphi\in[0,\pi]^m\).  For the classical baseline, we train standard SVMs on \(N_{\rm train}=600\) examples and validate on \(N_{\rm val}=60\), exploring linear, polynomial, and sigmoid kernels.  The linear kernel \(\kappa(\vec x_i,\vec x_j)=\langle \vec x_i,\vec x_j\rangle\) achieves the highest validation accuracy of \(90.00\%\), while the polynomial kernel \(\kappa(\vec x_i,\vec x_j)=(\gamma\langle \vec x_i,\vec x_j\rangle + c)^d\) and the sigmoid kernel \(\kappa(\vec x_i,\vec x_j)=\tanh(\gamma\langle \vec x_i,\vec x_j\rangle + c)\) each reach \(88.33\%\) under optimal hyperparameter settings determined by a five‐fold cross‐validation grid search.
\section{Leveraging the unitary dilation matrix for feature extraction}
\label{sec:lancelot_supp}
\subsection{Training the UDENN}
\begin{figure}[h]
    \centering
    \includegraphics[width=0.7\linewidth]{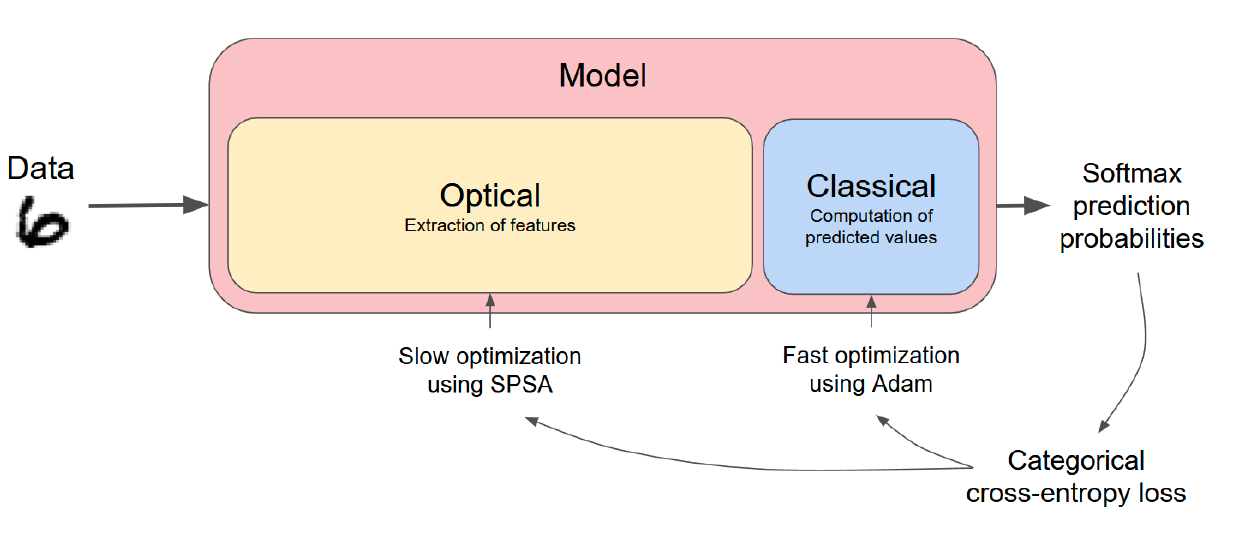}
    \caption{The hybrid model is trained in an alternating fashion}
    \label{fig:lancelot_model}
\end{figure}

In this part, we descrine how the UDENN is trained in an alternating fashion. During the challenge \verb|TensorFlow| was used for the classical part and the model was not trainable in an end-to-end manner. Therefore, the model is divided as two subsystems with an optical and classical components as depicted on Figure \ref{fig:lancelot_model}. 

Drawing from automatic control theory, our hybrid approach leverages the concept of time-scale separation found in singularly perturbed systems. Just as fast and slow subsystems with well-separated eigenvalues can be controlled independently without destabilizing the composite system, the quantum feature extraction and classical processing components operate on sufficiently different computational scales to permit independent optimization strategies.

For the UDENN, the training is performed as follows: for each batch of training data, a full optimization step of the classical parameters is performed while a slight update of the optical parameters is performed in order to ensure the convergence of the whole model.

For the optimization of the optical parameters, the stochastic sub-gradient method Simultaneous perturbation stochastic approximation (SPSA) \cite{spall2002multivariate} is used.
In the results presented in Table \ref{tab:lancelot_results}, the model was trained for 5 epochs.

\subsection{Discussion about the results}
It is important to note that the optical component of the hybrid model likely operates below its full potential due to limited parameter optimization. The SPSA algorithm, while suitable for derivative-free optimization in quantum systems, requires numerous iterations to achieve convergence due to its stochastic nature. In our implementation, the optical parameters underwent only a limited number of updates, potentially constraining the model's representational capacity. Future work should explore more efficient optimization strategies, such as implementing classical gradient-based methods where feasible, to fully realize the quantum component's learning potential.
\section{A Photonic Quantum Neural Network}
\label{sec:supp_pqNN}
In this approach, the goal is to implement a quantum NN. We want to create feature maps that map nonlinear data to a higher dimensional feature space in which a linear decision boundary can be found. Figure \ref{fig:CodeQalibur_pqNN} presents the overall architecture of the model, inspired by \cite{Schuld_2022,gan2022fock}.

Firstly, we want to investigate the encoding technique. Therefore, we propose an ablation study to define the best encoding function $S$ that maps the data from the MNIST dataset to the quantum circuit. Secondly, we investigate on the necessity of repeating this encoding $L$ times.
For these first studies, we will fix the number of modes $m$ of the circuit.
\subsection{The encoding strategy}
In this section $\vec{x}=(x_1,x_2,...,x_d)$ is the data we want to encode in our circuit. Firstly, we want to know what kind of data we want to encode in our circuit: should we encode the raw normalized data, partial data or PCA components?

In these experiments, $L=1$ and $m=10$. For different types of input (all of the images or PCA with $n_{components}$, we vary the encoding strategy:
\begin{enumerate}
    \item A phase/angular embedding: $\forall \vec{x} \in \mathbf{R}^d, S(\vec{x})=2\pi\vec{x}$
    \item A linear embedding: $\forall \vec{x} \in \mathbf{R}^d, S(\vec{x})=\vec{x}$
    \item A learnable scaling embedding: $\forall \vec{x} \in \mathbf{R}^d, S(\vec{x})=\vec{\lambda}.\vec{x}$ where $\vec{\lambda} \in \mathbf{R}^d$ is the vector of learnable scales.
\end{enumerate}
\subsubsection{Phase embedding}
This embedding provides a periodic representation of the data and effectively scales the normalized input to cover a full circle in radians. Figure \ref{fig:embs}  presents the best validation accuracy of the quantum NN with different encoding strategy, compared to the classical baseline. The number of trainable parameters in these models are also given.

Figure \ref{fig:embs} presents in red the validation accuracy of the qNN under different encoding strategies. Additionally, from the training curves, we observe that, for $n_{components}>10$ or the whole image, the model struggles to be trained and, even though the losses decrease slightly, the model plateaus. 

\subsubsection{Linear embedding}
This embedding provides a "cropped" angular representation of the data as it only projects to $[0,1]$. Figure \ref{fig:embs}  displays validation accuracy and number of parameters.
Figure \ref{fig:embs} presents in blue the validation accuracy of the qNN under different encoding strategies. Moreover, from the training curves, we observe that, for $n_{components}>10$ or the whole image, the model struggles to be trained and, even though the losses decrease slightly, the model plateaus but later in the training that with phase encoding. 

\subsubsection{Learnable scaling embedding}
Here, the model can determine the optimal scaling factor for the given task through the training process.  We can write the unitary matrix of the encoding layer such as \[U_e =
\begin{pmatrix}
 \prod_{k=1,k(\bmod m)=1}^{k=784} e^{i\lambda_kx_k} & 0 & ... & 0 \\
 0 & \prod_{k=1,k(\bmod m)=2}^{k=784}e^{i\lambda_kx_k} & ... & 0 \\
 0 & ... & ... & 0 \\
 0 & 0 & ... & \prod_{k=1,k(\bmod m)=0}^{k=784}e^{i\lambda_kx_k} 
\end{pmatrix}\] 

Figure \ref{fig:embs} presents in purple the validation accuracy of the qNN under different encoding strategies. We observe that these learnable embeddings provides a better way to represent the data on the interferometer.

\subsubsection{Frequency of apparition of the data $L$}
Here, we tune the frequency of apparition of the data. Table \ref{tab:cl-pca-frequency} presents the best validation accuracy for $L\in \{1,2,3,5\}$. From these results, it seems that the data encoding strategy does not benefit from multiple apparition of the data.
\begin{table}[h]
    \centering
    \begin{tabular}{c||c|c}
        Frequency apparition $L$ & Best Val. ACC. (quant.) & \#param (quant.) \\ \hline\hline
        1 & 61.95 & 21294 \\
        2 & 59.81 & 22238 \\
        3 & 51.15 & 23182 \\
        5 & 40.98 & 25070 \\ \hline
    \end{tabular}
    \caption{Best validation accuracy results for the quantum NNs with different frequency of apparition of the data}
    \label{tab:cl-pca-frequency}
\end{table}

\subsubsection{Conclusion}
\begin{figure}[h]
    \centering
    \includegraphics[width=0.5\linewidth]{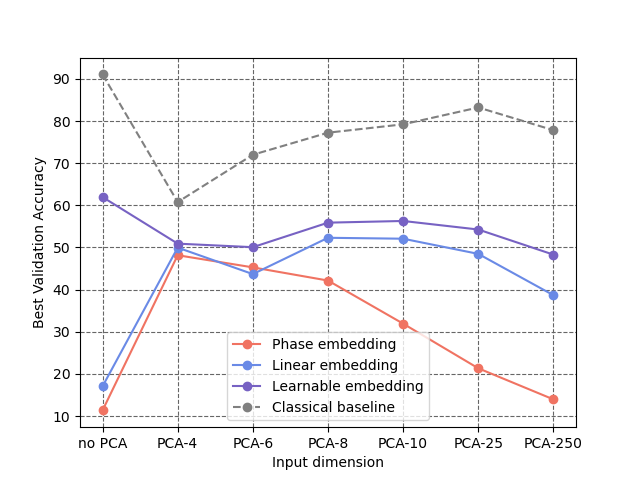}
    \caption{Validation accuracy with respect to the different embedding strategies and input type}
    \label{fig:embs}
\end{figure}

From Figure \ref{fig:embs}, we conclude that the learnable embedding provides the best accuracy throughout the different encoding strategies. Moreover, there is no benefit in repeating the data : we keep $L=1$.

\subsection{Training the quantum Neural Network}

\subsubsection{Training pipeline and hyperparameters}

\begin{figure}[h]
    \centering
    \includegraphics[width=0.7\linewidth]{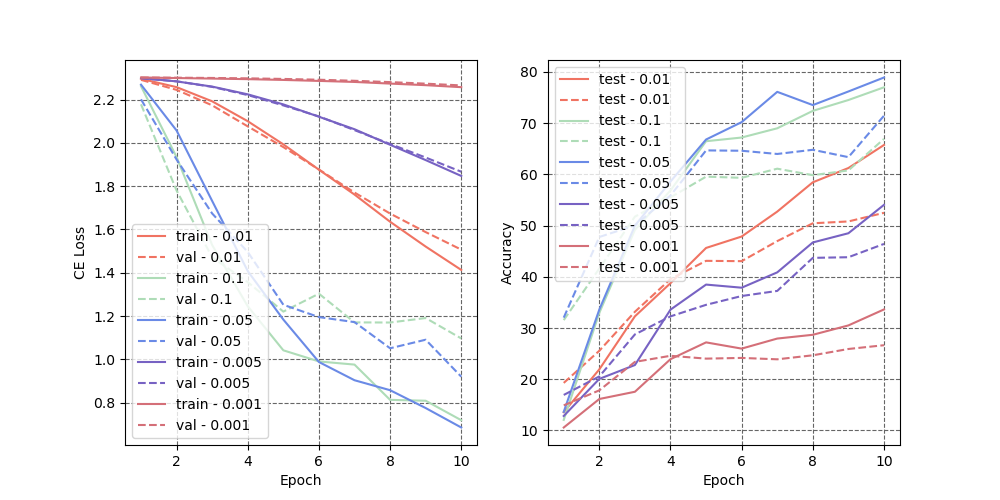}
    \caption{Training and validations losses and accuracies with respect to the learning rate}
    \label{fig:lr}
\end{figure}

\textbf{Influence of the learning rate:} in the following experiment, we study the influence of the learning rate in the dynamic of the training. Previous experiments were done with a learning rate \verb|lr=0.01|. The optimizer used is Adam vanilla: \verb|optimizer = torch.optim.Adam(model.parameters(),lr = lr)|
\begin{table}[h]
    \centering
    \begin{tabular}{c|c}
        \texttt{lr} & Best Val. Accuracy  \\ \hline\hline
        0.01 & 61.95 \\ 
        0.1 & 66.93 \\
        0.05 & 71.45 \\
        0.005 & 52.49 \\
        0.001 & 26.64 \\ \hline
    \end{tabular}
    \caption{Validation accuracy based on the learning rate}
    \label{tab:lr_rexp}
\end{table}
From Table \ref{tab:lr_rexp}, it seems that the Quantum Layer benefits from a larger learning rate. Figure \ref{fig:lr} comfirms this observation: with a small learning rate (\verb|lr = 0.001, lr = 0.005|), the convergence is very slow and even plateaus for \verb|lr = 0.001|. The best convergence occurs for \verb|lr = 0.05|.

\textbf{Influence of the weight decay}: for Adam optimizer, the weight decay is a regularization technique that aims at preventing overfitting by penalizing large weights. Here, we first use \verb|weight_decay = 0| and then decrease it but the best results are observed with \verb|weight_decay = 0|. One interpretation could be that there is no large gradients or weights to penalize here.

\textbf{Influence of $\beta_1,\beta_2$}: these are the coefficients used for computing running averages of gradient and its square. The default value is \verb|(0.9, 0.999)|. We obtain better results on this specific validation set using \verb|(0.8, 0.9999)| and :
\begin{itemize}
    \item a lower $\beta_1$ means a reduction of the momentum's influence, which makes the optimizer more responsive to recent gradients by allowing quicker changes.
    \item a higher $\beta_2$ means that we create more stable adaptive learning rates by taking longer history of squared gradients into account and that can prevent aggressive learning rate fluctuations
\end{itemize}

\subsubsection{Influence of the modes and number of photons}

\begin{figure}[h]
    \centering
    \includegraphics[scale = 0.5]{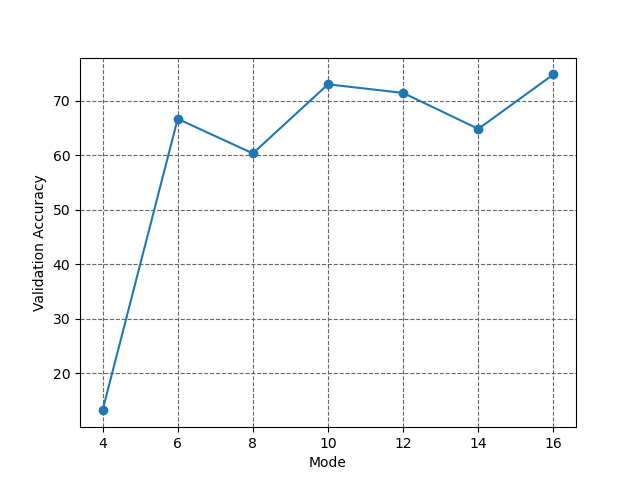}
    \caption{Validation accuracy with respect to number of modes}
    \label{fig:modes}
\end{figure}

\textbf{Influence of the number of modes}: Figure \ref{fig:modes} presents the validation accuracy with respect to the number of modes. Overall, increasing the number of modes seems to allows better generalization and therefore better accuracy.

\begin{figure}[h]
    \centering
    \includegraphics[scale = 0.5]{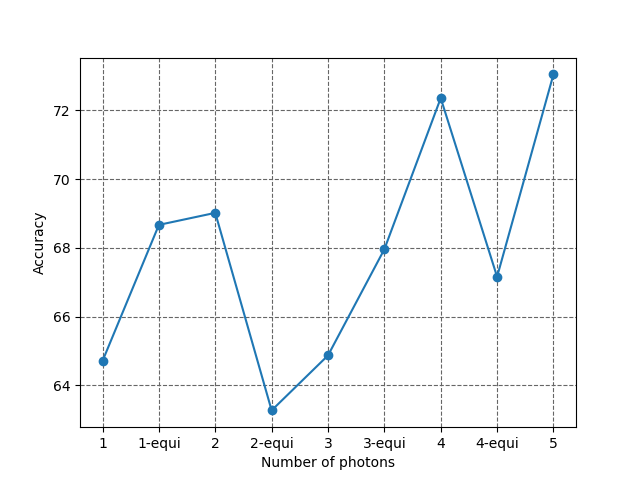}ma
    \caption{Validation accuracy with respect to number of photons and their entry in the interferometer, N-equi corresponds to N photons equi-placed at the entry, whereas N corresponds to an input state such as [1,0,1,0...,N,0]. Here, we use 10 modes.}
    \label{fig:photons}
\end{figure}

\textbf{Influence of the number of photons}: Figure \ref{fig:photons} shows the validation accuracy with respect to the number of photons for an interferometer with 10 modes. Overall, it seems that more photons, placed "one out of two" from the first mode allows better generalization

\subsubsection{Conclusion on the MNIST Dataset}
\begin{figure}[h]
    \centering
    \includegraphics[width=0.7\linewidth]{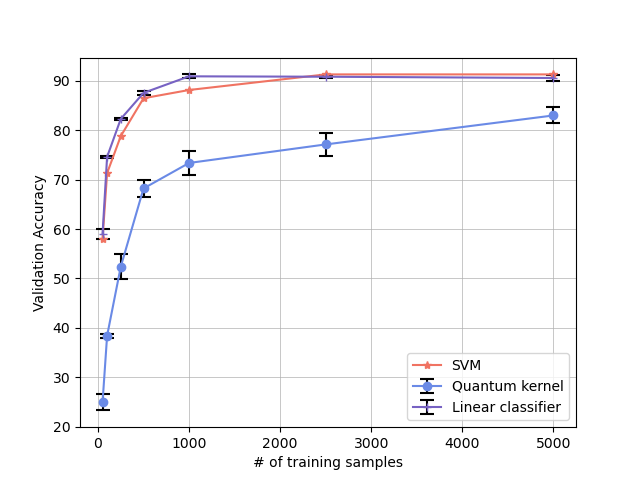}
    \caption{Validation accuracy for different sizes of training sets with different models}
    \label{fig:mnist_final}
\end{figure}

\begin{figure}[h]
    \centering
    \includegraphics[width=0.7\linewidth]{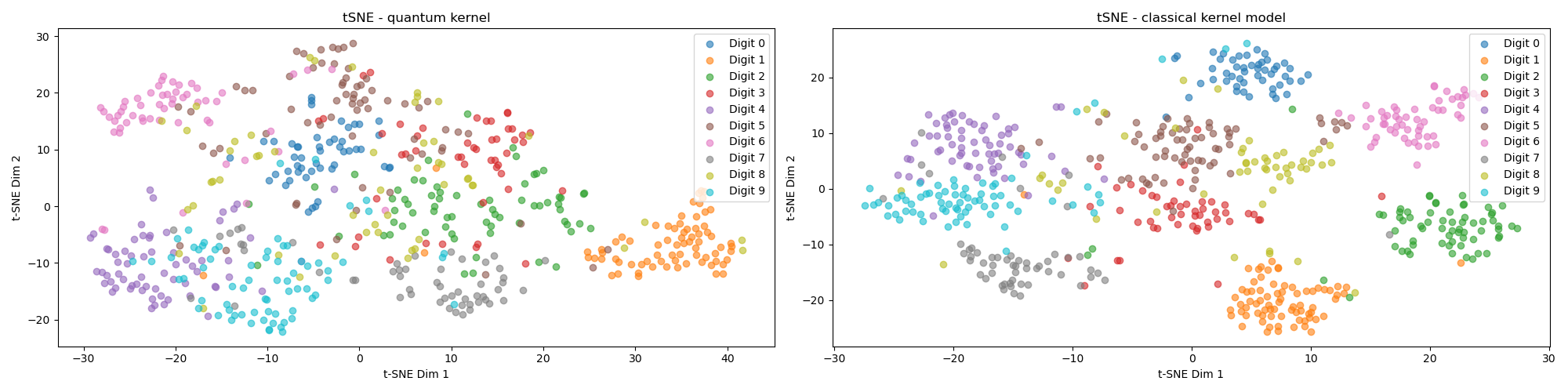}
    \caption{tSNE for the quantum NN and the linear kernel on 5000 training samples. The representations provided by the classical classifier are of higher quality and more discernable (distinct clusters) than the ones provided by the quantum NN}
    \label{fig:tsne_mnist}
\end{figure}

Using the circuit from Figure \ref{fig:CodeQalibur_pqNN} with $L=1$, 10 modes, \verb|input_state = [1,0,1,0,1,0,1,0,1,0]|. Table \ref{tab:final_mnist_results} presents the validation accuracy for different sizes of training sets compared to a linear classifier made of 2 linear layers, with similar number of parameters, and a SMV (using \verb|svm.SVC(kernel="linear")| with \verb|scikit-learn|). For better visualization, Figure \ref{fig:mnist_final} presents the same results. We observe that the quantum NN does not perform as well a the classical classifiers and needs more training samples to achieve good enough results. Additionaly, we can observe the t-SNE (t-distributed Stochastic Neighbor Embedding) of these classifiers. t-SNE is a dimensionality reduction technique used for visualizing high-dimensional data in 2D or 3D space. Unlike PCA, which preserves global structure, t-SNE emphasizes preserving the local relationships between points, making it particularly effective for visualizing complex datasets where clusters exist. tSNE for the classifiers trained on 5000 samples is shown in Figure \ref{fig:tsne_mnist}. We observe that the representations provided by the classical classifier are of higher quality and more discernable than the ones provided by the quantum NN.

\begin{table}[h]
    \centering
    \begin{tabular}{c|c|c|c}
        Model & Validation ACC & \# parameters & Training Samples \\\hline\hline
        quantum NN & $24\pm1.62$  & 21294 & 50 \\
        Linear Layer & $58.98\pm 1.07$ & 25450 & 50 \\
        SVM (linear kernel) & 58 & 7850 & 50 \\ \hline
        quantum NN & $38.39\pm 0.43$ & 21294 & 100 \\
        Linear Layer & $74.62\pm 0.23$ & 25450 & 100 \\
        SVM (linear kernel) & 71.33 & 7850 & 100 \\ \hline
        quantum NN & $52.38\pm 2.53$ & 21294 & 250 \\
        Linear Layer & $82.31\pm 0.19$ & 25450 & 250 \\
        SVM (linear kernel) & 79.83 & 7850 & 250 \\ \hline
        quantum NN & $68.23\pm1.8$  & 21294 & 500 \\
        Linear Layer & $87.59\pm 0.49$ & 25450 & 500 \\
        SVM (linear kernel) & 86.5 & 7850 & 500 \\ \hline
        quantum NN & $73.39\pm 2.39$ & 21294 & 1000 \\
        Linear Layer & $90.94\pm 0.37$ & 25450 & 1000 \\
        SVM (linear kernel) & 88.17 & 7850 & 1000 \\ \hline
        quantum NN & $77.15\pm 2.28$ & 21294 & 2500 \\
        Linear Layer & $90.86\pm 0.25$ & 25450 & 2500 \\
        SVM (linear kernel) & 91.33 & 7850 & 2500 \\ \hline
        quantum NN & $83.02\pm 1.61$ & 21294 & 5000 \\
        Linear Layer & $90.6\pm 0.54$ & 25450 & 5000 \\
        SVM (linear kernel) & 91.33 & 7850 & 5000 \\ \hline
        
    \end{tabular}
    \caption{Validation accuracy for different sizes of training sets for the quantum NN, a linear classifier and a SVM with linear kernel}
    \label{tab:final_mnist_results}
\end{table}

\begin{figure}[h]
    \centering
    \includegraphics[width=0.7\linewidth]{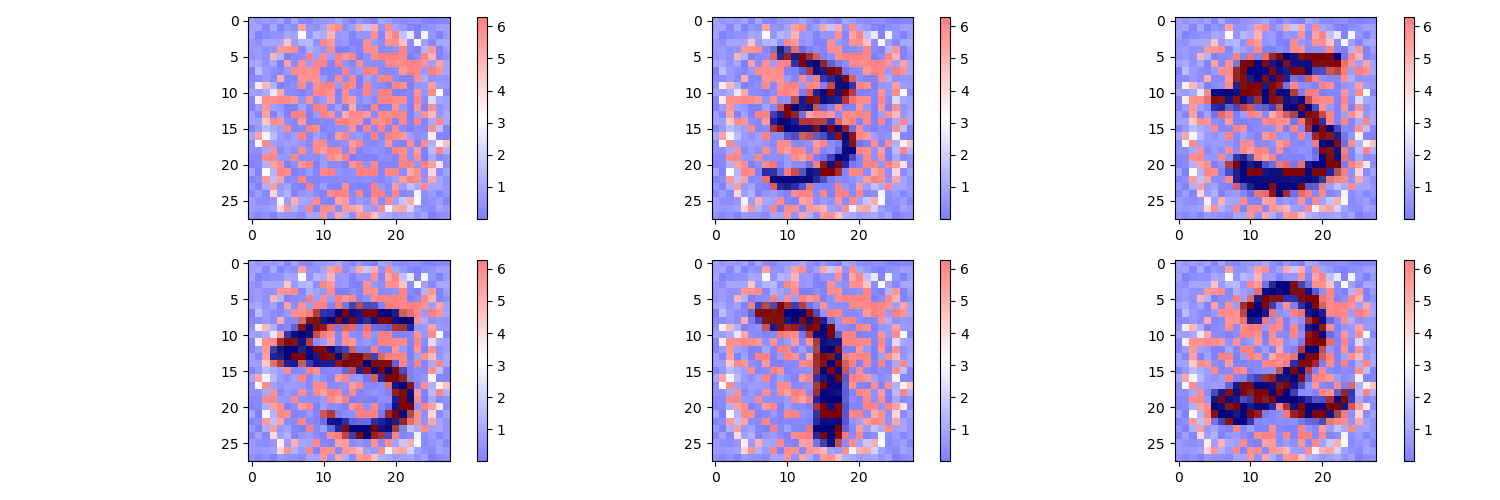}
    \caption{The learned scale parameters in the embeddings applied to different images in the validation set of MNIST}
    \label{fig:scale_mnist}
\end{figure}

\textbf{Influence of the learned scale embedding}: considering that we use the learnable embeddings, we could wonder if the learned scale parameters have meanings for the data. Figure \ref{fig:scale_mnist} displays the learned parameters (scaled between $[0,2\pi]$) overlayed on top of validation samples from the MNIST dataset. We do not observe any specific patterns highlighted, but it seems that the model draws more attention to the region at the center of the image.
\section{GLASE: Gradient-free Light-based Adaptive
Surrogate Ensemble}
\label{sec:supp_glase}
\subsection{Mathematical background}
Recall from Section \ref{sec:preliminaries} that, given a photon number state basis $\mathbf{n} = (n_1, \dots, n_M)$, the output probability distribution over multimode measurement outcomes is defined as
\[
p_{\mathbf{n}}(\boldsymbol{\phi}) = \frac{|\text{Perm}(U_{\mathbf{n}})|^2}{n_1! \cdots n_M!},
\]
where $U_{\mathbf{n}}$ is the submatrix of $U$ corresponding to the detected modes and $\text{Perm}(\cdot)$ denotes the matrix permanent. In practice, due to the exponential cost of computing the full distribution, we compute the expected photon count per mode:
\[
\langle \hat{n}_i \rangle = \sum_{\mathbf{n}} n_i \cdot p_{\mathbf{n}}(\boldsymbol{\phi}),
\]
which serves as the output signal from the photonic layer. To enable differentiable learning, we introduce a surrogate neural network $g_{\alpha}(\boldsymbol{\phi})$ trained to approximate this mapping:
\[
g_{\alpha}(\boldsymbol{\phi}) \approx \langle \hat{\mathbf{n}} \rangle.
\]

\subsection{Data encoding process}
We adopt a phase-based bosonic embedding approach. Each latent feature vector $\mathbf{z}$ is projected into the interferometer’s parameter space using a fixed learnable projection $\boldsymbol{\phi} = \Pi(\mathbf{z})$, where the phase shifts define the transformation matrix $U(\boldsymbol{\phi})$. This is implemented through Perceval's GenericInterferometer object. The resulting expectation values of photon numbers in each mode—collected from multi-shot simulations or real QPU executions—serve as the input to a downstream softmax layer for classification. 

\subsection{Discussion}
The experimental behavior of our GLASE architecture reinforces the importance of architectural alignment, surrogate modeling fidelity, and photonic system expressivity in hybrid quantum-classical pipelines. Our results show that surrogate-assisted optimization can effectively stabilize training and outperform purely classical models, but several subtleties emerge when analyzing surrogate interactions and QPU deployment.

One key insight lies in the update frequency of the surrogate model. While frequent updates ensure tighter alignment between the neural surrogate and the true photonic expectation values, they also introduce overhead and potential overfitting to intermediate simulation noise. We observed that updating every 5 epochs provides the best trade-off, too infrequent updates cause performance to degrade, and overly frequent updates yield diminishing returns.

The complexity of the surrogate model is another dimension of the trainability-expressivity trade-off. A deeper surrogate approximates the photonic behavior more accurately, particularly for larger circuits with more photons and modes. However, we found that a 3-layer MLP was sufficient to approximate most photonic behaviors while maintaining computational efficiency. Adding more depth did not yield further performance gains, suggesting it is not necessary for the surrogate to match the full expressivity of the quantum system, only provide a smooth local approximation for backpropagation.

We also investigated the scaling behavior with respect to photons and modes. Intuitively, we believe that increasing the number of modes allows the photonic network to capture higher-dimensional structure in the data, while adding more photons should provide richer interference patterns. Together, this can increase classification performance—especially for datasets like MNIST where digit class boundaries benefit from nonlinear transformations. However, practical constraints in QPU mode count limited real-hardware deployment to 16 modes, which led to a noticeable drop in performance due to noisy sampling and collision effects.

A significant takeaway from our experiments is the importance of preserving alignment between classical feature extractors and the structure of the photonic encoder. The GLASE approach relies on a linear map $\boldsymbol{\phi} = \Pi(\mathbf{z})$ from classical features to phase parameters. While this works well in simulation, misalignment or limited resolution in hardware (e.g., phase discretization or mode crosstalk) can severely degrade performance. 

Moreover, our method introduces a secondary optimization loop that must be tuned with care. If the surrogate fails to accurately approximate photon statistics, especially in regions of parameter space far from training samples, it can misguide gradient updates. Fortunately, in our experiments, the surrogate loss consistently correlated with downstream classification loss, providing a reliable signal for updating the front-end encoder.

Hardware deployment revealed another critical bottleneck: postselection. While our simulated circuits assume access to full probability distributions or expectation values, real QPU shots must be interpreted through postselected collision-free events, introducing sampling variance and limiting effective throughput. This challenge, combined with photon loss and phase instability, highlights the need for robust, noise-aware surrogate modeling and potentially hybrid training strategies that alternate between simulated and real data.

Finally, we note that GLASE avoids issues related to gradient-based quantum learning. Because it sidesteps gradient flow through the quantum layer entirely, training is governed solely by the behavior of the surrogate and the classical front-end. This makes it particularly well-suited for noisy intermediate-scale quantum (NISQ) devices, where gradient instability and sampling noise are major obstacles.

In summary, our findings confirm that surrogate-based learning offers a viable and scalable strategy for training photonic quantum neural networks. The surrogate acts not only as a practical tool for gradient approximation but also as a bridge that harmonizes classical learning dynamics with the structure of photonic computation. Future work may explore jointly learned encoding schemes, adaptive surrogate architectures, and the integration of noise models to further improve robustness on real hardware. Additionally, extending this approach to time-resolved photonic systems or continuous-variable encodings may expand its applicability to broader quantum machine learning domains.
\section{A photonic native quantum convolutional neural network}
We provide more details about the photonic QCNN architecture.
\subsection{Data encoding}
\paragraph{}In order to encoding classical 2D-structures of dimension $N_1\times N_2$, such as greyscale images, we make use of a strategy which uses 2 blocks of respectively $N_1$ modes and $N_2$ modes, each containing a single photon. This can be seen as encoding a path-encoded qudit, where:
\begin{align*}
    \ket{e_0} =& \ket{1, 0, 0, \ldots, 0}\\
    \ket{e_1} =& \ket{0, 1, 0, \ldots, 0}\\
    \vdots& \\
    \ket{e_{N_i - 1}} =& \ket{0, 0, 0, \ldots, 1}
\end{align*}
The values of the pixel are then encoded using the amplitude of the corresponding state. In addition, since we require the input quantum state to be normalised, we rescale the images such that for an image $\mathbf{x} = \left(x_{i,j}\right)_{i,j}$:
\begin{equation}
    \ket{\psi_{in}} = \sum_{i,j} \frac{x_{i,j}}{\left|\left|\mathbf{x}\right|\right|_2} \ket{e_i}\ket{e_j}
\end{equation}
where:
\begin{equation}
    \left|\left|\mathbf{x}\right|\right|_2 = \sqrt{\sum_{i, j}x_{i,j}^2}
\end{equation}

Since this is a state containing 2 photons, there exists a probabilistic procedure to prepare the input state using ancilla photons and mode, and heralding~\cite{de2024simple}.

This choice of encoding keeps the local features local and is therefore very useful for  image processing tasks. In addition, although we here only focus on 2D structures, this encoding can also easily be extended to arbitrary tensors by simply adding more registers (i.e. qudits) to the input state.

\subsection{Convolutional layer}\label{app:qloqroachConv}
\paragraph{}Given the encoding described in the previous section, it is possible to define translational invariant operations on the input data. We first define the kernel size $K$, which corresponds to the size of the receptive field (for simplicity, we will assume that $K$ is the same in both dimensions). Then, we define two operations $U_1$ and $U_2$ on $K$ modes each. The operation $U_1$ (resp. $U_2$) are then applied in parallel on $\left\lfloor\frac{N_1}{K}\right\rfloor$ (resp. $\left\lfloor\frac{N_2}{K}\right\rfloor$) distinct blocks of $K$ modes. We will take these unitaries $U_1$ and $U_2$ to be universal interferometers.

These operations are, by design, translation invariant with respect to horizontal and vertical shifts by $K$ pixels (but not arbitrary translations). In fact, a $K\times K$ quantum filter will produce $K\times K$ convolutions acting locally on each patch. This can be seen as follows. Each patch:
\begin{equation}
    \ket{\psi} = \sum_{i,j=0}^{K-1} \alpha_{i,j} \ket{e_{n_1K+i}}\ket{e_{n_2K+i}}
\end{equation}
is sent to the new state:
\begin{align}
    U_1\otimes U_2\ket{\psi} = \sum_{k,l = 0}^{K-1}\sum_{i,j=0}^{K-1} W^{(k,l)}_{i,j}\alpha_{i,j} \ket{e_{n_1K+k}}\ket{e_{n_2K+l}}
\end{align}
where for each $k, l=0,\ldots, K-1$ the filter $W^{(k,l)}$ is defined as:
\begin{equation}
    W^{(k,l)}_{i,j} = \left(U_1\right)_{i,k}\left(U_1\right)_{j,l}
\end{equation}

\subsection{Pooling layer}
\paragraph{}The aim of the pooling layer is to reduce the dimension of the image. Therefore, our approach is to use measurements in order to discard some of the mode. In particular, we will decide to measure every other mode, such that the dimension of the image after pooling is halved. 

However, if one of the photon is measured, we will leave the encoding space, as one of the register will no longer contain a photon. In order to stay within the encoding space, we will then \emph{adaptively inject a photon} to the corresponding mode whenever a photon is measured. Since there are 2 photons in total in the circuit, there is a maximum of 2 photons that need to be injected during a pooling layer. Then, an adaptive measurement will redirect a photon stores on an ancilla mode to the correct mode whenever a photon is measured.

\subsection{Dense layer}
\paragraph{} The dense layer consists simply of a generic interferometer over all the available modes. It is the only operations (after state preparation) where the photons are allowed to interfere. 
\section{A convolutional layer using a photonic quantum kernel}
\label{sec:supp_qubiteers}
This section aims at describing the different encoding strategies used in the photonic PQK described in Section \ref{sec:qubiteers_proposal}, the training parameters conducint to the results presented in Section \ref{sec:qubiteers_results} and an ablation study on the different hyperparameters.

\subsection{Encoding strategy}

The PQK acts like a convolutional kernel by processing each $k\times k$ image patch through a photonic interferometer. For a $k\times k$ kernel, $m = \left\lceil kernel\_size^2/2 \right\rceil$ modes are used. The first $m$ pixels are encoded in a phase shifter on each mode. This first layer of phase shifters is followed by a row of beam splitters and followed by the remaining $m-1$ pixels to be encoded.

\begin{figure}[!ht]
\centering
\includegraphics[width=0.5\columnwidth]{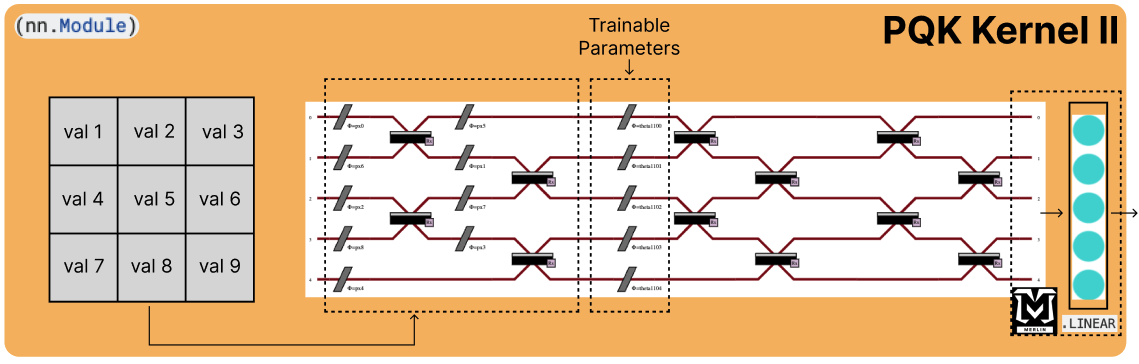}
\caption{Photonic quantum kernel circuit: Type~2 (delayed) encoding.}
\label{fig:type2_circuit_trainable}
\end{figure}

This encoding scheme is extended by introducing trainable interference parameters. After the pixel values are encoded, interference between modes is governed by these adjustable parameters. An example of this trainable kernel circuit is provided in Figure~\ref{fig:type2_circuit_trainable}.

\begin{figure}[!ht]
\centering
\includegraphics[width=0.5\columnwidth]{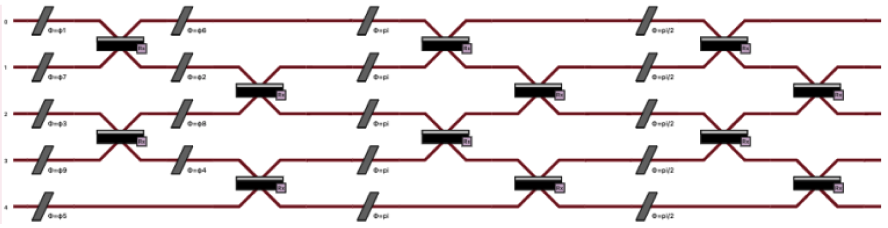}
\caption{Photonic quantum kernel circuit: Type~2 (delayed) encoding.}
\label{fig:type2_circuit}
\end{figure}

\subsection{Hybrid architecture components}
We provide implementation details about the different components of the hybrid framework:
\begin{itemize}
    \item \textbf{Classical Branch:} Applies standard CNN operations on the raw $28\times28$ grayscale image, extracting spatial features using convolution, pooling, and ReLU. We have two layers of classical CNN. The first layer has a kernel size of $3\times3$ with a padding of $0$ and outputs $16$ filters. The input for this layer is (1, 28, 28) and the output is of the shape (16, 26, 26). The second layer has the kernel of size $5\times5$, again with padding $0$ and outputs $32$ filters. For this layer, the input is (16, 26, 26) and output is (32, 22, 22).
    \item \textbf{Quantum Branch (Non-Trainable):} Applies PQK-based convolutions with stride 2 over $2\times2$ patches. Each patch yields a 5 or 20 channel feature vector, aggregated into a $14\times14\times N$ quantum feature map. These are passed through one or two small convolutional layers with ReLU to enhance representation. The PQK convolution was accelerated with the help of multi-threading and multiple sessions managed with the help of Scaleway. The input images are pre-processed with the help of these non-trainable kernel convolutions and then given as input to the classical post-NN to get the final class label.
    \item \textbf{Quantum Branch (Trainable Kernel):} We also implement a parallel-branched model where the quantum branch performs the convolution operations of the trainable Type~2 PQK. In this case, the original input images cannot be pre-processed and stored and used later when required. Since the kernel circuit in this case contains trainable parameters, the PQK convolutions need to be applied every time the model is called. To implement this, a custom convolution class is implemented and the kernek circuit can be trained.
    \item \textbf{Fusion:} Classical and quantum outputs are concatenated channel-wise. The combined tensor is flattened and passed through a dense network (128 hidden units, 10 output classes). This fusion allows the network to exploit both standard pixel features and high-order quantum-derived patterns. In a slightly different setup, the outputs from the final classical and PQK layers are concatenated to form tensors of shape (64, 22, 22). This is passed through a classical CNN layer of kernel size $3\times3$ with a padding of $1$. The output from this layer is (32, 22, 22). This tensor is then flattened and then passed through a classical feedforward neural network (FNN) (15488 $\rightarrow$ 512 $\rightarrow$ 64 $\rightarrow$ 10). This setup can be seen in Figure~\ref{fig:PQK_model}.
  \end{itemize}
\subsection{Training set-up}
All models are trained on the subset of MNIST dataset used in this challenge, consisting of 6,000 training and 1,000 test images. Training is performed with mini-batches of size 32, using the Adam optimizer with a learning rate of $10^{-3}$ and cross-entropy loss. The baseline CNN converges within 20 epochs. Hybrid models train up to 50 epochs with early stopping. For 5-channel PQK embeddings, convergence may require up to 100 epochs due to reduced input dimensionality.

\subsection{Training curves}

\begin{figure}[h]
  \centering
  \includegraphics[width=0.95\columnwidth]{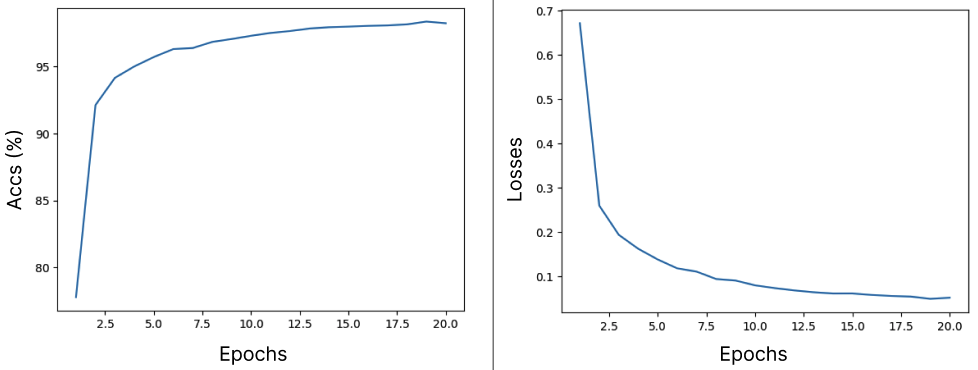}
  \caption{Training accuracies (left) and losses (right) for 20 epochs with the model having two convolution layers of Type~2 PQK, with kernel sizes 3 and 5 respectively.}
  \label{fig:extended_learning_curves}
\end{figure}

\begin{figure}[h]
  \centering
  \includegraphics[width=0.95\columnwidth]{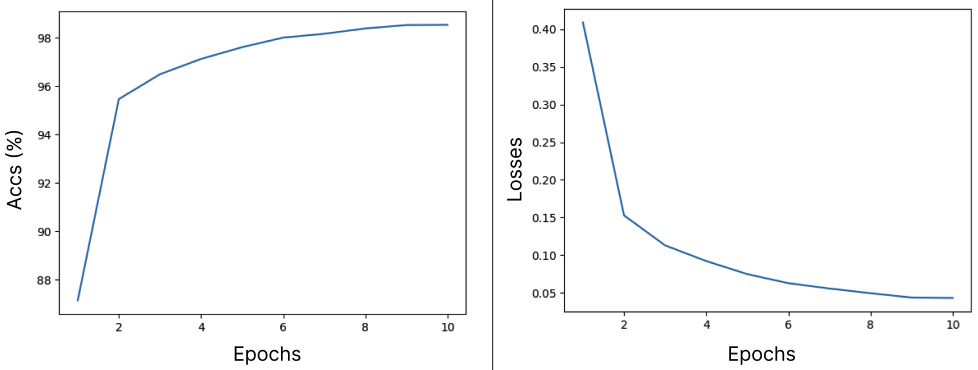}
  \caption{Training accuracies (left) and losses (right) for 10 epochs with the model having one convolution layer of Type~2 PQK, with kernel sizes 3.}
  \label{fig:layer1_curves}
\end{figure}

\begin{figure}[H]
  \centering
  \includegraphics[width=0.95\columnwidth]{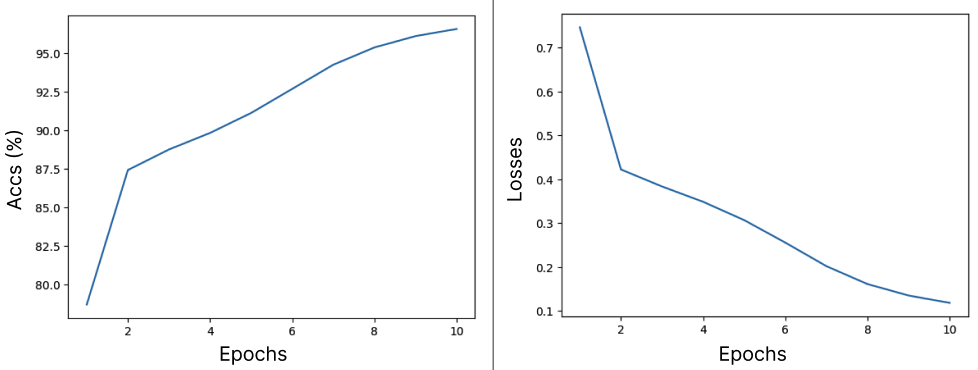}
  \caption{Training accuracies (left) and losses (right) for 10 epochs with classical-quantum parallel channel model; both channels with two convolution layers (Type~2 PQK for quantum) of kernel sizes 3 and 5.}
  \label{fig:parallel_curves}
\end{figure}

\begin{figure}[h]
  \centering
  \begin{minipage}{0.49\columnwidth}
    \centering
    \includegraphics[width=\linewidth]{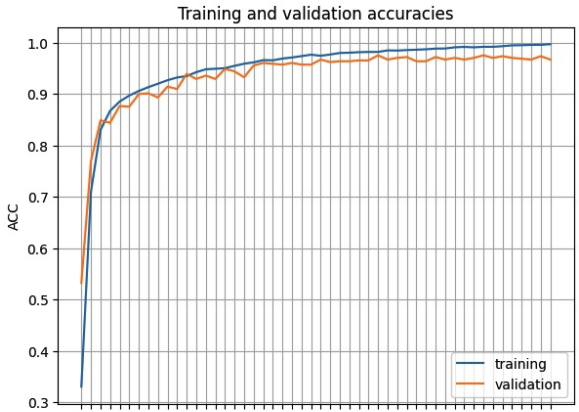}
  \end{minipage}
  \hfill
  \begin{minipage}{0.49\columnwidth}
    \centering
    \includegraphics[width=\linewidth]{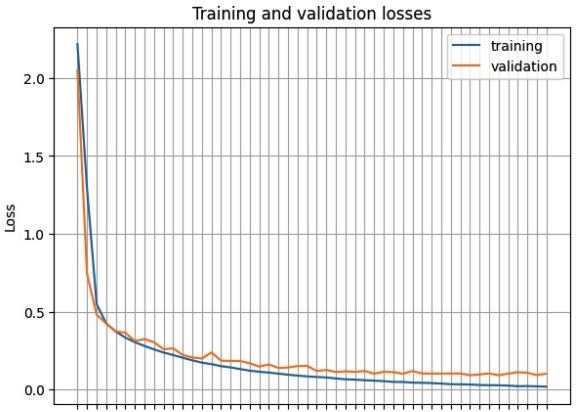}
  \end{minipage}

  \caption{Training and validation metrics for the Classical models with original dataset as
input.}
  \label{fig:p_11}
\end{figure}

\begin{figure}[H]
  \centering
  \begin{minipage}{0.49\columnwidth}
    \centering
    \includegraphics[width=\linewidth]{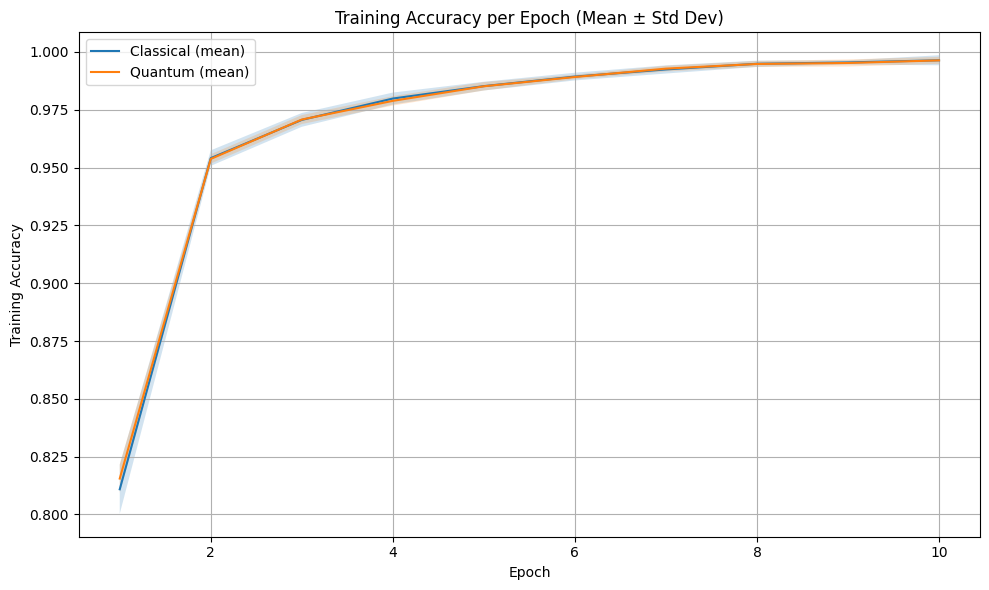}
  \end{minipage}%
  \hfill
  \begin{minipage}{0.49\columnwidth}
    \centering
    \includegraphics[width=\linewidth]{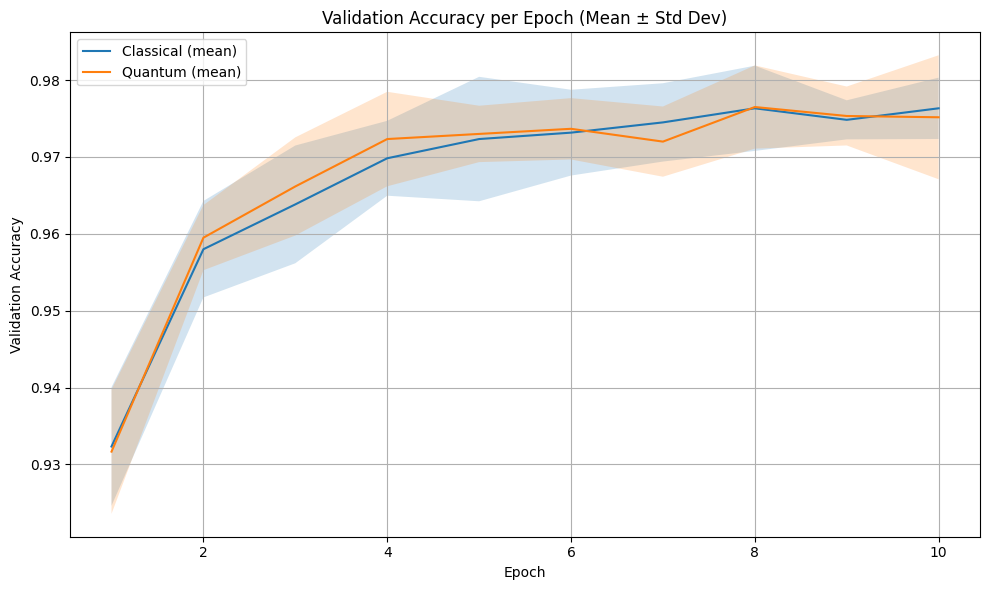}
  \end{minipage}

  \caption{Classical training accuracy over epochs}
  \label{fig:classical_curves}
\end{figure}

Here, we provide training losses and accuracies for the different types of PQK. First, those of the PQK with two convolutional layers of Type 2 with kernel sizes of 3 and 5, are presented in Figure \ref{fig:extended_learning_curves}. Then, the curves for the PQK with only one convolutional layer and with a kernel size of 3 are displayed in Figure \ref{fig:layer1_curves}. Then, the curves of the Hybrid model are given in Figure \ref{fig:parallel_curves}. For additional references, the curves of the classical CNN are displayed in Figure \ref{fig:p_11} and \ref{fig:classical_curves}.

\subsection{Formalization of the Unitary used} 
This encoding kernel implements delayed encoding of the pixels in the patch. The number of modes $m$ for a kernel of size $k$ is given by:

\begin{equation}
    m = \lceil k^2/2\rceil
\end{equation}

The encoding of pixels in the Type~2 kernel circuit is not a single layer of PS gate assignments. In this circuit, the encoding is done by alternate layers of PS gate and BS assignments. The encoding layer can be seen as
\begin{equation}
    U_{\text{enc}}^{(\text{T2})} = M_{L2}\cdot U_{enc}^{(2)}\cdot M_{L1}\cdot U_{enc}^{(1)}
\end{equation}
where,

\begin{equation}
    U_{enc}^{(1)} = \text{diag} \left( e^{ip_{0}},  e^{ip_{1}}, \dots,  e^{ip_{m-1}} \right)_{m\times m};
    \quad
    U_{enc}^{(2)} = \text{diag} \left( e^{ip_{m}},  e^{ip_{m+1}}, \dots,  e^{ip_{k^2-1}}, 1, \dots, 1 \right)_{m\times m};
\end{equation}

Here, $p_k$ is the $k^{th}$ pixel value from the kernel patch. $U_{enc}^{(1)}$ and $U_{enc}^{(2)}$ are the matrices for the first and second layers of pixel encoding.

Finally, for a model with \(L \) such layers, the overall PQK unitary transformation becomes: 

\begin{equation}
    U_{\text{PQK}} = \left(\prod_{\ell=1}^{L} \left( U_{BS}^{(\ell)} \cdot U_{\text{trainable}}^{(\ell)}  \right)\right) \cdot U_{\text{enc}}^{(\text{T2})}
\end{equation}

We began by varying the quantum feature dimensionality. Comparing 5-channel and 20-channel PQK embeddings under identical training settings revealed that the richer 20-channel representation substantially accelerates learning and improves accuracy. Specifically, the 20-channel models reached 90\% validation accuracy within about 15 epochs, while their 5-channel counterparts often required over 80 epochs to achieve a similar level. This confirms that a higher number of quantum-derived features provides greater expressivity, enabling the classifier to capture finer-grained correlations in the image patches.

Next, we examined the encoding strategies. Entropy measurements of the circuit outputs showed that the encoding yields near-maximal entropy (~0.99), indicating noise-like outputs. 

\begin{figure}[h]
  \centering
  \begin{minipage}{0.48\textwidth}
    \centering
    \includegraphics[width=\linewidth]{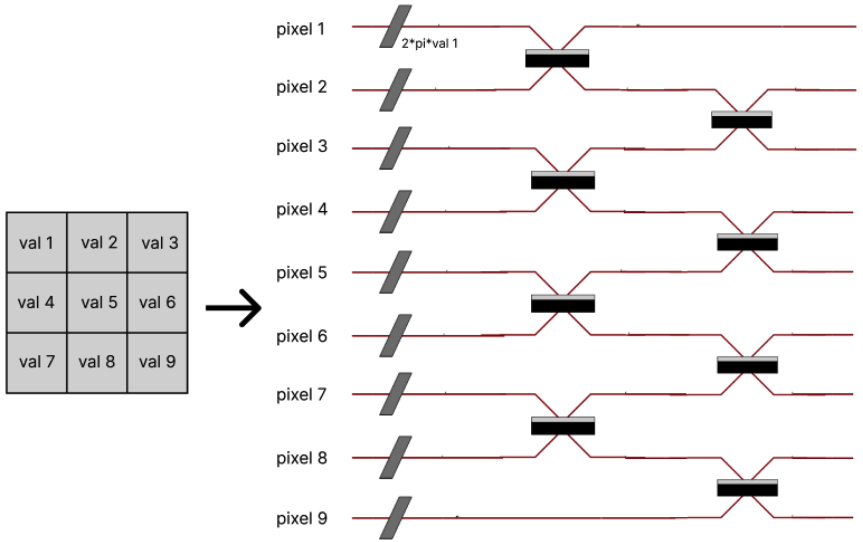}
    \caption{Type~1 PQK circuit: simultaneous encoding (one mode per pixel).}
    \label{fig:pqk_type1}
  \end{minipage}\hfill
  \begin{minipage}{0.48\textwidth}
    \centering
    \includegraphics[width=\linewidth]{figures/qubiteers/PQK_Kernel_Type_2_cropped.png}
    \caption{Type~2 PQK circuit: delayed encoding (half the modes at a time).}
    \label{fig:pqk_type2}
  \end{minipage}
\end{figure}



To assess the necessity of the classical branch, we trained a PQK-only variant by removing the CNN backbone. Despite attaining 98.4\% accuracy on the training set, this model collapsed to 67.5\% on validation, demonstrating severe overfitting. The drop underscores that quantum features alone lack sufficient structure for generalization and that their integration with classical pixel-based features is essential for robust classification.

We also tested the convolution stride used for PQK scanning. Our default stride-2 configuration processes 196 patches per image, while stride-1 scanning generates 729 overlapping patches. Although stride-1 yields slightly smoother quantum feature maps, it does not offer any meaningful boost in validation accuracy but incurs roughly 3.7 times  greater computational cost. As a result, stride-2 remains the optimal choice for balancing performance with efficiency.

Finally, we probed the depth of the PQK interferometer by sweeping the number of beam-splitter layers. Shallow circuits (1–2 layers) underutilize quantum interference and register lower validation accuracy (96.5\%) with low output entropy (0.21). In contrast, overly deep circuits (7–8 layers) produce almost random embeddings (entropy ~0.99) and also degrade accuracy (97.0\%). A middle ground of 3–5 layers achieves both structured entanglement (entropy ~0.49) and peak performance (~99.0\%), confirming that a moderate circuit depth best balances complexity and information preservation.


\section{A convolutional layer using a photonic feature map}
\label{sec:qaradoq_supp}
\subsection{More details on the feature maps and ansatz}
The two-dimension quantum convolution is made of a photonic circuit with two parts: 
\begin{itemize}
    \item a \textbf{feature map} to encode the input using a fixed input Fock state and containing beam splitters (\textit{BS}) and phase shifters (\textit{PS}). The parameters of the phase shifters were fixed or used to encode the input. Two architectures were considered: \textit{Achilles}, which is made of a fixed set of \textit{BS} to dispatch 2 photons over all modes, followed by one \textit{PS} on each mode whose angles encode the input pixels $x_i$ using $(x_i-0.5)\times\frac{\pi}{2}$, and \textit{Odysseus}, made of 27 components (\textit{(BS.H, BS.Ry, and PS)}) parametrized by inputs $x_i$ using three variants: $\{-2\pi x,2\pi x, \sin(2\pi x)\}$
    \item the \textbf{ansatz} consisting of \textit{BS} and \textit{PS} whose parameters are learned during the training. Two architectures were considered here as well: the \textit{Penarddun} architecture consists of a rectangle arrangement of Mach-Zender interferometers with a depth of 6, totalling 48 learnable parameters, and the \textit{Gofanon} architecture which is made up of repeating \textit{BS.H + 2PS and BS.Ry + 2PS} for a total of 96 variable parameters.
\end{itemize}

\begin{figure}[h]
    \centering
    \includegraphics[width=0.8\linewidth]{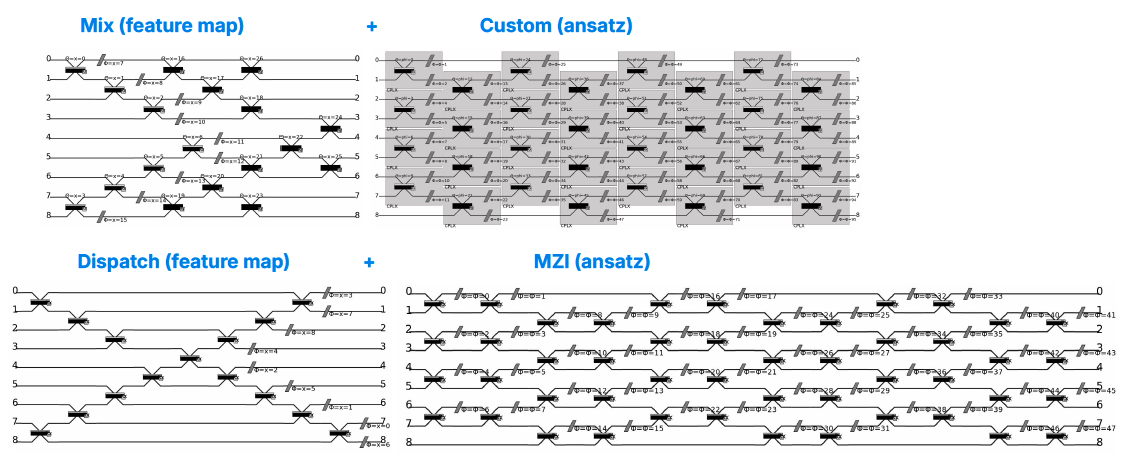}
    \caption{Views of the photonic circuit for feature maps + ansatz}
    \label{fig:qaradoq_implementation}
\end{figure}

\subsection{Study of the output mapping}
\begin{figure}[h]
    \centering
    \includegraphics[width=0.5\linewidth]{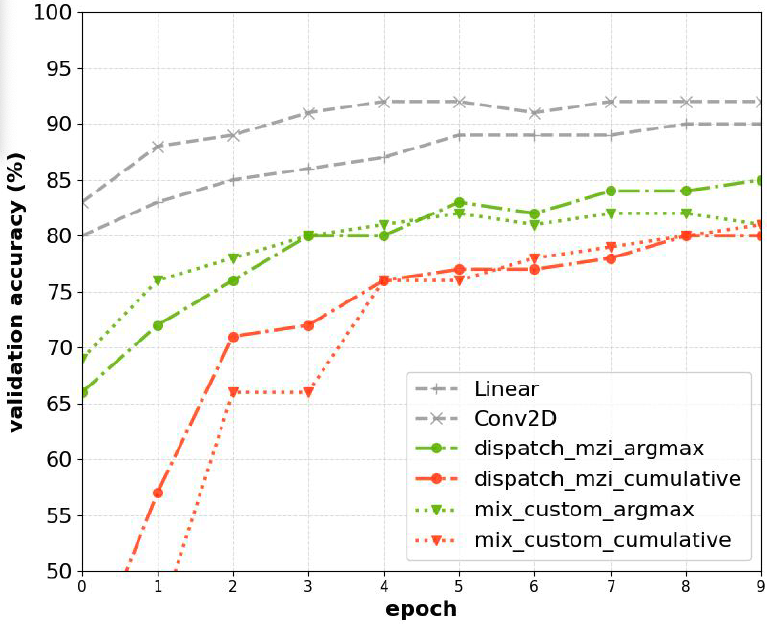}
    \caption{Comparison of training dynamics for the different output mapping strategies}
    \label{fig:qaradoq_output_results}
\end{figure}

Two output mapping methodologies were employed to interface the quantum convolution layer with classical processing stages. In both mapping strategies, the quantum layer produces $m$ matrices related to the photon number distributions over Fock state. The first mapping strategy implements maximum likelihood estimation by selecting the most probable Fock state (seen as an $m$-dimensional vector) for each image patch, generating discrete photon count matrices with integer entries bounded by the number of photons in circuit. The second strategy computes expectation values by performing probability-weighted summation over all Fock states, yielding continuous-valued matrices representing average photon occupancy per mode. From Figure \ref{fig:qaradoq_output_results}, we observe that the first method (argmax) converges faster than the second method in terms of epochs. Moreover, the first method converges faster in terms of training time as well (30 to 50\% faster).

The proposed \verb|qconv2d| achieves efficient training with minimal circuit depth while maintaining spatial localization capabilities. A key finding is that effective MNIST classification can be accomplished using compact feature maps and a single quantum ansatz, whereas classical 2D convolution typically requires multiple kernels. This indicates potential representational efficiency gains in the quantum convolution paradigm.
\section{Photonic Quantum-Train}
\label{sec:qtx_supp}
\subsection{Combinatorial capacity of the architecture and mapping strategy}
\label{sec:qtx_supp_math}

\paragraph{Architecture and combinatorial capacity.}
Consider a target NN with parameter vector $w_{\mathrm{CNN}}=(w_1,\dots,w_m) \in \mathbb{R}^m$. We instantiate two photonic quantum neural networks (QNNs),
\(
\mathrm{QNN}_{1}(\vec{\theta}^{(1)}) \) and \( \mathrm{QNN}_{2}(\vec{\theta}^{(2)})
\),
with $M_1$ and $M_2$ optical modes, respectively. Each device is operated in a fixed-excitation (Hamming-weight) subspace with $N_1$ and $N_2$ excitations.\footnote{Equivalently, one may view $M_i$ two-level modes measured in the Hamming-weight-$N_i$ sector; this yields the binomial dimension $\binom{M_i}{N_i}$. In photonic number-state language, this corresponds to a hard-core (no-bunching) model.} The corresponding numbers of distinct measurement events are
\(
C(M_i,N_i)=\binom{M_i}{N_i}
\).
Let $P_1 \in \Delta_{C(M_1,N_1)}$ and $P_2 \in \Delta_{C(M_2,N_2)}$ denote the outcome-probability vectors (elements of the probability simplices). We form the joint vector by the Kronecker product
\begin{equation}
    P_w \equiv P_1 \otimes P_2 \in \Delta_{C(M_1,N_1)\,C(M_2,N_2)} ,
\end{equation}
and choose the sector sizes to satisfy
\begin{equation}
\label{eq:nm_condition}
    C(M_1,N_1)\, C(M_2,N_2) \;\ge\; m .
\end{equation}
Thus, by steering the QNN controls one can populate at least $m$ effective degrees of freedom for the target NN. In the interferometer meshes used here, the number of continuous controls scales quadratically with the mode count (e.g., $O(M_i^2)$ tunables per device), so that a comparatively small number of quantum parameters governs a combinatorially large set of probabilities.

\paragraph{Mapping probabilities to real-valued weights.}
Because $P_w$ is supported on $[0,1]$ and normalized, while $w_{\mathrm{CNN}} \in \mathbb{R}^m$, we introduce a learnable \emph{mapping model} $G_{\boldsymbol{v}}$ based on a matrix-product state (MPS) \cite{liu2024quantumTN}:
\begin{equation}
    G_{\boldsymbol{v}}:\; [0,1]^{C(M_1,N_1)C(M_2,N_2)} \longrightarrow \mathbb{R}^{C(M_1,N_1)C(M_2,N_2)} .
\end{equation}
Let $\Pi_m$ denote the projection onto the first $m$ components. The target parameters are then defined by
\begin{equation}
    w_{\mathrm{CNN}} \;=\; \Pi_m \bigl( G_{\boldsymbol{v}}(P_w) \bigr),
\end{equation}
i.e., any surplus components beyond $m$ are discarded once the target vector is filled. The task loss $\mathcal{L}=\mathcal{L}(w_{\mathrm{CNN}})$ is evaluated by the classical model, while being implicitly a function of the quantum and mapping parameters $\bigl(\vec{\theta}^{(1)},\vec{\theta}^{(2)},\boldsymbol{v}\bigr)$. Table \ref{tab:qtx_config} shows the configuration of the mapping model.

\begin{table}[t]
    \centering
    \begin{tabular}{lcc}
        \hline\hline
        \textbf{Hyperparameter} & \textbf{Description} & \textbf{Value} \\
        \hline
        Input size & $\left(\ket{\phi_i},\, |\!\braket{\phi_i|\psi(\vec{\theta}^{(i)})}\!|^2\right)$ features \cite{liu2024quantumTN} & $\lceil \log_2 m \rceil + 1$ \\
        Bond dimension & MPS internal dimension & $1$–$10$ \\
        \hline\hline
    \end{tabular}
    \caption{Configuration of the mapping model $G_{\boldsymbol{v}}$.}
    \label{tab:qtx_config}
\end{table}

\subsection{Gradient propagation}
\label{sec:qtx_supp_gradient}

\paragraph{Gradients via the chain rule.}
Let $x\in\{\vec{\theta}^{(1)},\vec{\theta}^{(2)},\boldsymbol{v}\}$ collectively denote the quantum and mapping parameters. Differentiating $\mathcal{L}$ through the generation pipeline yields
\begin{equation}
\label{eq:grad}
\nabla_{x}\,\mathcal{L}
\;=\;
\left( \frac{\partial w_{\mathrm{CNN}}}{\partial x} \right)^{\!\!T}
\;\nabla_{w_{\mathrm{CNN}}}\mathcal{L},
\end{equation}
where $\partial w_{\mathrm{CNN}}/\partial x$ is the Jacobian capturing the sensitivity of the classical weights to the underlying quantum controls and to the mapping parameters. For hardware execution, the entries involving quantum controls are estimated with parameter-shift rules (and variants) for gates with suitable generators \cite{mitarai2018quantum, schuld2019evaluating}.

\paragraph{Parameter updates.}
With learning rate $\eta>0$, a first-order update reads
\begin{equation}
\label{eq:update}
\vec{\theta}^{(i)}_{t+1} \;=\; \vec{\theta}^{(i)}_{t} \;-\; \eta\, \nabla_{\vec{\theta}^{(i)}} \mathcal{L},
\qquad
\boldsymbol{v}_{t+1} \;=\; \boldsymbol{v}_{t} \;-\; \eta\, \nabla_{\boldsymbol{v}} \mathcal{L}.
\end{equation}
In our implementation, $\boldsymbol{v}$ is optimized with ADAM, while $\vec{\theta}^{(i)}$ are tuned with COBYLA (derivative-free) when gradients are noisy or costly to evaluate. Figure~\ref{fig:qtx_scheme} provides a schematic of the photonic QT pipeline; detailed hyperparameters are given below.

\subsection{More details on the photonic implementation}
\label{sec:qtx_supp_photons}

We implement the photonic QNN with a programmable multi–mode interferometer realized as a rectangular mesh of nearest–neighbour two–mode Mach–Zehnder Interferometers (MZIs), each composed of two balanced beam splitters and internal/external phase shifters. This architecture follows the decomposition of Clements \emph{et al.}~\cite{clements2016optimal}, which provides an efficient, fully parameterized factorization of any $m\times m$ unitary $U$ into $m(m-1)/2$ two–mode blocks with interleaved single–mode phases, arranged in $O(m)$ layers (linear optical depth).

Formally, write
\begin{equation}
    U = \left[
        \prod_{\ell=L}^{1}
        \left(
            D_{\mathrm{out}}^{(\ell)}
            \prod_{(i,j) \in \mathcal{P}_\ell}
            B_{(i,j)}(\theta_\ell, \phi_\ell)
        \right)
    \right]
    D_{\mathrm{in}}.
\end{equation}

where $B_{(i,j)}(\theta,\phi)$ acts nontrivially only on modes $i$ and $j$, $D_{\text{in}}$ and $D_{\text{out}}^{(\ell)}$ are diagonal phase shifts, and $\{\mathcal{P}_\ell\}_{\ell=1}^{L}$ is a sequence of disjoint nearest–neighbour pairs implementing the rectangular mesh. Each two–mode block is an SU(2) transformation with real angle $\theta$ (effective reflectivity) and phase $\phi$:
\begin{equation}
\label{eq:bs_block}
    B(\theta,\phi)
    \;=\;
    \begin{pmatrix}
        \cos\theta & -e^{-i\phi}\sin\theta \\
        e^{i\phi}\sin\theta & \cos\theta
    \end{pmatrix},
\end{equation}
and adjacent single–mode phase shifters provide full U(2) freedom on each pair. The construction uses $m(m-1)/2$ two–mode blocks, guaranteeing universality for any target $U$ at fixed mode count, with an optical depth that scales linearly in $m$.

\paragraph{Experimental workflow.}
We initialize $m$ input modes in QNN–specified photonic states (e.g., single–photon Fock states for fixed–excitation sectors). The state then propagates through alternating layers of $B(\theta_\ell,\phi_\ell)$ blocks and diagonal phase shifters in a checkerboard pattern. If $\hat a_i^\dagger$ and $\hat a_j^\dagger$ create photons in modes $i$ and $j$, the action of a single two–mode unit in layer $\ell$ is
\begin{equation}
\begin{pmatrix}
\hat a_i^\dagger \\[2pt] \hat a_j^\dagger
\end{pmatrix}
\longrightarrow
\begin{pmatrix}
\cos\theta_\ell & e^{-i\phi_\ell}\sin\theta_\ell \\
-e^{i\phi_\ell}\sin\theta_\ell & \cos\theta_\ell
\end{pmatrix}
\begin{pmatrix}
\hat a_i^\dagger \\[2pt] \hat a_j^\dagger
\end{pmatrix},
\end{equation}
followed by mode–local phase shifts. Repeating across all layers realizes the global $U$ \emph{in situ}, enabling arbitrary multi–mode transformations required for QNN training.

\subsection{Parameter efficiency in Photonic QT}
\label{sec:qtx_supp_parameter}
To evaluate the parameter efficiency of the photonic QT framework, we implement a classification task based on the Quandela challenge using a subset of the MNIST dataset. The baseline target model is a classical convolutional neural network (CNN) comprising $m = 6690$ trainable parameters. Following the quantum parameter generation scheme described previously, we employ two photonic QNNs with configurations $(M_1 = 9, N_1 = 4)$ and $(M_2 = 8, N_2 = 4)$. These yield $C(9,4) = 126$ and $C(8,4) = 70$ distinct measurement outcomes, respectively. Thus, the joint space produces $126 \times 70 = 8820$ candidate parameters, from which the first 6690 values are selected to initialize the classical CNN weights.

The total number of trainable quantum parameters is $108 + 84 = 192$, corresponding to the internal degrees of freedom in the two interferometers. Additionally, the matrix product state (MPS) mapping model contributes further trainable parameters, governed by its bond dimension $D$. We vary the bond dimension from $D = 1$ to $D = 10$ to examine its effect on model performance.

\begin{figure}[t]
\centering
\includegraphics[width=\linewidth]{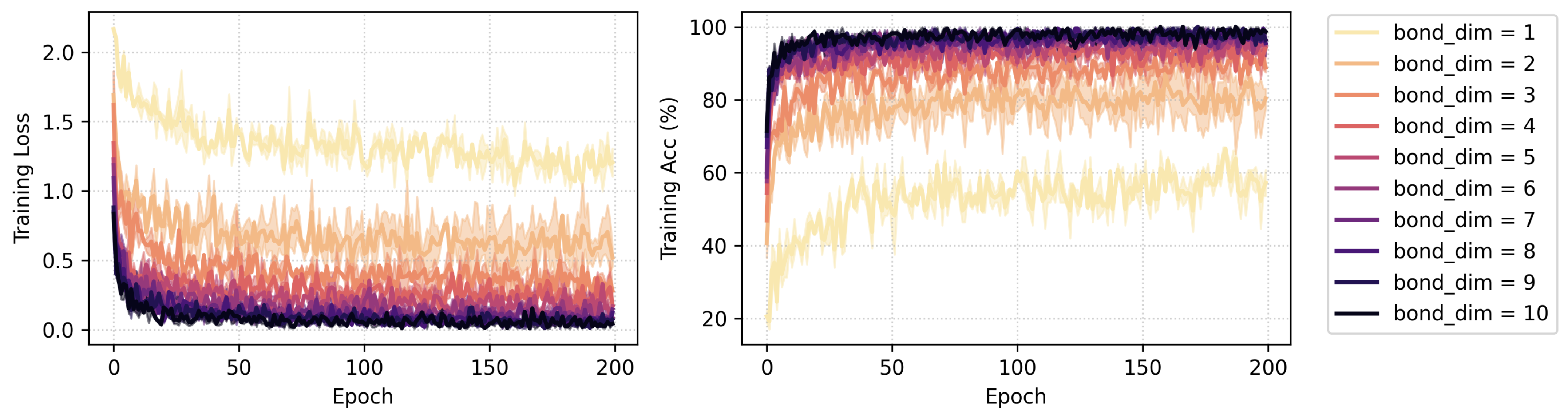}
\caption{Training loss (left) and accuracy (right) over 200 epochs for various MPS bond dimensions. Higher bond dimensions achieve better optimization and accuracy\cite{chen2025distributed}}.
\label{fig:train_loss_acc}
\end{figure}

Figure~\ref{fig:train_loss_acc} illustrates the training loss and accuracy over 200 epochs for various bond dimensions. The left panel shows that models with higher bond dimensions achieve consistently lower training loss, indicating enhanced expressiveness and optimization. The right panel confirms this trend, as training accuracy improves with increasing bond dimension and saturates at high performance.

The last row of Table \ref{tab:diff_bd_results} summarizes the performance of the reference classical CNN model. It achieves near-perfect training accuracy ($99.98\%$) and high testing accuracy ($96.89\%$) using all 6690 parameters. This establishes a benchmark against which we compare photonic QT and classical compression baselines.

\begin{table}[htbp]
\centering
\begin{tabular}{cccc}
\toprule
\textbf{\# of Parameters} & \textbf{Train Acc. (\%)} & \textbf{Test Acc. (\%)} & \textbf{Gen. Error} \\
\midrule
6690 & $99.983 \pm 0.02$ & $96.890 \pm 0.31$ & $0.1690 \pm 0.005$ \\
\bottomrule
\end{tabular}
\caption{Performance of the original classical CNN.}
\label{tab:original_cnn}
\end{table}

Table~\ref{tab:original_cnn} reports the performance of the photonic QT framework across bond dimensions $D = 1$ to $10$. As $D$ increases, the total number of parameters grows from 223 to 3292. Testing accuracy improves substantially, from $55.8\%$ to $95.5\%$, approaching the classical baseline. However, the generalization error also increases, reflecting a trade-off between model capacity and overfitting. For example, the lowest bond dimension ($D=1$) yields the smallest generalization error ($0.0219$), while $D=10$ gives the highest ($0.2552$).

\begin{table}[htbp]
\centering
\begin{tabular}{ccccc}
\toprule
\textbf{Bond Dim.} & \textbf{\# Params} & \textbf{Train Acc. (\%)} & \textbf{Test Acc. (\%)} & \textbf{Gen. Error} \\
\midrule
1 & 223 & $58.26 \pm 2.34$ & $55.78 \pm 3.27$ & $0.0219 \pm 0.007$ \\
2 & 316 & $83.34 \pm 2.77$ & $81.38 \pm 2.28$ & $0.0462 \pm 0.032$ \\
3 & 471 & $88.69 \pm 1.67$ & $87.06 \pm 2.66$ & $0.0364 \pm 0.016$ \\
4 & 688 & $93.92 \pm 0.45$ & $93.29 \pm 0.62$ & $0.0679 \pm 0.002$ \\
5 & 967 & $95.45 \pm 0.39$ & $93.04 \pm 0.77$ & $0.0950 \pm 0.010$ \\
6 & 1308 & $96.95 \pm 0.02$ & $94.92 \pm 0.60$ & $0.1135 \pm 0.013$ \\
7 & 1711 & $97.77 \pm 0.22$ & $94.96 \pm 0.82$ & $0.1315 \pm 0.031$ \\
8 & 2176 & $97.87 \pm 0.78$ & $94.71 \pm 0.47$ & $0.1399 \pm 0.007$ \\
9 & 2703 & $98.37 \pm 0.12$ & $94.84 \pm 0.48$ & $0.1624 \pm 0.021$ \\
10 & 3292 & $98.99 \pm 0.34$ & $95.50 \pm 0.84$ & $0.2552 \pm 0.053$ \\
\bottomrule
\end{tabular}
\caption{Performance of photonic QT with varying MPS bond dimensions\cite{chen2025distributed}}
\label{tab:diff_bd_results}
\end{table}
\section{Enrich classical CNN representations}
\label{sec:supp_q2pi}
To understand what the benefit may be of using a boson-sampler-based embedding, we explored how well it separates data classes in high-dimensional feature space. In our analysis, we observed that photon count distributions resulting from images of different classes tend to be highly distinct — often nearly orthogonal — in the embedding space. Even without any classical training, a simple nearest-centroid classifier based on these distributions could perform well above a random baseline.

This behavior is supported by theory: let $U$ be an $m \times m$ unitary interferometer, and suppose we inject $n$ photons into specified input modes. Recall from Section \ref{sec:preliminaries}  that the output distribution over Fock states is given by:
\[
P_U(\vec{n}) = \frac{|\text{Perm}(U_{\vec{n}})|^2}{n_1! \cdots n_m!}
\]
where $U_{\vec{n}}$ is a submatrix of $U$ corresponding to the input/output configuration $\vec{n}$. Variations in image features correspond to variations in the encoded phases in the circuit, which thus define different matrices $U$. Through the permanent function, this can yield very different output photon-count distributions.

A key question is whether boson-sampling embeddings naturally cluster data by class. Because output probabilities are governed by matrix permanents of interferometer submatrices, even small phase changes (from input features) lead to sharp variations in the photon-count distribution. Intuitively, this should map different classes to nearly orthogonal regions of Fock space.

For each sample $x$ with class label $c$, let $P^c_x$ denote the corresponding output distribution. We define the class-average prototype distribution as 
\begin{equation}\label{class_avg}
    \langle P_c\rangle := \frac{1}{|X_c|}\sum_{x\in X_c} P^c_x,
\end{equation}
where $X_c$ is the set of validation samples belonging to class $c$. To quantify separability, we first introduce the general Kullback–Leibler (KL) divergence \cite{Kullback1951} for two discrete distributions $P,Q$ over the same outcome space:
\begin{equation}\label{kl}
KL(P\|Q):=\sum_{i} P(i)\ln\frac{P(i)}{Q(i)}.
\end{equation}
To avoid singularities due to zero probabilities, we compute a smoothed KL divergence by adding a small constant $\epsilon=10^{-12}$ to the two probabilities.
Using this measure, we define for each sample:
\begin{align}
KL_{\text{true}} &= KL(P^c_x \parallel \langle P_c\rangle),\\
KL_{\text{best-wrong}} &= \min_{c'\neq c} KL(P^c_x \parallel \langle P_{c'}\rangle),\\
KL_{\text{diff}} &= KL_{\text{best-wrong}} - KL_{\text{true}}.
\end{align}
If $KL_{\text{true}} \ll KL_{\text{best-wrong}}$, then $x$ is closer (in the KL sense) to its own class centroid than to any other. 
This unsupervised clustering effect suggests that the boson-sampling embedding intrinsically preserves class structure, offering interpretability advantages compared to classical embeddings.

\begin{figure}[ht]
  \centering
  \includegraphics[width=0.75\textwidth]{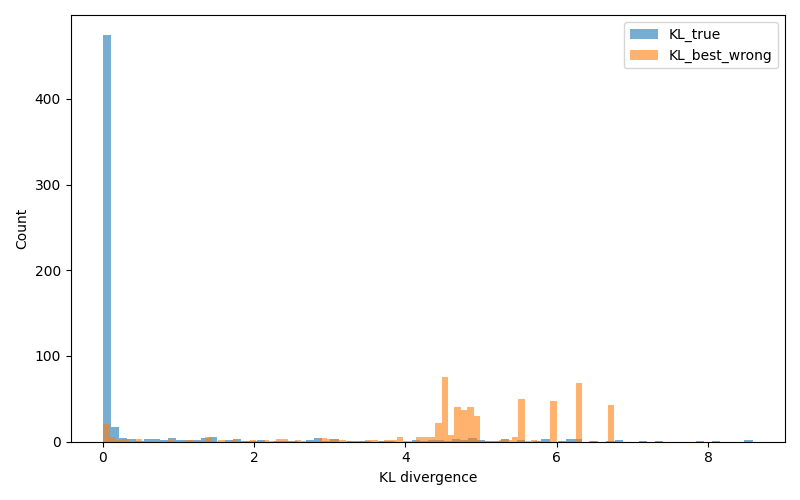}
  \caption{The histogram compares $KL_{\text{true}}$ (blue) and $KL_{\text{best-wrong}}$ (orange). The blue distribution is concentrated near zero, clearly left-shifted relative to orange, confirming that samples are consistently closer to their true prototype.}
  \label{fig:kl_hist}
\end{figure}

\begin{figure}[ht]
  \centering
  \includegraphics[width=0.5\textwidth]{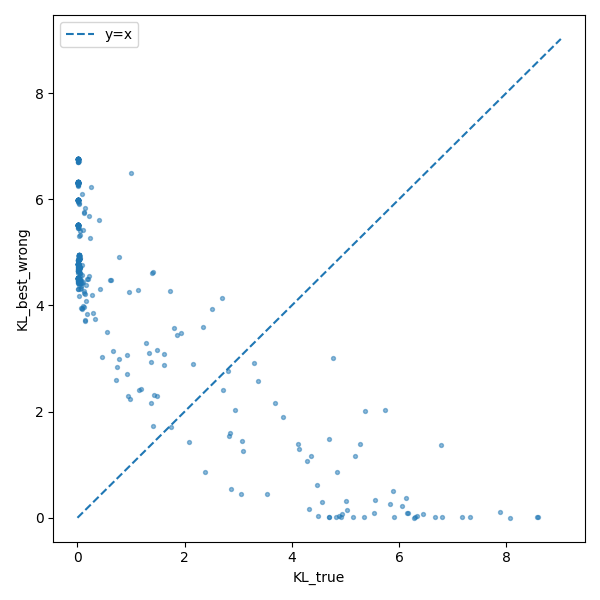}
  \caption{Comparison of the KL-based measures for the validation samples. Most points lie in a region of low $KL_{\text{true}}$ and high $KL_{\text{best-wrong}}$.}
  \label{fig:kl_scatter}
\end{figure}

Across $N=600$ validation samples we computed the following:
\begin{itemize}
\item $89.3\%$ satisfied $KL_{\text{true}} < KL_{\text{best-wrong}}$,
\item mean $KL_{\text{true}} = 0.63$,
\item mean $KL_{\text{best-wrong}} = 4.69$,
\item mean margin $KL_{\text{diff}} = 4.06 \pm 3.14$ nats (95\% CI $[3.80,4.31]$),
\item effect size: Cohen’s $d=1.29$, Wilcoxon $p \approx 3.6\times10^{-70}$.
\end{itemize}
We expand on this further in Figures~\ref{fig:kl_hist} and~\ref{fig:kl_scatter}, together with Table~\ref{tab:per_class_summary}. We note that the unsupervised KL nearest centroid accuracy of $\sim89\%$ is competitive with certain classical kernels of random features applied without supervised training. 
Nevertheless, separability is not uniform across all classes as shown in Table \ref{tab:per_class_summary}. This suggests that combining the boson-sampling embedding with lightweight supervised fine-tuning could further enhance performance.

\begin{table}[ht]
  \centering
  \caption{Per-class KL summary.}
  \label{tab:per_class_summary}
  \small 
  \begin{tabular}{@{} c r c r r r @{}}
    \toprule
    class & n\_samples & acc\_by\_KL & mean\_KL\_true & mean\_KL\_best\_wrong & mean\_KL\_diff \\
    \midrule
    0 & 49 & 0.9388 & 0.3088 & 6.3825 & 6.0737 \\
    1 & 74 & 0.9730 & 0.1678 & 6.1394 & 5.9716 \\
    2 & 67 & 0.9254 & 0.5033 & 5.0455 & 4.5422 \\
    3 & 49 & 0.8980 & 0.6241 & 4.1241 & 3.5001 \\
    4 & 64 & 0.9375 & 0.4253 & 4.1344 & 3.7091 \\
    5 & 66 & 0.8485 & 0.9255 & 3.9167 & 2.9913 \\
    6 & 55 & 0.9636 & 0.2881 & 5.7302 & 5.4422 \\
    7 & 54 & 0.8148 & 1.0155 & 3.9621 & 2.9466 \\
    8 & 52 & 0.7692 & 1.2285 & 3.7839 & 2.5554 \\
    9 & 70 & 0.8429 & 0.9182 & 3.7045 & 2.7863 \\
    \midrule
    \multicolumn{2}{l}{\textbf{Global KL-based accuracy:}} & \multicolumn{3}{c}{\textbf{0.8933}} & \\
    \bottomrule
  \end{tabular}
\end{table}

If $P_x(\mathbf{n})$ is the output distribution for image $x$, and let $\langle P_c \rangle$ be the average distribution for class $c$. Then, for a given test input $x$ from class $c$, we typically found:
 \[
 \mathrm{KL}(P_x \parallel \langle P_c \rangle) \ll \mathrm{KL}(P_x \parallel \langle P_{c'} \rangle), \quad \forall c' \neq c,
 \]
 which indicates that the quantum embedding clusters inputs of the same class around distinct modes in the output space.

 Empirically, we also measured cosine similarities between feature vectors of different classes and found low overlap. This suggests that the boson sampler projects images from different classes into nearly orthogonal directions, a property often desired in kernel methods.

In summary, our empirical findings indicate that the fixed boson sampling embedding offers a powerful mechanism for \emph{unsupervised} class separation, effectively simplifying the task for the downstream classical classifier.



\section{Hybrid Feature Extractor }
\label{sec:supp_solal}

\subsection{Data Preprocessing}
First of all, we process the input MNIST images using Principal Component Analysis (PCA), which allows us to project each image to a smaller dimension $d$, where $d \leq m$, with $m$ being the number of optical modes. This projection is essential as feeding all of the image features ($784$ in our case) is not feasible in practice. 

Following this PCA, we scale the resulting $d$-dimensional vector using a min-max normalization. This maps the vector features to a range $[0,1]$ which meets the requirements of the quantum layer. 

\subsection{Training set-up}
As for the \textbf{training set-up}, all experiments were carried out in simulation with a GPU. Each experiment was repeated 25 times to ensure reliable averages.

Each considered architecture was trained on 6000 MNIST images for 10 epochs, while the testing set contains 1000 MNIST images. During the training, parameter updates were conducted using Adam optimizer with categorical cross-entropy loss.
\section{Transfer Learning}
\label{sec:supp_nomad}
\subsection{Feature Encoding Technique} 
Our amplitude-encoding strategy involves two distinct approaches, designed to test the effectiveness of embedding classical image data into a bosonic quantum system. The main difference between the two is a classical pre-processing step that is applied to the features of the dataset before they are encoded in the quantum circuits.
\begin{enumerate}
    \item \textbf{Classical Linear Encoding.} A linear classical layer is applied on the representation space to transform the representations before injecting them into the bosonic system. This layer is specifically trained to reshape and re-map the input data (e.g., MNIST) to better match the target input space. Unlike feature vectors extracted from pretrained models such as those trained on CIFAR-10, this custom layer adapts to the unique structure and feature distribution of MNIST. This approach attempts to provide a better inductive bias for subsequent classification, yet it still remains fully classical.
    \item \textbf{Quantum Feature Embedding via Linear Optics.} Classical features are mapped into a quantum photonic state using phase shifters applied to specific optical modes. These phase-encoded modes then propagate through a linear optical network composed of MZIs. The motivation for this strategy is to exploit the natural statistical structure and expressivity of boson samplers. We hypothesized that encoding information in this way would allow the beam splitter network to naturally uncover useful correlations in the representation space, due to its high-dimensional interference pattern. However, due to the linearity and passive nature of the optical circuit, this encoding may not optimally preserve class-separability or inject the necessary nonlinear transformations for effective learning.
\end{enumerate}

\noindent\textbf{Retrospect:} The results of our experiments provide a clear and consistent picture of the limitations faced when incorporating static boson sampling layers into a transfer learning pipeline aimed at MNIST digit classification. Across the most difficult transfer learning setups (where the dataset contains 10 classes)—whether using ResNet18 pretrained on ImageNet or CIFAR-10, or using shallow vanilla CNNs—the quantum models with boson sampling consistently underperformed, yielding classification accuracy worse than classical models but significanlty better than random guessing. For simpler tasks of binary classification of both distant (e.g., 1 vs 8) and similar (e.g., 3 vs 5) digit classes the performance of the two models (classical or quantum) was comparable with the achieved validation accuracies being identical.


In stark contrast, classical transfer learning models not only achieved accuracies well above chance but often approached or reached 100\% on MNIST, even without fine-tuning. This points to a clear expressivity mismatch between classical convolutional features and the fixed transformation implemented by the boson sampling layer.

A central insight from these observations is the trade-off between expressivity and trainability. Classical architectures like ResNet18 have been honed to extract hierarchically rich features from image data. In contrast, boson sampling circuits apply fixed, non-trainable linear optics transformations, which operate in a vastly different representation space based on quantum interference. Without the capacity for alignment with the learned classical features, these quantum layers often fail to act meaningfully, instead injecting randomness that disrupts rather than enhances classification.

Moreover, reducing the classical architecture to a minimal vanilla CNN further exacerbates this misalignment. With only one or two convolutional layers and a linear head, the feature extraction is significantly limited, making it even less likely that the quantum layer will find anything useful to amplify or transform meaningfully.


Additionally, the effectiveness of the quantum encoding strategy cannot be overstated. If classical-to-quantum encoding fails to preserve the structure embedded in classical features, then the boson sampler effectively operates on noise. In our experiments, both encoding strategies—(1) a linear classical layer to reshape data into a format expected by the quantum model and (2) direct encoding of classical features into optical phases—did not lead to performance improvement. This suggests that the encoding stage plays a critical role and may act as a limiting factor for the performance of the model.

Noise is another concern, especially when considering potential implementation on near-term quantum hardware. Even in simulations, the lack of error correction or regularization mechanisms may lead to performance degradation. While classical models benefit from robust training mechanisms, over-parametrization, and redundancy, quantum models are fragile and prone to performance collapse from small perturbations.



Future work should consider more expressive and adaptive quantum architectures. Trainable quantum circuits could allow gradient-based optimization and better alignment with classical layers. Another promising direction involves improving classical-to-quantum encoding, perhaps by learning the encoding itself via an auxiliary network. Furthermore, analyzing gradient flow across the hybrid model could uncover bottlenecks and suggest architectural changes to improve synergy. Finally, reevaluating the role of the boson sampler—not as a classifier but as a pre-processing feature extractor—may uncover new roles for static quantum optics in machine learning pipelines.


\end{document}